\documentclass[12pt,a4paper]{article}
\usepackage{graphicx,color}
\usepackage{amsmath}
\usepackage{amssymb}
\usepackage{url}
\usepackage{subfigure}
\usepackage{authblk}
\usepackage{fullpage}
\usepackage{lscape}

\begin{document}

\title{\LARGE{Technical Design Report (TDR):\\
Searching for a Sterile Neutrino at J-PARC MLF (E56, JSNS$^2$)}}
\author[1]{\normalsize{S.~Ajimura}}
\author[2]{M.~K.~Cheoun}
\author[3]{J.~H.~Choi}
\author[4]{H.~Furuta}
\author[5]{M.~Harada}
\author[5]{S.~Hasegawa}
\author[4]{Y.~Hino}
\author[1]{T.~Hiraiwa}
\author[6]{E.~Iwai}
\author[7]{S.~Iwata}
\author[8]{J.~S.~Jang}
\author[9]{H.~I.~Jang}
\author[10]{K.~K.~Joo}
\author[6]{J.~Jordan}
\author[11]{S.~K.~Kang}
\author[7]{T.~Kawasaki}
\author[5]{Y.~Kasugai}
\author[12]{E.~J.~Kim}
\author[10]{J.~Y.~Kim}
\author[13]{S.~B.~Kim}
\author[14]{W.~Kim}
\author[4]{K.~Kuwata}
\author[13]{E.~Kwon}
\author[10]{I.~T.~Lim}
\author[15]{T.~Maruyama\footnote{{Spokesperson:(takasumi.maruyama@kek.jp)}}}
\author[4]{T.~Matsubara}
\author[5]{S.~Meigo}
\author[15]{S.~Monjushiro}
\author[10]{D.~H.~Moon}
\author[1]{T.~Nakano}
\author[16]{M.~Niiyama}
\author[15]{K.~Nishikawa}
\author[1]{M.~Nomachi}
\author[3]{M.~Y.~Pac}
\author[15]{J.~S.~Park}
\author[17]{H.~Ray}
\author[18]{C.~Rott}
\author[5]{K.~Sakai}
\author[5]{S.~Sakamoto}
\author[13]{H.~Seo}
\author[13]{S.~H.~Seo}
\author[7]{A.~Shibata}
\author[1]{T.~Shima}
\author[6]{J.~Spitz}
\author[19]{I.~Stancu}
\author[4]{F.~Suekane}
\author[1]{Y.~Sugaya}
\author[5]{K.~Suzuya}
\author[15]{M.~Taira}
\author[20]{W.~Toki}
\author[7]{T.~Torizawa}
\author[21]{M.~Yeh}
\author[18]{I.~Yu}

\affil[1]{\it{\it{\footnotesize {Research Center for Nuclear Physics, Osaka University, Osaka, JAPAN}}}}
\affil[2]{\it{\footnotesize {Department of Physics, Soongsil University, Seoul 06978, KOREA}}}
\affil[3]{\it{\footnotesize{Department of Radiology, Dongshin University, Chonnam 58245, KOREA}}}
\affil[4]{\it{\footnotesize {Research Center for Neutrino Science, Tohoku University, Sendai, Miyagi, JAPAN}}}
\affil[5]{\it{\footnotesize {J-PARC Center, JAEA, Tokai, Ibaraki JAPAN}}}
\affil[6]{\it{\footnotesize {University of Michigan, Ann Arbor, MI, 48109, USA}}}
\affil[7]{\it{\footnotesize{Department of Physics, Kitasato University, Sagamihara 252-0373, Kanagawa, JAPAN}}}
\affil[8]{\it{\footnotesize {Gwangju Institute of Science and Technology, Gwangju, 61005, KOREA}}}
\affil[9]{\it{\footnotesize {Department of Fire Safety, Seoyeong University, Gwangju 61268, KOREA}}}
\affil[10]{\it{\footnotesize {Department of Physics, Chonnam National University, Gwangju, 61186, KOREA}}}
\affil[11]{\it{\footnotesize{School of Liberal Arts, Seoul National University of Science and Technology, Seoul, 139-743, KOREA}}}
\affil[12]{\it{\footnotesize {Division of Science Education, Physics major, Chonbuk National University, Jeonju, 54896, KOREA}}}
\affil[13]{\it{\footnotesize{Department of Physics and Astronomy, Seoul National University, Seoul 08826, KOREA}}}
\affil[14]{\it{\footnotesize{Department of Physics, Kyungpook National University, Daegu 41566, KOREA}}}
\affil[15]{\it{\footnotesize{High Energy Accelerator Research Organization (KEK), Tsukuba, Ibaraki, JAPAN}}}
\affil[16]{\it{\footnotesize{Department of Physics, Kyoto University, Kyoto, JAPAN}}}
\affil[17]{\it{\footnotesize {University of Florida, Gainesville, FL, 32611, USA}}}
\affil[18]{\it{\footnotesize{Department of Physics, Sungkyunkwan University, Gyeong Gi-do, KOREA}}}
\affil[19]{\it{\footnotesize {University of Alabama, Tuscaloosa, AL, 35487, USA}}}
\affil[20]{\it{\footnotesize {Colorado State University, Tuscaloosa, AL, 35487, USA}}}
\affil[21]{\it{\footnotesize {Brookhaven National Laboratory, Upton, NY, 11973-5000, USA}}}   
\maketitle
\vspace*{-1.5in}
\thispagestyle{empty}

\renewcommand{\baselinestretch}{2}
\large
\normalsize

\setlength{\baselineskip}{5mm}
\setlength{\intextsep}{5mm}

\renewcommand{\arraystretch}{0.5}

\newpage

\tableofcontents
\vspace*{0.5in}
\setcounter{figure}{0}
\setcounter{table}{0}
\indent

\newpage

\section{Introduction}
\indent

In this document, the technical details of the JSNS$^2$ (J-PARC Sterile Neutrino
Search at J-PARC Spallation Neutron Source) experiment are described.

The search for sterile neutrinos is currently one of the hottest topics in neutrino physics. The JSNS$^2$ experiment aims to search 
for the existence of neutrino oscillations with $\Delta m^2$ near 1 
eV$^2$ at the J-PARC Materials and Life Science Experimental Facility
(MLF). A 1 MW beam of 3 GeV protons incident on a spallation neutron 
target produces an intense neutrino beam from muon decay at rest. 
Neutrinos come predominantly from $\mu^+$ decay:
$\mu^{+} \to e^{+} + \bar{\nu}_{\mu} + \nu_{e}$.
The experiment will search for $\bar{\nu}_{\mu}$ to $\bar{\nu}_{e}$
oscillations which are detected by the inverse beta
decay interaction $\bar{\nu}_{e} + p \to e^{+} + n$, followed by
gammas from neutron capture on Gd.
The detector has a fiducial volume of 17 tons and is located 24 meters away from the mercury target. JSNS$^2$ offers the ultimate 
direct test of the LSND anomaly.\\

In addition to the sterile neutrino search, the physics program includes cross section measurements with
neutrinos with a few 10's of MeV from muon decay at rest and with 
monochromatic 236 MeV neutrinos from kaon decay at rest. These cross 
sections are relevant for our understanding of supernova explosions and 
nuclear physics.
\section{\setlength{\baselineskip}{4mm} Physics of the JSNS$^2$ experiment}

\subsection{\setlength{\baselineskip}{6mm} Search for $\bar{\nu_{\mu}} \to \bar{\nu_{e}}$ oscillation as a direct and an ultimate test for LSND}

\subsubsection{Experimental status}
~~
Experimental evidence for sterile neutrinos would come from disappearance or appearance of active flavors with a new $\Delta m^2$ inconsistent with $\Delta m^2_{12}$ or $\Delta m^2_{23}$. 
Table \ref{tab:LSNDetc} gives a summary of observed experimental anomalies and their significance.
\begin{table}[h]
\begin{center}
	\begin{tabular}{|l|c|c|c|}
	\hline
	Experiment     & neutrino source      & Signal                                & $\sigma$ \\ \hline \hline
	LSND               & $\pi$ decay at rest & $\bar{\nu}_{\mu}\rightarrow\bar{\nu}_e$  & $3.8\sigma$     \\ \hline
	MiniBooNE       & $\pi$ decay in flight & $\nu_{\mu}\rightarrow\nu_e$         & $3.4\sigma$         \\ \hline
	MiniBooNE       &  $\pi$ decay in flight & $\bar{\nu}_{\mu}\rightarrow\bar{\nu}_e$ & $2.8\sigma$ \\ \hline
	Gallium/SAGE  & e capture         & $\nu_e\rightarrow\nu_x$   & $2.7\sigma$  \\ \hline
	Reactor           &  $\beta$ decay           & $\bar{\nu}_e\rightarrow\bar{\nu}_x$ & $3.0\sigma$   \\ 
	\hline
	\end{tabular}
	\caption{Possible large $\Delta m^2$ anomalies}
        \label{tab:LSNDetc}
\end{center}
\end{table}

The first indication of a possible sterile neutrino was reported by the LSND experiment. LSND reported an excess of $87.9\pm22.4\pm6.0$ $\bar{\nu_e}$ events ($3.8\sigma$) in 1998 \cite{LSND}. The MiniBooNE experiment recently reported observed excesses of $\nu_e, \bar{\nu}_e$ candidates in the 200-1250 MeV energy range in neutrino mode ($3.4\sigma$) and in anti-neutrino mode ($2.8\sigma$) respectively. The combined excess is $240.3\pm34.5\pm52.6$ events, which corresponds to $3.8\sigma$~\cite{cite:MiniBooNE}. It is not clear whether the excesses are due to oscillations, but if they are, both LSND and MiniBooNE indicate a flavor conversion of $\bar{\nu}_{\mu}$ to $\bar{\nu}_e$ at a probability of about 0.003 with a $\Delta m^2$ of $\sim 1$ eV$^2$.

The second indicatation is a deficit observed in the calibration of low energy radio-chemical solar neutrino experiments. The results indicate a deficiency in neutrino event rates.  Monoenergetic neutrino sources ($^{51}$Cr and $^{37}$Ar) were used in these experiments. Their results were presented in terms of the ratio of the observed and predicted rates. The predictions are based on theoretical calculations of neutrino cross sections by Bahcall and by Haxton. The quoted numbers are $R_{obs}/R_{pred}=0.86 \pm 0.05 (\sigma_{Bahcall}), 0.76\pm 0.085 (\sigma_{Haxton})$~\cite{GaAnomaly}.

The final anomaly is the so-called reactor anomaly, a 6\% deficit of detected $\bar{\nu}_e$ from nuclear reactors at baselines less than 100 m. The ratio of the observed and expected rates is $0.927\pm 0.023$, and is based entirely on the re-analysis of existing data. The deficit is caused by three independent effects which all tend to increase the expected neutrino event rate. There have been two re-evaluations of reactor anti-neutrino fluxes and both indicate an increase of flux by about
3\%. The neutron lifetime decreased from 887-899s to 885.7s and thus the inverse $\beta$-decay cross section increased by a corresponding amount. The contribution from long-lived isotopes to the neutrino spectrum was previously neglected and enhances the neutrino flux at low energies~\cite{cite:ReactorAnomaly}.

All these hints have a statistical significance around $3 - 3.8\sigma$ and may be caused by one or more sterile neutrinos with a mass of roughly 1 eV.  If they are due to neutrino oscillation with new mass state $m_4(\sim$eV), the disappearance and the appearance of active neutrinos are related by 
($m_4\gg m_{1,2,3}$ and $U_{s4} \sim 1 \gg U_{e \mu \tau,4}$. )
\begin{eqnarray}
P(\nu_e, \nu_{\mu}\rightarrow\nu_s)&=&-4\sum_{i > j}Re(U_{si}U^*_{\mu,e i} U^*_{sj}U_{\mu,e j})\sin^2\Delta_{ij}                \nonumber \\
 &-& 2\sum_{i > j}Im (U_{si}U^*_{\mu,e i}U^*_{sj}U_{\mu,e j})\sin2\Delta_{ij}
\nonumber \\
P(\nu_{\mu}\rightarrow\nu_e)&=&-4\sum_{i > j}Re(U_{ei}U^*_{\mu i} U^*_{ej}U_{\mu j})\sin^2\Delta_{ij}  \nonumber\\
 &-& 2\sum_{i > j}Im (U_{ei}U^*_{\mu i}U^*_{ej}U_{\mu j})\sin2\Delta_{ij} \nonumber \\
\Delta _{ij}&=&(m^2_j-m^2_i)L/4E_{\nu}  \nonumber 
\end{eqnarray}
For a short baseline experiments $(L(m)/E(MeV)\sim 1)$ and if only one sterile neutrino involved in mixing,\\
\begin{eqnarray}
P(\nu_{e,\mu}\rightarrow\nu_s)&\sim&-4\sum_{j}Re(U_{s4}U^*_{\mu,e 4} U^*_{sj}U_{\mu,e j})\sin^2(m^2_4L/4E_{\nu})  \nonumber\\
 &-& 2\sum_{j}Im (U_{s4}U^*_{\mu,e 4}U^*_{sj}U_{\mu,e j})\sin2(m^2_4L/4E_{\nu})
 \nonumber\\
 &=&4\mid{U_{s4}}\mid^2\mid{U_{\mu,e4}}\mid^2\sin^2(m^2_4L/4E_{\nu})   
\nonumber\\
P(\nu_{\mu}\rightarrow\nu_e)&\sim&-4\sum_{i}Re(U_{e4}U^*_{\mu 4} U^*_{ei}U_{\mu i})\sin^2(m^2_4L/4E_{\nu})  \nonumber\\
 &-& 2\sum_{j}Im (U_{e4}U^*_{\mu 4}U^*_{ej}U_{\mu j})\sin2(m^2_4L/4E_{\nu})
 \nonumber\\
 &=&4\mid{U_{e4}}\mid^2\mid{U_{\mu4}}\mid^2 \sin^2(m^2_4L/4E_{\nu})   
\nonumber\\
\end{eqnarray}
 Thus
$P(\nu_{\mu}\rightarrow\nu_s) \cdot P(\nu_e\rightarrow\nu_s)\sim P(\nu_{\mu}\rightarrow\nu_e)$.

In order for the LSND and MiniBooNE data to be consistent with the sterile neutrino hypothesis,  $\nu_{\mu}$ disappearance at $\Delta m^2\sim$ eV$^2$ should exist in addition to the observed $\nu_e$ deficiencies. So far only several \% level upper limits exist for $\nu_{\mu}$ disappearance and thus some tensions exist~\cite{cite:tension}.\\

\begin{figure}[htbp]
 \centering
 \includegraphics[width=1.0 \textwidth]{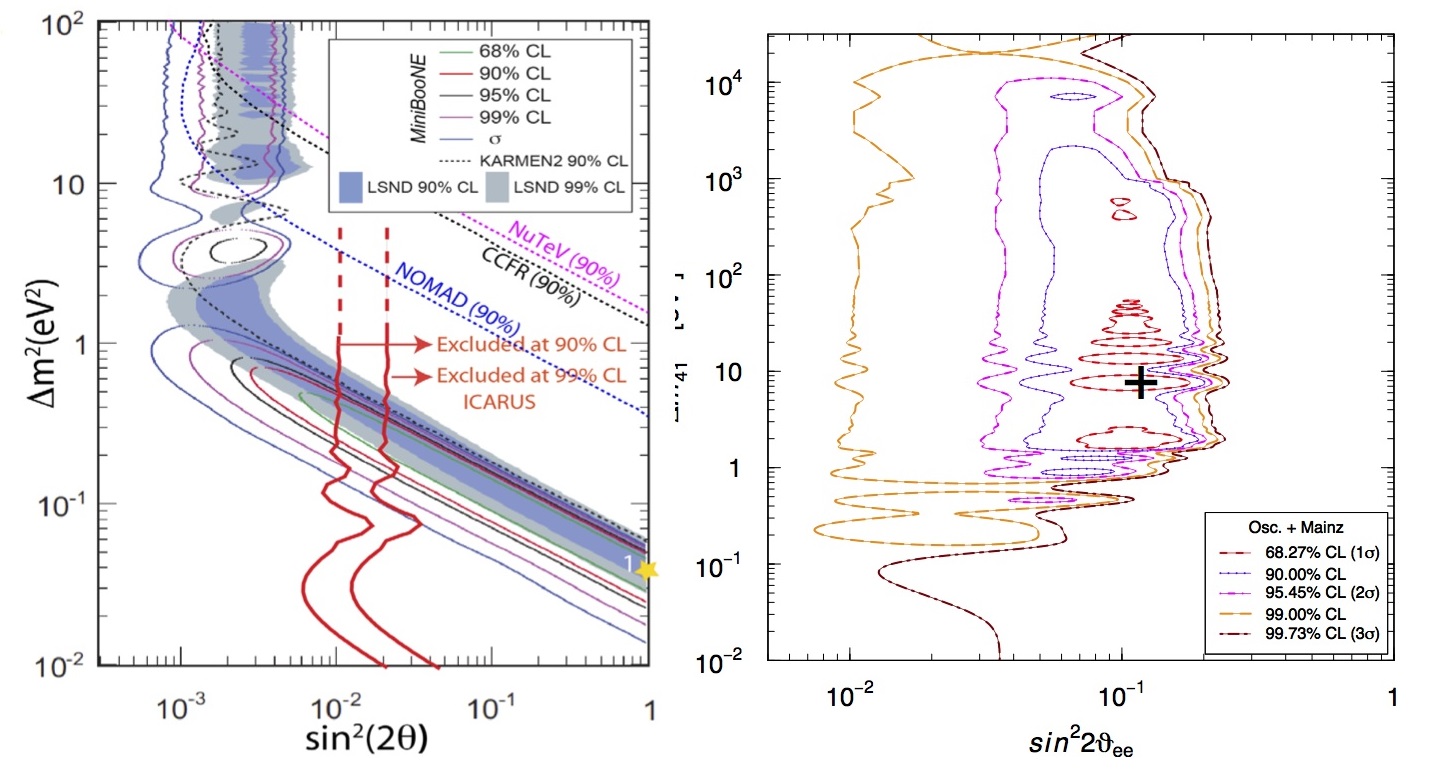}
 \caption{\setlength{\baselineskip}{4mm} 
   Left figure : Allowed region for $\bar{\nu}_{\mu}\rightarrow\bar{\nu}_e$ appearance channel as a result of combining LSND, MiniBooNE and ICARUS~\cite{cite:contmue}.
Right figure : Allowed region for disappearance channel with Reactor and $\beta$ source anomalies, taken into account the KATRIN and neutrino-less double $\beta$ decay limits~\cite{cite:contee}.
 }
 \label{fig:LSNDALL}
 \end{figure}

The allowed regions are shown in Fig. \ref{fig:LSNDALL} for the appearance channel (left figure) ($\bar{\nu}_{\mu}\rightarrow\bar{\nu}_e$)  and for the disappearance channel (right figure) ($\nu_e\rightarrow\nu_s$).

Recently, several results on the sterile
neutrino search have been
updated~\cite{CITE:DB, CITE:NEOS, CITE:SK, CITE:MINOS, CITE:IC} for
$\nu_{\mu}$ and $\nu_{e}$ disappearance modes.
In short, the Daya Bay and NEOS experiments have excluded the region $\Delta m^2 \le 1$ eV$^2$  of the Fig.~\ref{fig:LSNDALL}, and the  MINOS,
Super-Kamiokande, and IceCube experiments have
crucial null results for the $\nu_{\mu}$ disappearance at $\Delta m^2\sim$ eV$^2$.

These results prefer higher $\Delta m^2$ in the global
fit~\cite{CITE:GARIAZZO} and motivate the JSNS$^2$ experiment which is sensitive in the higher $\Delta m^2$ region favored by the fits.

\subsubsection{The JSNS$^2$ Experiment}
\indent

\begin{figure}[htbp]
 \centering
 \includegraphics[width=0.83\textwidth]{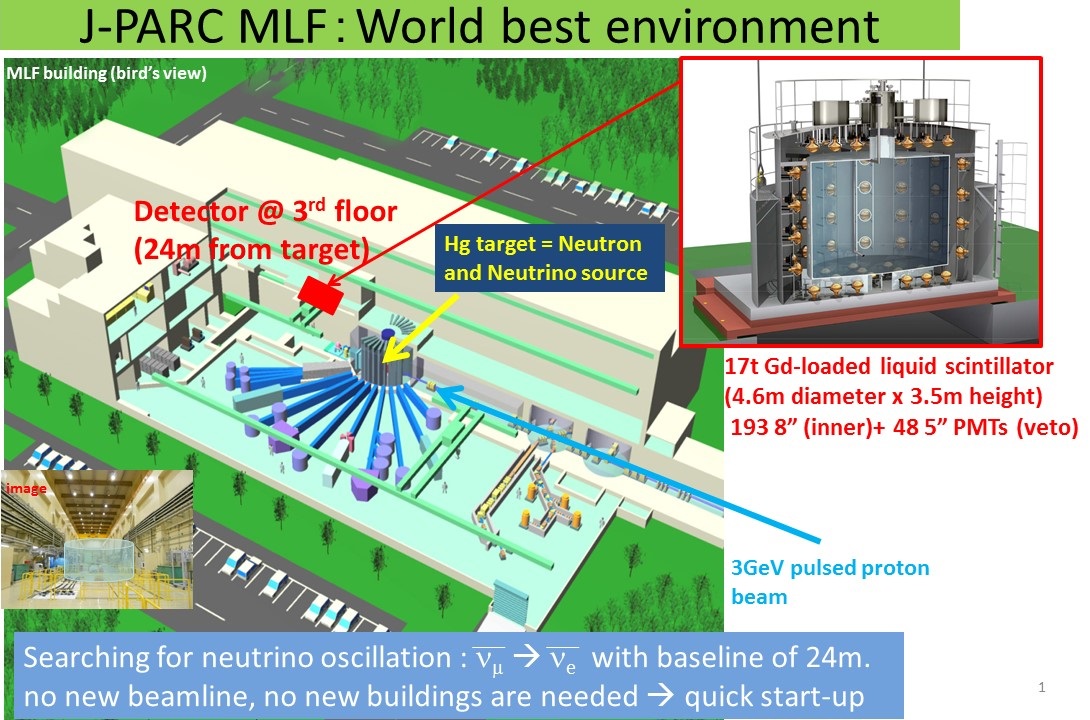}
 \caption{\setlength{\baselineskip}{4mm}
 The MLF building and the overview of the JSNS$^2$ experimental setup. 
}
 \label{fig:setupJSNS2}
\end{figure}

In the context of these global results,
we proposed a search for the existence of neutrino oscillations
with $\Delta m^2$ near 1~eV$^2$ at the J-PARC MLF: JSNS$^2$ (J-PARC Sterile Neutrino Search
at J-PARC Spallation Neutron Source) experiment in 2013~\cite{CITE:PROPOSAL}.  
With the 3 GeV proton beam from the Rapid Cycling Synchrotron (RCS),  and a spallation neutron target, an intense neutrino beam from muon
decay at rest ($\mu DAR$) is available. Neutrinos come predominantly from
$\mu^+$ decay: $\mu^+\rightarrow e^+ +\bar{\nu}_{\mu} +\nu_e$.
We will search for $\bar{\nu}_{\mu} \rightarrow \bar{\nu}_e$ oscillations which are detected via the inverse $\beta$ decay (IBD) 
interaction $\bar{\nu}_e+p \rightarrow e^++ n$ followed by gammas from
neutron capture inside the liquid scintillator.
Figure~\ref{fig:setupJSNS2} shows the overall setup
of the JSNS$^2$ experiment. The detector will be 
placed at a baseline of 24 meters and will contain 17 tons of gadolinium (Gd) loaded
liquid scintillator (LS) inside an inner acrylic vessel, and $\sim$30 tons
unloaded LS in the space between the acrylic vessel and an outer stainless steel tank.
193 8-inch PMTs between the acrylic vessel and stainless tank will view the sensitive inner volume and 48 5-inch
PMTs will be placed in the outer veto region.

This experiment is the ultimate direct test of the LSND experiment because 
it uses the same neutrino source ($\mu$DAR), and the
same neutrino target and interaction (IBD) as LSND,
but with several imrpovements. The signal-to-noise ratio is much better due to the
lower duty factor of the proton beam (0.7 $\mu$s beam pulse / 40000 $\mu$s pulse separation) with repetition at
25 Hz. The use of Gd-loaded liquid scintillator in the target volume also lowers the neutron capture time which lowers accidental backgrounds and improves the signal-to-noise ratio.

\begin{figure}[htbp]
 \centering
 \includegraphics[width=0.7\textwidth]{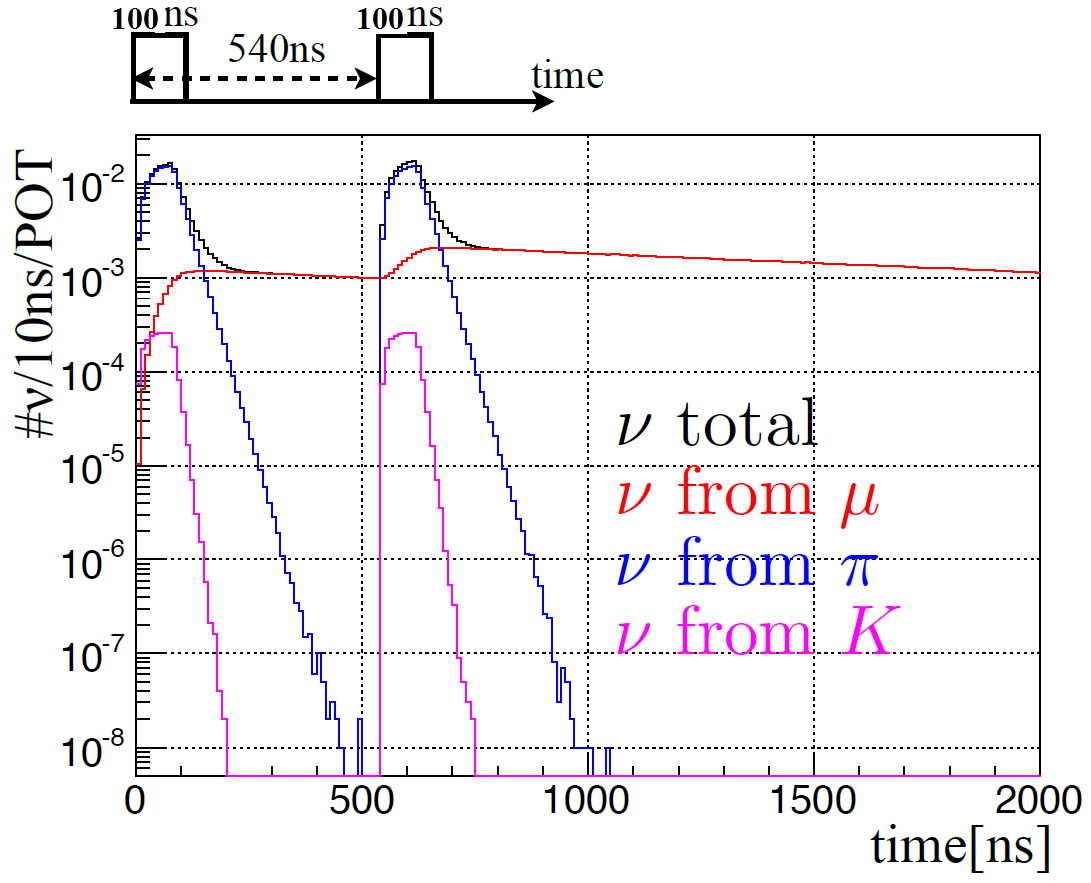}
 \caption{\setlength{\baselineskip}{4mm}
Time distribution of neutrinos from pion, muon and kaon decays is shown. Only neutrinos from muon decay at rest survive after 1 $\mu$s from the start of proton beam.
}
 \label{fig:muDARtime}
\end{figure}

Figure~\ref{fig:muDARtime} shows the merit of using the short pulse, low duty factor beam.
The black square pulse corresponds to the proton beam bunch timing, and 
the time distribution of neutrinos from pion, muon and kaon decays is shown.
Only neutrinos from muon decay at rest survive after 1 $\mu$s from the
start of proton beam. Thus, neutrinos from pion and kaon decays as well as
the beam fast neutrons are eliminated with the 1 $\mu$s timing gate. \\
In addition to the on-bunch timing cuts, a timing gate from 1 to 10 $\mu$s
from the beam starting time is applied for event selection because the
next beam bunch comes after 40 ms (25 Hz). This reduces the cosmic ray
induced background by factor of 9/40000. 
Note that the LSND had 600 $\mu$s beam bunches with 120 Hz operation from
the LINAC beam, and therefore the on-bunch neutrinos and neutrons could not be
removed. Also the beam duty factor was 600 $\mu$s $\times$ 120 Hz = 7.2$\%$,
which is higher than that of the MLF by factor of $\sim$14400.

The Gd-loaded LS is also quite a strong tool for eliminating the accidental
background. The IBD delayed signal from neutron capture gammas in Gd-doped liquid
scintillator gives a shorter coincidence gate (about 30 $\mu$s) due to 
higher capture cross section and higher gamma energy (8 MeV) than
capture on hydrogen ($\sim$ 200 $\mu$s and 2.2 MeV).
The coincidence gate timing reduces the background by a factor 6-7 and the higher
energy of the capture gammas reduces the background by factor of more than 100 due to the 
presence of environmental gammas up to 2.6 MeV. 

\vspace*{0.5cm}

Compared to experiments using the horn-focused beams (e.g.\ the
SBN experiments~\cite{cite:SBN}),
JSNS$^2$ has a few advantages.
First, using neutrinos from the spallation neutron source reduces the intrinsic background from $\bar{\nu}_e$ by a factor of $\sim$10. Second, the energy reconstruction for neutrinos from $\mu$DAR is simple which allows for a precise determination of the energy of candidate oscillation events.

The spallation neutron source is the mercury target, which is a high-$Z$
material, surrounded by thick iron and concrete in the target enclosure as shown in
Fig.~\ref{fig:MLF}.  
Due to strong nuclear absorption of $\pi^-$ and $\mu^-$ in the mercury
target, neutrinos from $\mu^-$ decay are strongly suppressed to about the
$10^{-3}$ level. The resulting neutrino beam is predominantly $\nu_{e}$ and
$\bar{\nu}_{\mu}$ from $\mu^{+}$ with contamination from other neutrino
species at the level of $10^{-3}$. 
For horn-focused neutrino beams, however, it is well-known that the  $\nu_{\mu}$ beam from pions contains $\nu_e$ background at the 1$\%$ level from muon contamination. 

The energy of neutrinos from $\mu DAR$ is quite well known: it is
the Michel spectrum. The energy reconstruction of IBD is also very easy:
$E_{\nu} \sim E_{visible} + 0.8$ MeV where $E_{visible}$ is the visible energy of positron. These two features make the energy information
available for the final analysis which is important because the neutrino oscillation is
a function of the neutrino energy.
On the other hand, a horn-focused beam has large uncertainty on the
neutrino energy spectrum because the parent pion production at the target is not well
understood and it propagates to uncertainty in the energy spectrum. In addition, the reconstruction of neutrino
interactions in the sub-GeV and multi-GeV neutrino energy regions suffers from
uncertainty in nuclear effects, which again gives large uncertainty in the neutrino energy spectrum.

\begin{figure}
 \centering
 \includegraphics[width=1.\textwidth]{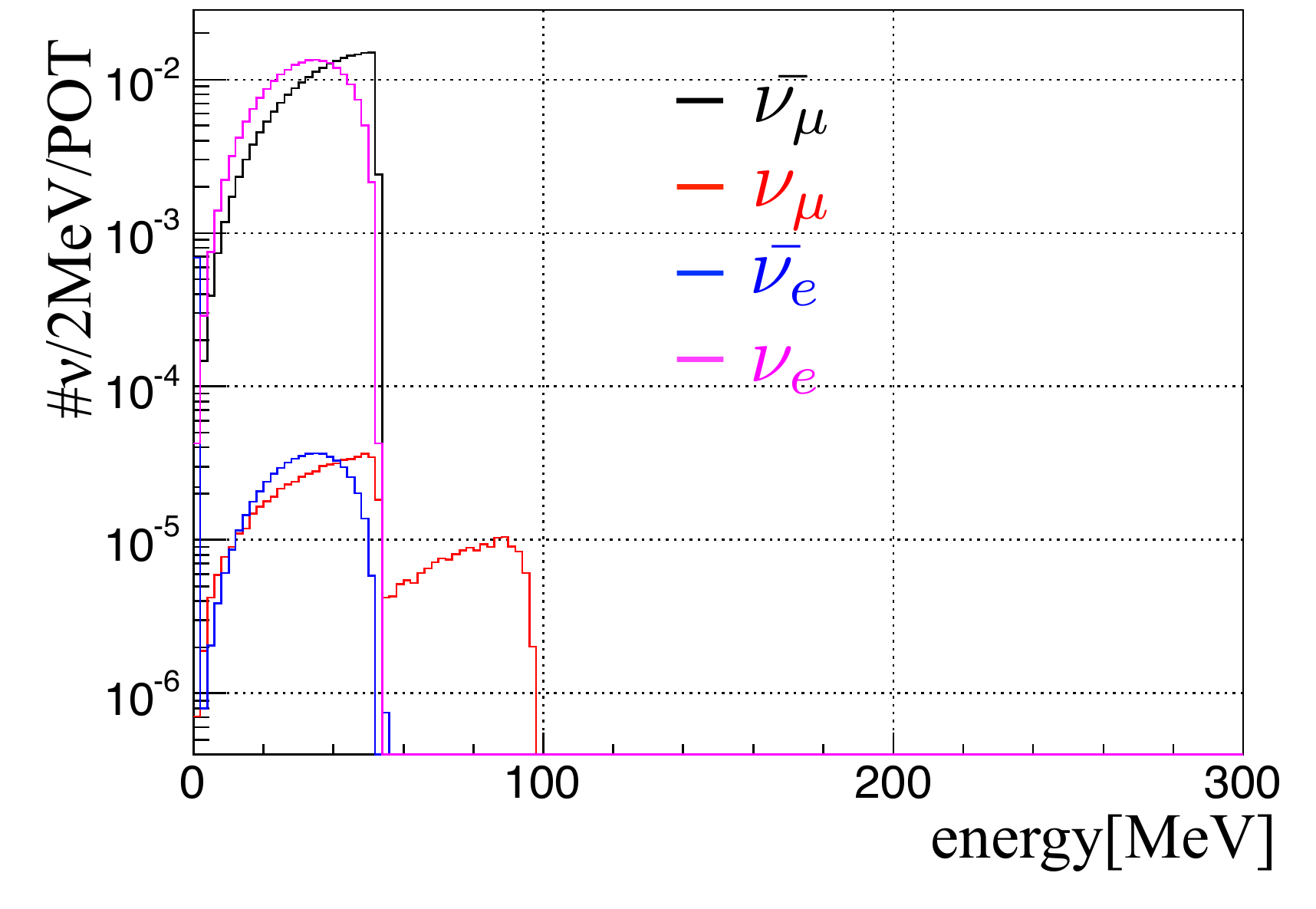}
 \caption{\setlength{\baselineskip}{4mm}
Estimated neutrino flux after 1 $\mu$s from the start of proton beam. The $\mu^{+} DAR$ components are selected and main background come from $\mu^-$ decays. 
}
 \label{fig:muDARflux}
 \end{figure}

The $\mu$ DAR component of the neutrino flux can be selected by gating out the first 1 $\mu$s from
the start of the proton beam. The resulting neutrino fluxes for each type of
neutrino species are shown in Fig.~\ref{fig:muDARflux}.
Note that the resulting $\bar{\nu}_{\mu}$ and $\nu_e$ fluxes
have different spectra with endpoint energy of 52.8 MeV. A possible survived
$\mu^-$ decay will be at the level of $10^{-3}$ and produce $\nu_{\mu}$ and
$\bar{\nu_e}$ with same spectrum as those of $\bar{\nu}_{\mu}$ and $\nu_e$,
respectively.

\subsubsection*{Signatures of the oscillation}
~~

A sensitive search for $\bar{\nu}_e$ appearance ($\bar{\nu}_{\mu} \rightarrow \bar{\nu}_e$ from $\mu^+ DAR$) can be performed by searching for the two-fold signature of $\bar{\nu}_e + p\rightarrow e^{+} + n$ which produces a prompt positron signal with an endpoint of 52.8 MeV followed by gammas due to neutron capture on Gd.

The main background coming from $\mu^-$ decays (as shown in 
Fig.~\ref{fig:muDARflux}) is highly suppressed by $\pi^-$  and $\mu^-$ capture in heavy metals like Hg. However, $\mu^-$ which stop in a light metal such as Be, usually decay before absorption. This background can be estimated
from the $E_{\nu}$ reconstructed distribution, which is well defined and
distinct from oscillated events. 

\begin{figure}
 \centering
 \includegraphics[width=1.0\textwidth]{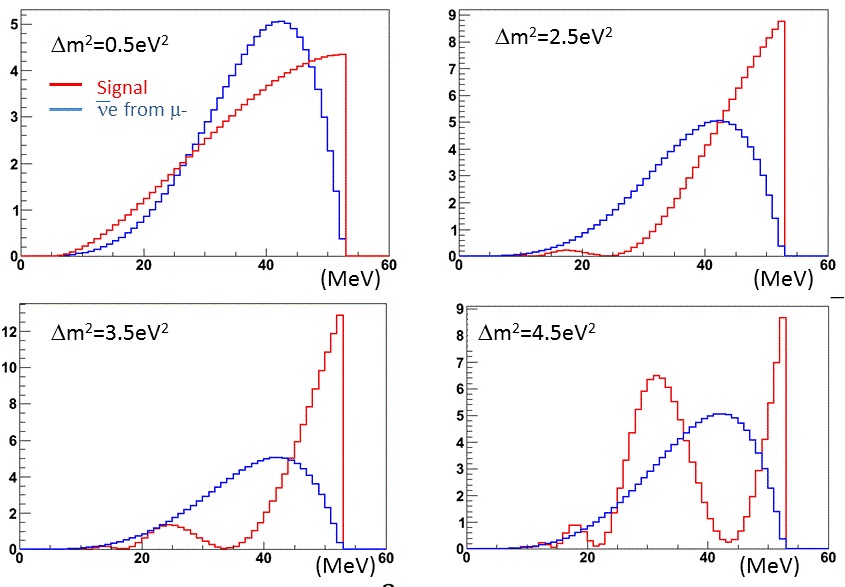}
 \caption{\setlength{\baselineskip}{4mm}
Examples of oscillation signals at typical $\Delta m^2$ for a baseline of 24 m. The red graphs are shapes of $\bar{\nu}_e$ appearance signal and the blue graphs are shapes of signal from $\mu^-$ decays.
All plots are normalized by area. 
 }
 \label{fig:Examposc}
 \end{figure}

Since the oscillation probability is given by
\begin{eqnarray}
P=\sin^2 2\theta \sin^2(\frac{1.27 \Delta m^2(eV^{2}) L(m)}{E_{\nu} (MeV)} )\nonumber ,
\end{eqnarray}
there are two distinct signatures of oscillation signal.
One is the energy spectrum of the oscillated signal, which is a convolution of
the energy spectrum of the original neutrino (in this case, $\bar{\nu}_{\mu}$) and the oscillation probability. The other signature is the distribution of
events as a function of distance from the source. The background $\bar{\nu}_e$ from $\mu ^{-}$ decay has a different spectrum from that of $\bar{\nu}_{\mu}$ oscillations which can be used to distinguish background from signal.
Figure \ref{fig:Examposc} shows $E_{\bar{\nu}}$ distributions of oscillation signals at some typical $\Delta m^2$ values for a baseline of 24 m.

\subsubsection*{Signal identification}
\label{sec_signatures}
~~
The signal for $\bar{\nu_e}$ appearance from 
$\bar{\nu}_{\mu} \rightarrow \bar{\nu}_e$ 
is a primary positron signal followed by delayed signal from neutron capture. 
The primary signal is $\bar{\nu_e} + p \rightarrow e^+ +n$ (IBD) 
and the delayed signal consists of gammas from neutron capture on Gd.
For the normalization of $\mu^+$ decay, $\nu_e + C \rightarrow e+N_{gs}$ events will be measured. The primary signal is an electron and the delayed signal is a positron from $N_{gs}$ $\beta$ decay.

The time gate for the primary signal should be from 1 $\mu$s to 10 $\mu$s, 
corresponding to several muon lifetimes and avoiding pion decay from both 
decay at rest and decay in flight. 
Table~\ref{tab:signal} is a summary of the primary and delayed signals.
\begin{table}[h]
\begin{center}
	\begin{tabular}{|l|c|c|c|c|}
	\hline
	~          & primary   &   primary  & delayed & delayed \\ 
       ~          &  timing         &  energy       & timing      &  energy \\ \hline \hline
$\bar{\nu}_{\mu} \rightarrow \bar{\nu}_e$ & 1-10 $\mu$s & 0-53 MeV & 10-100 $\mu$s & 8 MeV \\ \hline
$\nu_e C\rightarrow e N_{gs} $, $N_{gs}\rightarrow C e^+ \nu_e$&  1-10 $\mu$s & 0-37 MeV & 100 $\mu$s-10 ms & 0-16 MeV \\ \hline \hline
	\end{tabular}
	\caption{The timing and energy of the prompt and delayed signals for IBD events and $\nu_e$ scattering of C.}
	\label{tab:signal}
\end{center}
\end{table}

\subsubsection{Experiment Strategy}
\indent

Our strategy is to put a detector with 17 tons of Gd-loaded LS at a 
baseline of $\sim$24 meters on the third floor of the MLF (maintenance 
area of the mercury target). We aim to start the experiment with one 17-ton 
detector in early 2019 to have world-competitive results soon.
This TDR is dedicated to the construction of one detector, and
its sensitivity only. If the construction of another detector is 
realized with the budget, we will submit an amended or another TDR.

\subsection{\setlength{\baselineskip}{6mm} Measurement of Neutrino-Induced 
Nuclear Reaction Cross Sections}

\subsubsection{Physics Motivations}
\indent

Stars with initial masses greater than $\sim$8-10 times the
mass of the sun are expected to end their lives with core-collapse 
supernovae (Type-II SNe). When such a massive star 
exhausts the nuclear fuel at its center, it produces no more heat 
to sustain its own weight, and materials in the outer layers begin to 
fall into the central core. As the density and the temperature of 
the core rapidly increase, the core is photo-dissociated into 
a mixture of nucleons and light nuclei which absorb free electrons 
by emitting neutrinos, forming a proto-neutron star. 
The core becomes so hard that material falling into it bounces at the 
surface and collides with  still falling matter, generating an outgoing shockwave. 

It has been found through a number of simulations on Type-II SNe 
that the kinetic energy supplied by the core bounce is not 
sufficient for the shockwave to travel to infinity, and the 
additional energy supplied by the interactions between the 
neutrinos and the nuclei contained in the shockwave should 
play a critical role in successful explosion~\cite{SN1}. 
This effect, called neutrino-heating, was studied 
by one-dimensional and two-dimensional simulations 
with different neutrino luminosities, and it was found that 
enhancement of the neutrino luminosity by $\sim$10-15$\%$ 
leads to an increase of the kinetic energy of the shockwave by 
$\sim$10$^{50}$-10$^{51}$ erg/s, which is sufficient for a 
successful explosion~\cite{SN5}. 
Since enhancement of the neutrino-nucleus reaction 
rates are expected to give the same effect as that of the 
neutrino luminosity, the neutrino-nucleus cross sections
should be known with uncertainties smaller than 
$\sim$10-15$\%$. 

Another important role of neutrino-induced nuclear 
reactions is in r-process nucleosynthesis. 
A recent scenario of the r-process assumes 
the formation of a high-entropy, neutron-rich gas, called 
the neutrino-driven wind, in the atmosphere of a nascent 
neutron star by neutrino-induced spallation reactions, 
and the synthesis of heavy elements from protons and 
neutrons up to the nuclides with mass number of 
$\sim$200-250 within about one second~\cite{SN6,SN7,SN8}. 
This scenario is preferred, because it does not require 
the existence of seed nuclei like iron, and naturally 
explains the universality of the r-elements (i.e.\ the similarity 
in the r-element abundances observed in stars with different 
metallicities). More recently, it was pointed out that the 
neutrino-induced spallation reactions on light nuclei such as 
$^{4}$He and $^{12}$C may efficiently produce lithium and 
boron in the oxygen/carbon layer of a Type-II SNe~\cite{SN9}. 
Such a process is interesting as a possible source of Li-Be-B 
in addition to spallation by cosmic rays, and also as a 
new probe to constrain the parameters of flavor oscillation 
in the neutrino sector. For a precise analysis of those light element 
abundances, it is necessary to carry out detailed simulations using 
accurate data of the neutrino-nucleus reaction rates. 

\subsubsection{\setlength{\baselineskip}{6mm} Measurement Plan for the $^{12}$C($\nu_{e}$,e$^{-}$)$^{12}$N Cross Section in JSNS$^2$}
\indent

\begin{figure}[htpb]
 \centering
 \includegraphics[width=0.6 \textwidth]{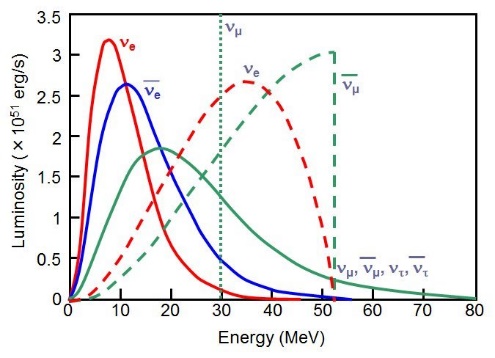}
\caption{\setlength{\baselineskip}{4mm}
Energy spectra of neutrinos from Type-II SNe (solid curves) 
and DAR pions and muons (dashed curves, arbitrary units in 
luminosity). 
}
 \label{Fig:SN_DAR}
\end{figure}

So far, experimental data on the neutrino-induced nuclear reaction 
cross sections has been obtained by using neutrinos produced by 
accelerators or radioactive isotopes. DAR 
neutrinos from stopped pions and muons generated with high-energy 
accelerators are very useful for studies of nuclear reactions 
induced by supernova (SN) neutrinos, because their energy spectra 
overlap with those of SN neutrinos as shown in
Fig.~\ref{Fig:SN_DAR}. 

Table~\ref{TAB:SN1} shows the list of the presently available 
experimental data for neutrino-induced nuclear reaction 
cross sections. 

\begin{table}[htbb]
  \caption{\setlength{\baselineskip}{4mm}
    Summary of the existing data for neutrino-induced 
    nuclear reaction cross sections.}
\begin{center}
\begin{tabular}{|l|c|c|c|} \hline
Reaction & Neutrino Source & Accuracy & Reference \\\hline
$^{12}$C($\nu_e$,e$^-)^{12}$N$_{g.s.}$ & Accelerator $\nu$ & $\sim$10$\%$ & \cite{SN10}\cite{SN11} \\\hline
$^{12}$C($\nu_e$,e$^-)^{12}$N$^{*}$   & Accelerator $\nu$ & $\sim$15$\%$ & \cite{SN10}\cite{SN11} \\\hline
$^{12}$C($\nu$,$\nu^{'})^{12}$C$(1^+1)$ & Accelerator $\nu$ & $\sim$20$\%$ & \cite{SN10}\cite{SN11} \\\hline
$^{13}$C($\nu_e$,e$^-)^{13}$N & Accelerator $\nu$ & 76$\%$ & \cite{SN10} \\\hline
$^{56}$Fe($\nu_e$,e$^-)^{56}$Ni & Accelerator $\nu$ & 37$\%$ & \cite{SN10} \\\hline
$^{71}$Ga($\nu_e$,e$^-)^{71}$Ge & RI ($^{51}$Cr)     & 11$\%$ & \cite{SN12}\cite{SN13} \\\hline
$^{127}$I($\nu_e$,e$^-)^{127}$Xe & Accelerator $\nu$ & 33$\%$ & \cite{SN14} \\\hline
\end{tabular}
\end{center}
\label{TAB:SN1}
\end{table}

As shown in Table~\ref{TAB:SN1}, even in best case, the accuracy of the experimental data on $^{12}$C cross sections is not better than 
what is required for SN simulations, and therefore new experimental 
data with better accuracy is still needed. JSNS$^2$ is expected to 
provide an experimental opportunity to measure the neutrino-induced 
nuclear reaction cross sections with better accuracy thanks to a 
high-intensity neutrino beam from the J-PARC/MLF and a detector 
with excellent sensitivity as well as a high signal-to-noise ratio. 
According to the result of our realistic Monte Carlo simulation based 
on the background data measured near the site of 
the JSNS$^{2}$ detector, a very small statistical error of 6.0$\%$
is expected with three years of running and a fiducial detector mass of 17 tons.
For more details on the estimation of the sensitivity, 
the reader is referred to \cite{CITE:SR_16JAN7}.

\subsection{Physics with neutrinos from charged kaon decay-at-rest}
\indent

JSNS$^2$ has the unique ability to precisely measure monoenergetic 236~MeV neutrinos from charged kaon decay-at-rest (KDAR) ($K^+ \rightarrow \mu^+ \nu_\mu$; BR=63.5\%~\cite{pdg}) for the first time. These neutrinos represent (1) an unprecedented weak-interaction-only, known energy probe of the nucleus, (2) a standard candle for developing a thorough understanding of the neutrino interaction and cross sections critical for future long baseline neutrino experiments, and (3) a source for a sterile neutrino search using electron neutrino appearance~\cite{kdar1,kdar2}. These neutrinos have also been cited as important for probing muon neutrino disappearance at short baseline~\cite{kdar3} and as a possible dark matter annihilation signature~\cite{kumar}. 

Despite the importance of the KDAR neutrino across multiple aspects of particle and nuclear physics, these neutrinos have never been studied or even identified before. The decay-in-flight neutrino ``background" in conventional beamlines drowns out the KDAR signal in such experiments. Decay-at-rest sources of neutrinos, most notably spallation neutron sources, are excellent locations for studying KDAR due to their minimal decay-in-flight background and intense beams. However, the historically most intense spallation sources have been too low energy to produce kaons readily. The J-PARC MLF 3~GeV primary proton energy is sufficient to produce kaons efficiently and, also in consideration of the facility's beam intensity (eventually 1~MW, currently 500~kW~\cite{mlf_news}), represents the best facility in the world to accomplish this physics. The KDAR neutrino can easily be seen in Fig.~\ref{fig:sense}, which shows the neutrino flux at the J-PARC MLF source.  

JSNS$^2$ expects to collect a sample of between 30,000 and 60,000 $\nu_\mu$ charged current events in 17~tons of fiducial volume in its 3 year run\footnote{
  \setlength{\baselineskip}{4mm} The large variation in the expected number of events is due to the highly uncertain kaon production at this energy. The lower and upper bounds come from Geant4~\cite{geant4} and MARS~\cite{mars} predictions, respectively.}. These events ($\nu_\mu n \rightarrow \mu^-  p$ or $\nu_\mu  \mathrm{^{12}C} \rightarrow \mu^- X$) are easily identifiable due to the characteristic double coincident signal of the prompt muon plus proton(s)/nucleus followed by the muon decay electron ($\mu^- \rightarrow e^- \overline{\nu}_e \nu_\mu$) a few $\mu s$ later. 

\begin{figure}[htpb]
\centering
\begin{minipage}{.75\linewidth}
  \includegraphics[width=\linewidth]{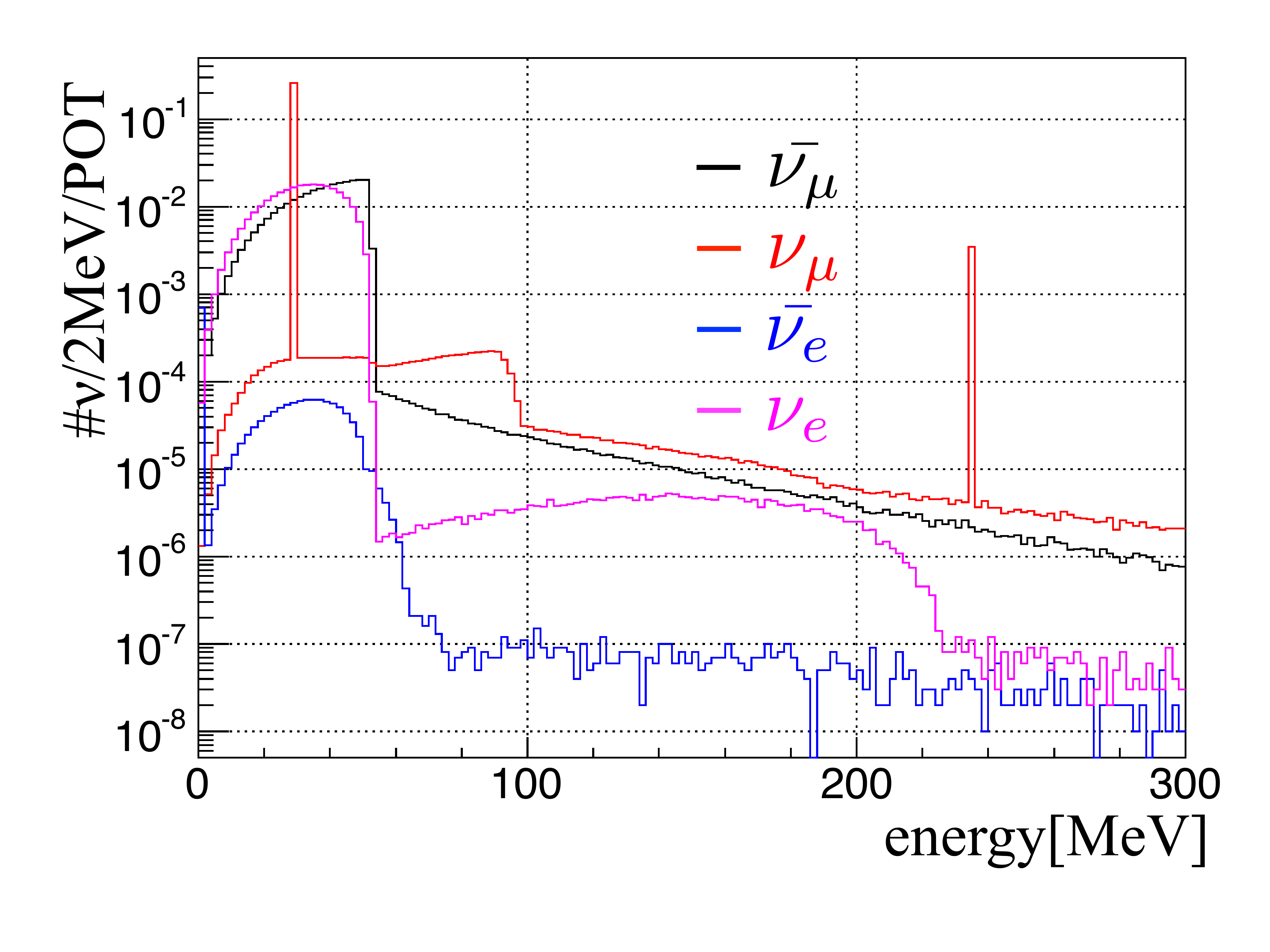}
\end{minipage}
\vspace{-.3cm}
\caption{\setlength{\baselineskip}{4mm}
  The neutrino flux at the J-PARC MLF source without timing cuts. The 236~MeV muon neutrino from charged kaon decay-at-rest can easily be seen.}
\label{fig:sense}
\end{figure}

\begin{figure}[htpb]
\begin{centering}
\includegraphics[height=4.8in]{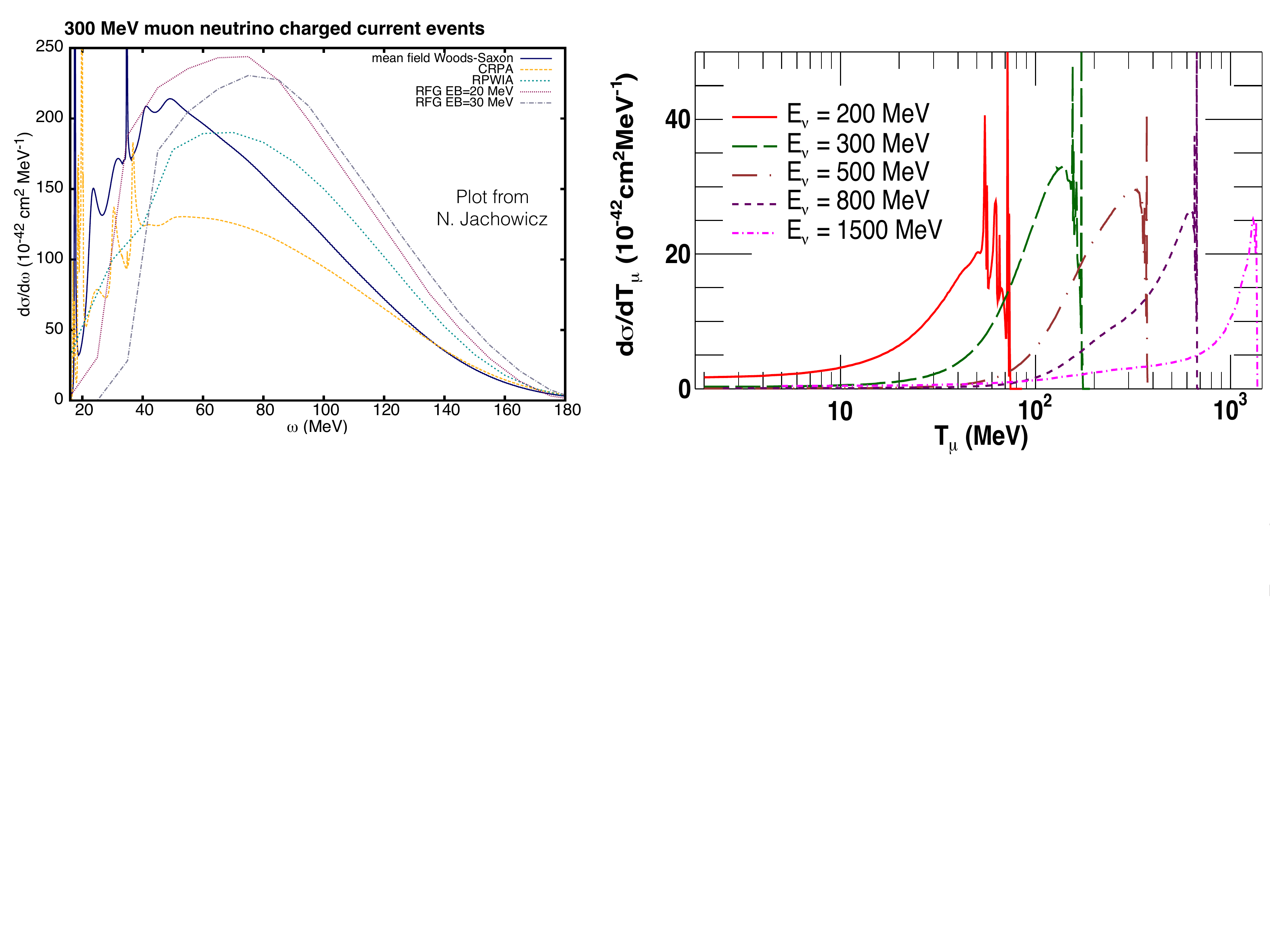} 
\vspace{-6.7cm}
\caption{\setlength{\baselineskip}{4mm}
  Left: The differential cross section in terms of energy transfer ($\omega=E_\nu-E_\mu$) for 300~MeV $\nu_\mu$ CC scattering on carbon. Predictions from various models are shown. This plot is adapted from Ref.~\cite{natalie}. Right: The differential cross section in terms of muon kinetic energy for various neutrino energies within the Continuum Random Phase Approximation model; this plot is adapted from Ref.~\cite{crpa}.}
  	\label{fig:kinematics1}
\end{centering}
\end{figure}

The known energy KDAR neutrinos provide the exclusive tool, for the first time, to study nuclear structure and the axial vector component of the interaction using electron scattering variables such as $\omega$ ($\omega=E_\nu-E_\mu$). The importance of this unique access to the nucleus is potentially far-reaching. For example, a double differential cross section measurement in terms of $\omega$ vs. $Q^2$ allows one to distinguish effects of the form factors, which depend only on $Q^2$, and of the nuclear model, which depends on both. Figure~\ref{fig:kinematics1} (left) shows a number of model predictions for the differential cross section in terms of energy transfer for 300~MeV $\nu_\mu$ CC scattering on carbon. The disagreement between the models, in terms of both shape and normalization, is striking. Notably, the JSNS$^2$ muon energy resolution may allow the nuclear resonances, easily seen in Fig.~\ref{fig:kinematics1} (right), to be measured via neutrino scattering. The KDAR neutrino is likely the only way to study these excitations with neutrino scattering and, in general, to validate/refute these models in the $<400$~MeV neutrino energy range (see, e.g., Ref.~\cite{teppei}).

Along with studying nuclear physics relevant for future neutrino experiments, the large sample of KDAR muon neutrinos collected with JSNS$^2$ will provide a standard candle for understanding the neutrino energy reconstruction and outgoing lepton kinematics in the 100s-of-MeV neutrino energy region. While the KDAR neutrino is simply not relevant for experiments featuring significantly higher neutrino energies, most notably for MINOS, NOvA and DUNE~\cite{minos,nova,dune}, it is highly relevant for experiments with a large or majority fraction of few-hundred-MeV neutrinos, for example, T2K~\cite{t2knim}, MOMENT~\cite{moment}, the European Spallation Source Neutrino Super Beam (ESS$\nu$SB)~\cite{essnu}, and a CERN-SPL-based neutrino beam CP search~\cite{beta_spl}. In particular, MOMENT and ESS$\nu$SB both feature $\nu_\mu$ spectra which peak at about 200-250~MeV.

The KDAR neutrino can also be used to search for electron neutrino appearance ($\nu_\mu \rightarrow \nu_e$) for providing a probe of the sterile neutrino that will be highly complementary to the JSNS$^2$ IBD search ($\overline{\nu}_\mu \rightarrow \overline{\nu}_e$). The advantage of the KDAR technique over other sterile neutrino searches is that the signal energy (236~MeV) is known exactly. A background measurement on either side of the signal energy window around 236~MeV can allow an interpolated determination of the expected background in the signal region with high precision. However, as compared to $\nu_\mu$ CC interactions, $\nu_e$ events ($\nu_e n \rightarrow e^-  p$ or $\nu_e  \mathrm{^{12}C} \rightarrow e^- X$) are more challenging to identify over background, since they do not feature a double coincidence signal. While the KDAR $\nu_e$ events are expected to be distinct, in the sense that their reconstructed energy will lie close to 236~MeV, beam-induced neutrons can interact inside of the detector to produce an energetic single flash of light (e.g. a proton), mimicking a 236~MeV $\nu_e$ event. Pulse shape discrimination can be used to mitigate this background, but the background event rate expectation remains significant. This is worrisome because the oscillated signal expectation is $<100$~events in consideration of the global best fit region at high-$\Delta m^2$. The possibility of probing $\nu_e$ appearance using KDAR neutrinos at the MLF remains an intriguing possibility, however, especially given strong pulse shape discrimination.

The KDAR neutrino opens up new avenues for research in neutrino oscillation, interaction, and nuclear physics, and the importance of these measurements is clear. Perhaps most intriguing, the nucleus has simply never been studied using a known energy, weak-interaction-only probe and KDAR provides the exclusive technique to explore this frontier. JSNS$^2$ represents the world's best hope to take advantage of the KDAR neutrino in the near future. Other existing facilities worldwide simply cannot match the large expected KDAR signal and small expected decay-in-flight background rate at the J-PARC MLF.

\section{Experimental Components}

\subsection{The J-PARC MLF as a DAR Neutrino Source}
\label{sec_beam}
\indent

The J-PARC MLF is the best suited facility 
to search for neutrino oscillations using neutrinos from stopped muon decay
in the mass range $\Delta m^2\sim$ eV$^2$ 
for the following reasons:
\begin{enumerate}
\itemsep-10pt
\item High available beam power (1 MW) \\
\item Suppression of $\mu^-$ free decay through absorption by the mercury target \\
\item A low duty factor, pulsed beam which enables elimination of decay-in-flight components and separation of $\mu$DAR from other background sources. The resulting  $\nu_e, \bar{\nu}_e$ have well-defined spectra and known cross sections.
\end{enumerate}

\subsubsection{The RCS beam and the target}
~~
The proton intensity is expected to reach 0.33 mA (1 MW) after major upgrades 
to the mercury target. The protons are produced with a repetition rate of 25 Hz, where each spill contains two 100 ns wide pulses of protons spaced 540 ns apart. The 1 MW beam provides 3.8$\times$10$^{22}$ protons-on-target (POT) during 5000 hours / year operation (i.e.
4.5$\times$10$^{8}$ spills are provided during one year).
The short pulsed beam provides the ability to distinguish between neutrinos from pion decay and those from muon decay.

\begin{figure}
 \centering
 \includegraphics[bb=0 0 885 661, width=0.7 \textwidth]{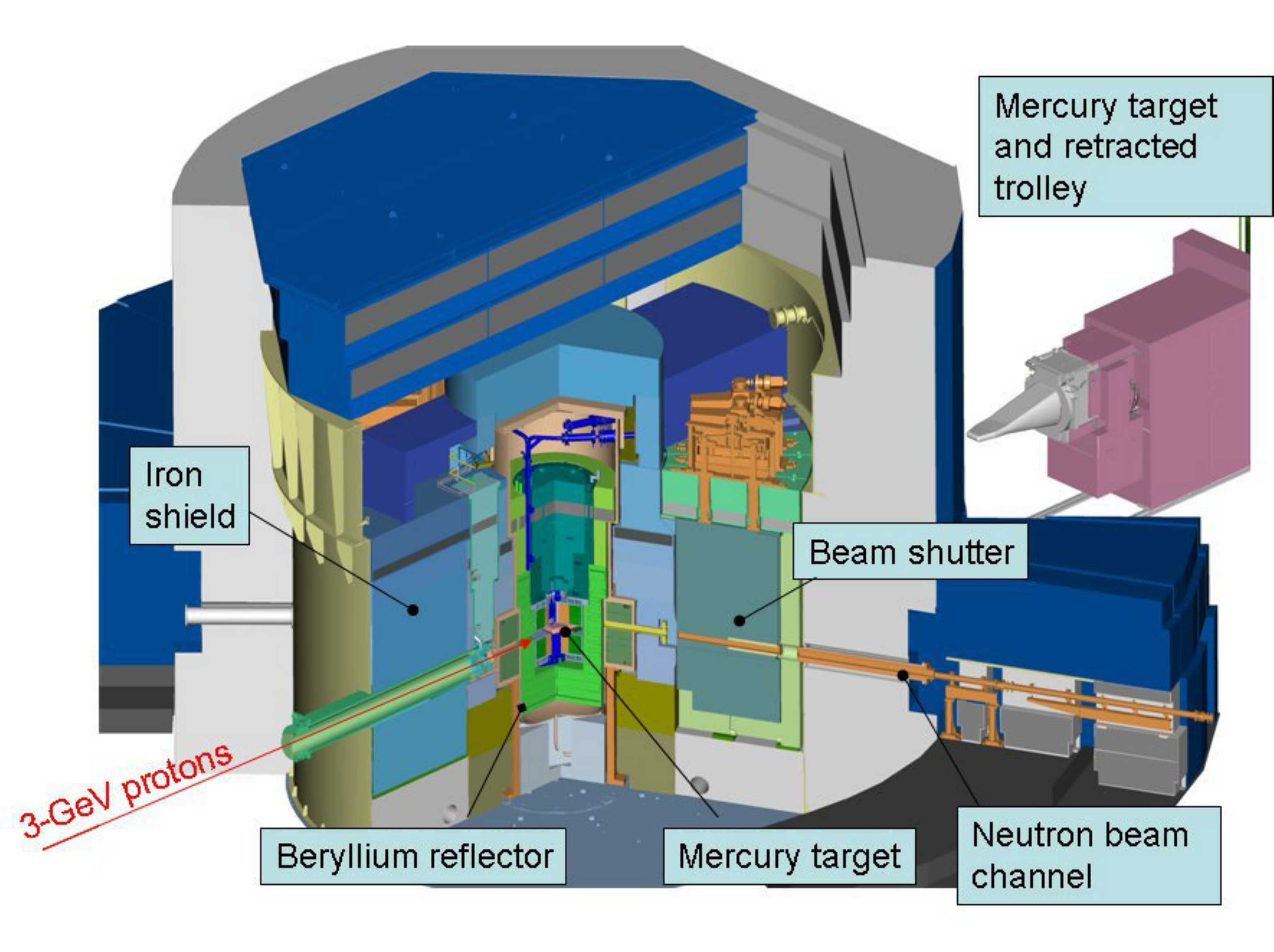}
\caption{\setlength{\baselineskip}{4mm}
A schematic drawing of the J-PARC spallation neutron source.
}
 \label{fig:MLF}
\end{figure}

Figure \ref{fig:setupJSNS2} shows a bird eye view of MLF in J-PARC. After penetrating a 2 cm thick muon production target made of carbon graphite, protons are introduced to the mercury target. A schematic drawing of the J-PARC spallation neutron source is shown in Fig. \ref{fig:MLF}. 3 GeV protons interact in the mercury spallation target, producing pions and kaons that decay into 
$\nu_e$ and $\nu_{\mu}$ and their anti-neutrinos after heavy shielding. Surrounding the target are cooling pipes, beryllium reflectors, and steel shielding.

\begin{figure}
 \centering
 \includegraphics[bb=0 0 590 296, width=0.7 \textwidth]{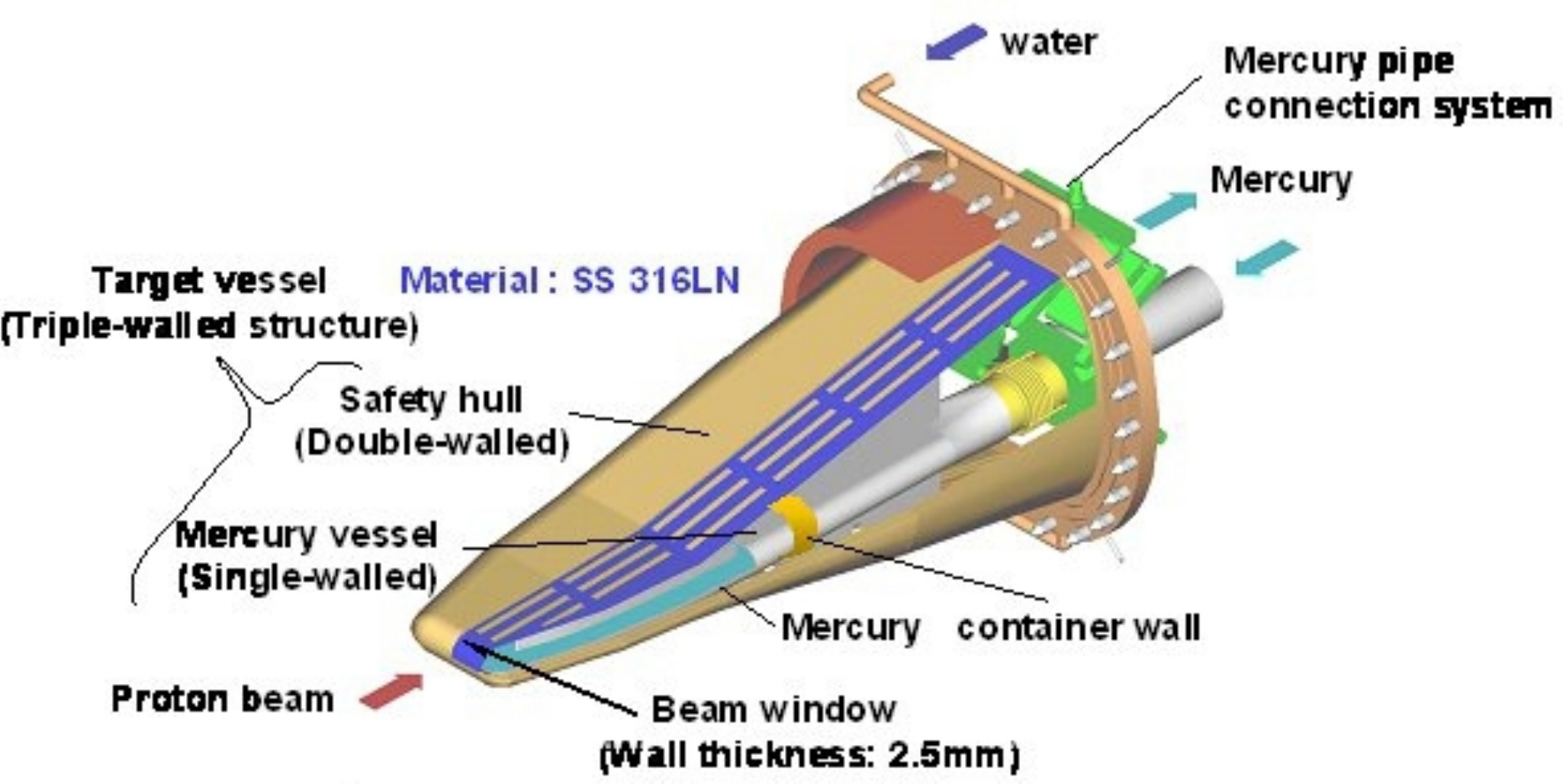}
\caption{\setlength{\baselineskip}{4mm}
A schematic drawing of the mercury target in the J-PARC MLF.
}
 \label{fig:MLFTarget}
\end{figure}

A beam of protons enters from the left and strikes the target. The beam has a wide spot size (e.g.: 3.3 cm by 1.3 cm in root mean square) for reduction of the local heat load in the target. The target, shown in Fig.~\ref{fig:MLFTarget}, has dimensions of 54 cm in width by 19 cm in height by 210 cm in length. Mercury is contained within a multiple wall structure made of stainless steel. To remove heat, the mercury of the target is constantly circulated at a rate of 154 kg/sec. Cryogenic liquid hydrogen moderators are located at the top and bottom of the target. The target and moderators are surrounded by a beryllium reflector and iron shielding which extends to a minimum radius of 5 m around the target. There are 23 neutron channels looking at the moderators, rather than at the target. Shutters are provided on each channel.

\subsubsection{Neutrino Beam}
~~
There are two time structures in the neutrino beam. One is the `on-
bunch' component (neutrinos produced during the proton bunch and within the pion or kaon lifetime) and features neutrinos from:
\begin{itemize}
\item $\pi^+ \rightarrow \mu^+~\nu_{\mu}$ decay at rest with monochromatic neutrino energy of 30 MeV
\item $\mu^- +A \rightarrow \nu_{\mu} +A$ with a 105 MeV endpoint
\item $K^+ \rightarrow \mu^+~\nu_{\mu}$ decay at rest with monochromatic energy of 236 MeV
\item $K^+ \rightarrow \mu^+~\pi^0 ~\nu_{\mu}$ decay at rest with an endpoint energy of 215 MeV
\item $K^+ \rightarrow e^+~\pi^0 ~\nu_e$ decay at rest with an endpoint energy of 228 MeV
\item Small components from $\pi$ and K decay in flight
\end{itemize}
The other time structure is the `off-bunch' component (during the muon lifetime), which is produced by muon decay at rest:
\begin{itemize}
\item $\mu^+ \rightarrow e^+~\nu_e~\bar{\nu_{\mu}}$
\item If $\mu^-$ stops in a light material, $\mu^-$ also decays by $\mu^- \rightarrow e^+~\bar{\nu_e}~\nu_{\mu}$
\end{itemize}
This `off-bunch' component can be selected by gating out the first 1 $\mu$s from the start of the proton beam. 
Figures \ref{fig:muDARflux} and \ref{fig:sense} show the expected neutrino energy spectrum from the MLF target with and without the timing cut (T $>$ 1$\mu$s). Figure~\ref{fig:muDARtime} shows the time distributions from various sources.

Tables~\ref{tab:pionprod} and \ref{tab:pionprod2} are summary tables for the production of neutrinos from $\mu$ decays.The $ \mu$ decay at rest neutrino beam was simulated using the following steps.
\begin{enumerate}
\item Particle production by 3 GeV protons \\
The interaction of the 3 GeV proton beam with the mercury target and beam line components has been simulated with FLUKA~\cite{FLUKA} and QGSP-BERT (in Geant4~\cite{geant4}) hadron interaction simulation packages.
 \item $\pi ^{\pm}$ interactions and decay \\
After production, both $\pi^+$ and $\pi^-$ lose their energy mainly by ionization. In addition, they disappear by the charge exchange reaction $\pi^{\pm} (n,p) \rightarrow \pi^0 (p, n)$, $\pi^0\rightarrow \gamma \gamma$.
The survived $\pi^+$ stop and decay with a 26 ns lifetime.
On the other hand, the survived $\pi^-$ are absorbed by forming a $\pi$-mesic atoms and getting absorbed promptly. Decay-in-flight takes place with very suppressed rate of about $\sim 8\times 10^{-3}$ of produced $\pi^{\pm}$.
\item $\mu ^{\pm}$ absorption and decay \\
All $\mu^+$ decay by $\mu^+\rightarrow e^+\nu_e \bar{\nu_{\mu}}$. Because of the muon lifetime and energy loss process, the decay-in-flight is negligible.
$\mu^-$ are captured by nuclei by forming a mu-mesic atom and eventually produce $\nu_{\mu}$ with an endpoint energy of 100 MeV. The absorption rate depends on the nucleus and becomes faster for heavier nuclei.
The total nuclear capture rates for negative muons have been measured in terms of effective muon lifetime~\cite{mudecay}.
\end{enumerate}
The resulting neutrino fluxes for each type of neutrino species are shown in Figure \ref{fig:muDARflux}.
Tables~\ref{tab:pionprod} and \ref{tab:pionprod2} are expected production rates of $\pi ^{\pm}$ by 3 GeV protons on the mercury target and the resulting $\mu^+$ and $\mu^-$ decay neutrinos per proton, based on a pion production model.  

\begin{table}[htb]
\begin{center}
	\begin{tabular}{|c|c|c|}
\hline
 ~  &  $\pi^+\rightarrow \mu^+ \rightarrow \bar{\nu_\mu}$ &   $\pi^-\rightarrow \mu^- \rightarrow \bar{\nu_e}$   \\ \hline \hline
$\pi$/p  &   $6.49\times 10^{-1} $ &  $4.02\times 10^{-1}$   \\ \hline
 $\mu$/p   &  $3.44\times 10^{-1} $ &  $3.20\times 10^{-3}$ \\ \hline
$\nu$/p  & $3.44\times 10^{-1} $ & $7.66\times 10^{-4}$  \\ \hline
 $ \nu$ after $1\mu$s & $2.52\times 10^{-1}$ & $4.43\times 10^{-4}$
\\ \hline \hline
	\end{tabular}
	\caption{\setlength{\baselineskip}{4mm}
	An estimate of $\mu$DAR neutrino production by 3 GeV protons using the FLUKA hadron simulation package.}
	\label{tab:pionprod}
\end{center}
\end{table}

\begin{table}[htb]
\begin{center}
	\begin{tabular}{|c|c|c|}
\hline
 ~  &  $\pi^+\rightarrow \mu^+ \rightarrow \bar{\nu_\mu}$ &   $\pi^-\rightarrow \mu^- \rightarrow \bar{\nu_e}$   \\ \hline \hline
$\pi$/p  &   $5.41\times 10^{-1} $ &  $4.90\times 10^{-1}$   \\ \hline
 $\mu$/p   &  $2.68\times 10^{-1} $ &  $3.90\times 10^{-3}$ \\ \hline
$\nu$/p  & $2.68\times 10^{-1} $ & $9.34\times 10^{-4}$  \\ \hline
 $ \nu$ after $1\mu$s & $1.97\times 10^{-1}$ & $5.41\times 10^{-4}$
\\ \hline \hline
	\end{tabular}
	\caption{\setlength{\baselineskip}{4mm}
	An estimate of $\mu$DAR neutrino production by 3 GeV protons using the QGSP-BERT hadron simulation package.
	}
	\label{tab:pionprod2}
\end{center}
\end{table}

Needless to say, there are many sources of ambiguities in pion production. For example, the production rates are sensitive to production by secondary particles in the thick target, target geometrical modeling, and uncertainty in pion production from mercury. We use these calculations as estimates
and the actual $\mu^-$ backgrounds should be determined from the data based on their known spectrum and known cross section. 

For this TDR, numbers from Table~\ref{tab:pionprod} are used 
to estimate the central values, and those in Table~\ref{tab:pionprod2} are used for the cross checks.

\subsubsection{Estimated Neutrino Flux}
\label{ENER}
~~
The proton intensity is assumed to be 0.33 mA, delivering $3.8\times 10^{22}$
protons on target (POT) per 5000 hour operation in one year.
The stopping $\nu$/p ratio is estimated from the FLUKA simulations to be 0.344.
The $\bar{\nu}_{\mu}$ flux from
the $\pi^+\rightarrow \nu_{\mu}+\mu^{+}; \mu^{+} \rightarrow e^{+} + \nu_{e} + \bar{\nu}_{\mu}$ chain at 24 m is then equal to 1.8$\times$10$^{14}\ \nu$ year/cm$^2$.

\subsection{Detector Location and Constraints}
\indent

\begin{figure}[htbp]
\centering
\includegraphics[width=120mm]{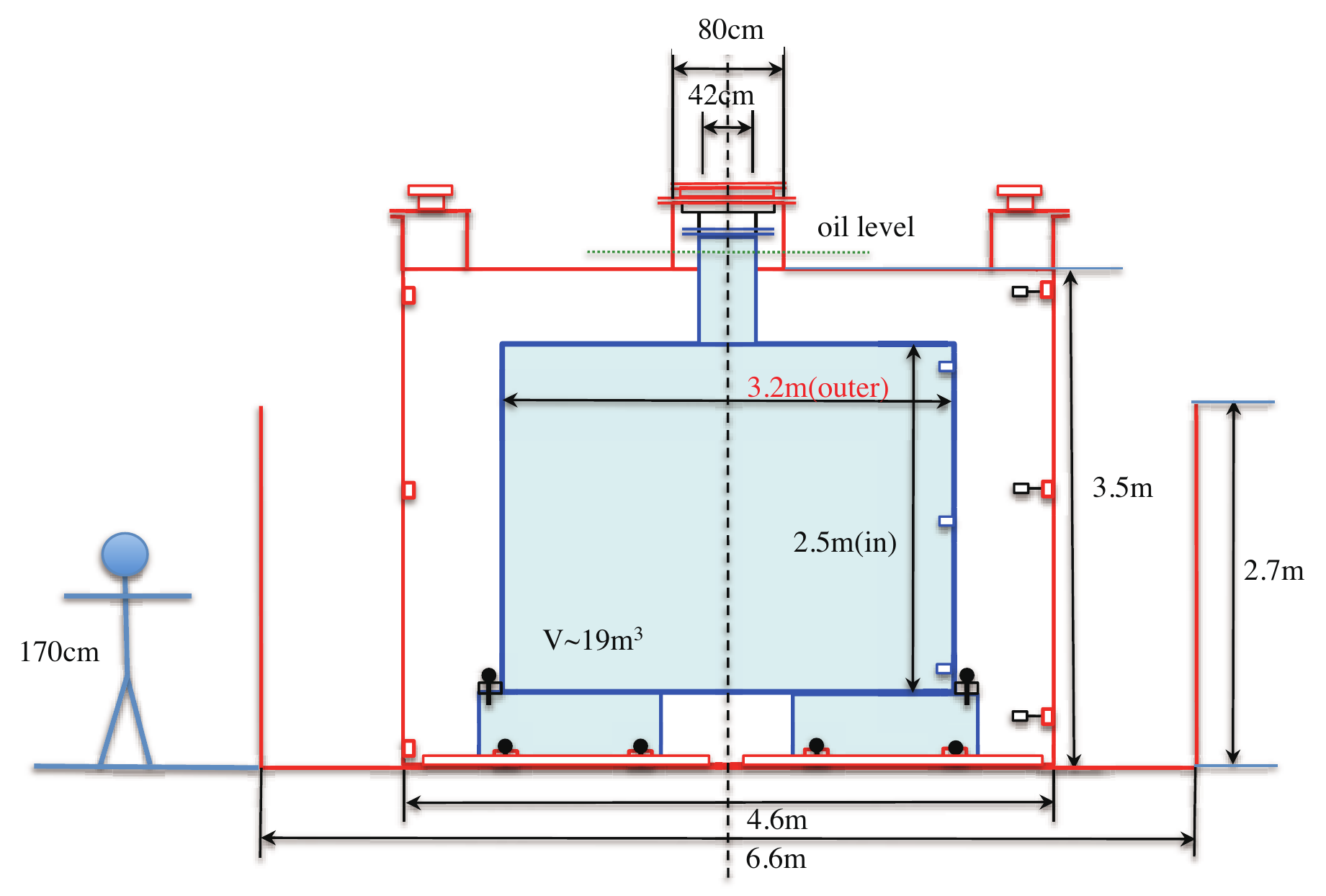}
\caption{\small{Conceptual structure of the JSNS$^2$ detector tanks.  }}
\label{fig:Tank_Overall}
\end{figure}

We plan to place the detector shown in Fig.~\ref{fig:Tank_Overall} on the third floor of the MLF.
As previously noted, the detector is at a 24 meter baseline and contains 17 tons of Gd-loaded
LS inside the inner acrylic vessel, and $\sim$30 tons
unloaded LS in the space between the acrylic vessel and stainless tank.
There are 193 8-inch PMTs between the acrylic and stainless tanks and 48 5-inch
PMTs in the veto region. The specific PMT locations will be described later.

\begin{figure}[h]
 \centering
 \includegraphics[width=0.8 \textwidth]{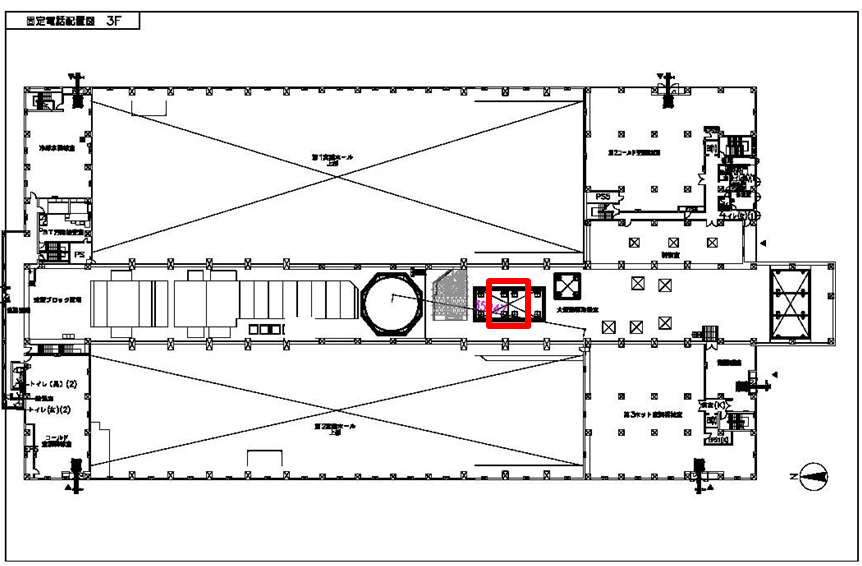}
 \caption{\setlength{\baselineskip}{4mm} 
   Two dimensional drawings of the MLF 3rd floor. The red box shows the
   detector location based on the 2014 background measurements
   described in Section~\ref{SEC:BKG2014}}
 \label{fig:LCHR}
\end{figure}

Figure~\ref{fig:LCHR} shows two dimensional drawings of the MLF 3rd floor,
and the red box shows the detector location based on the
background measurements which were performed in 2014 (see
Section~\ref{SEC:BKG2014} for more details). In this area,
we need to put iron plates under the detector in order
to provide shielding from beam gammas coming through the hatches.  
The left-hand side of the figure is
upstream of the beam and the circle around the center of the figure corresponds
to the location of the mercury target. You can see the hatches in the red
box, which correspond to the concrete hatches used to exchange the mercury target.
Under the hatches, there is a radiation hot room as shown in Figs.~\ref{fig:OP}
and \ref{fig:gum}, and therefore we have to
prepare safety systems so that LS does not leak into the radiation hot room. (Oil spill prevention system is described
in Section~\ref{SEC:SPS}).

The experts of MLF call this area "Large Component Handling Room", that is
the maintenance area for the mercury target, the target cooling system,
the beamline equipment and so forth.
Therefore, the JSNS$^2$ detector must be moved outside the MLF
during the maintenance period which is held typically from July to
September (three months) each year in order to avoid interference
between the maintenance work and JSNS$^2$ experiment.

To put the JSNS$^2$ detector on the MLF 3rd floor, the withstand load and the sinkage around the area of the building are quite important. Thus, we estimated the former with the Mitui-Zosen company and the results will be shown later.

For the latter, we have also estimated how much the MLF building
will settle when the JSNS$^2$ detectors are placed inside the room. 
The total weight of the MLF building is about 40,000 tons.
The Neutron Source Group in JAEA has measured the amount of subsidence of the
MLF building continuously from the construction phase,
and a summary of the findings
has been published in Reference ~\cite{settlement}.
According to the publication, the current sinkage has substantially stopped,
and it estimated that a sinkage of 0.3 mm/1000 tons occurs when the weight load
is increased in the MLF building. 
The weight of JSNS$^2$ detector is about $\sim$190~tons,  
thus the sinkage effect of the JSNS$^2$ detector is less than 0.1~mm,
which can be considered to be negligible.

\subsubsection{Withstand Weight Load}
\indent

The JSNS$^2$ detector consists of liquid scintillator, containment tanks,
support frames, and iron plates ($\gamma$ shields) that have a total weight of about 190 tons.
The position of the JSNS$^2$ detectors is currently planned to be around the large hatch (12 m $\times$6 m) as shown in Fig.~\ref{fig:LCHR}.
The hatches are moved to access to the irradiated components handling room
as shown in Fig.~\ref{fig:gum}. (The room has orange colored light in this picture.)  

\begin{figure}[htbp]
\centering
\subfigure[The detector location (top view)]{
\includegraphics[width=0.45\textwidth,angle=0]{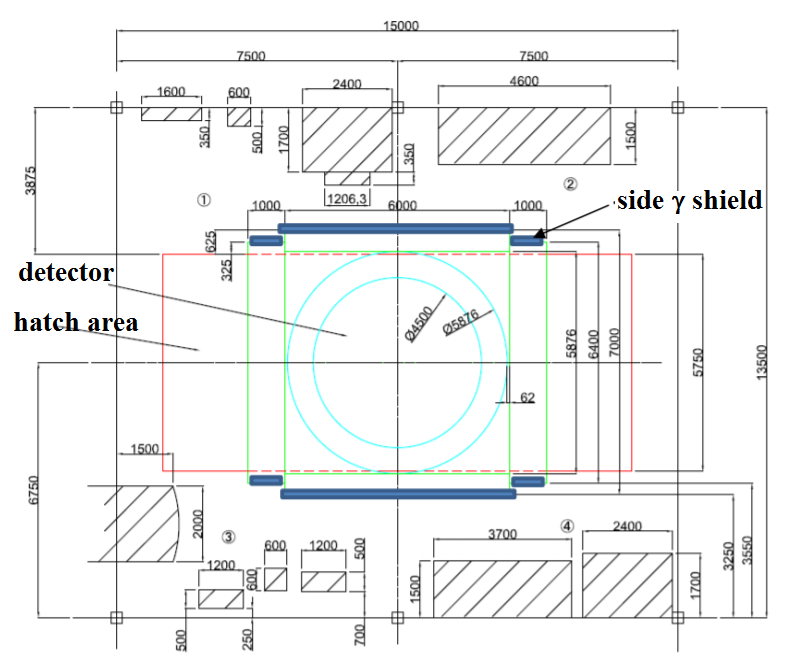}
}
\subfigure[The detector location (side view)]{
\includegraphics[width=0.45\textwidth,angle=0]{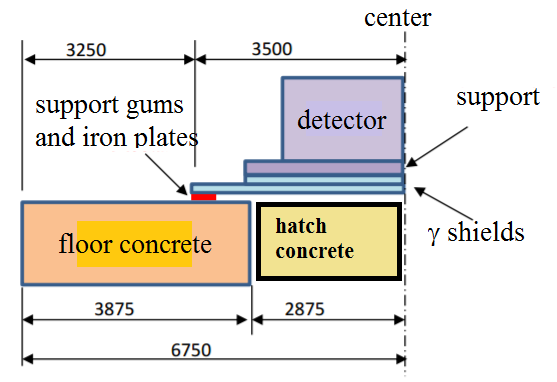}
}
\caption{\setlength{\baselineskip}{4mm}
  Top (left) and side (right) views around the detector. Blue bands on the left
  show the gums and iron support plates which reduce the bearing pressure on the floor.
  This structure is described in Appendix A. The green area (left) shows  the
  $\gamma$ shield regions, and the light blue corresponds to the detector
  and anti oil-leak tanks.
  Red area (left) indicates the concrete hatches
  to separate the 3rd floor and radiation hot area under the detector. 
} 
\label{FIG:WL}
\end{figure}

The hatches themselves cannot sustain 190 tons.
For this reason, a detector support structure using
side areas around the large hatches is necessary.
Figure~\ref{FIG:WL} shows the top and side views around the detector including
the $\gamma$ shields (iron plates) and detector support structure.
This structure is designed to be as light as possible. The following considerations are taken into account by this design:

\begin{itemize}
\item The detector including stainless, acrylic vessels, anti oil-leak tanks,
  PMTs and LS is supported by I-beams (height of 25 cm)
  under the detector, and is
  transported from the MLF entrance to the 3rd floor by a 130 ton crane.
  Note again that the support structure is transported with the LS and needed
  for the transportation to support LS, and
  the iron plates are separated from the detector.
\item To shield the detector from beam $\gamma$s coming through the hatches, iron plates are placed underneath. To reduce the weight as much as possible, two iron plates are used. One plate has dimensions 6.0 (beam direction) $\times$ 5.88 (orthogonal direction to the beam) $\times$ 0.15 (thickness) m$^3$ and is placed just below the detector and the other has dimensions
  6.0 $\times$ 7.0 $\times$ 0.15 m$^3$ and is put under the first plate.
\item According to the MC simulation, a total coverage of 8.0 m in the beam direction is preferrable, and therefore we put additional plates of dimensions
  1.0 (beam direction) $\times$ 6.4 (orthogonal direction to the beam) $\times$ 0.3 (thickness) m$^3$ on both sides of the detector along the beam direction.
\end{itemize}
Table~\ref{Tab:weight_all} shows the weight of the detector components.
\begin{table}[htbp]
\begin{center}
	\begin{tabular}{|c|c|c|}
	  \hline
        Components & Weight (tons) & comments \\ \hline \hline    
	Supporting structure & 10 & \\
        Acrylic vessel + LS &  20.3 & 17.0 (LS) + 3.3 (tank)\\ 
        Stainless tank + LS &  36.9 & 30.0 (LS) + 6.9 (tank)\\
        $\gamma$ shield 1 & 41.5 & 5.88 $\times$ 6.0 m$^2$ $\times$ 15 cm (t)\\
        $\gamma$ shield 2 & 49.5 & 7.0 $\times$ 6.0 m$^2$ $\times$ 15 cm (t)\\
        Side $\gamma$ shields & 30.1 & 6.4 $\times$ 1.0 m$^2$ $\times$ 30 cm (t) $\times$ 2\\ \hline 
        Total & 188.3 & \\
	\hline
	\end{tabular}
	\caption{Weight for the detector components.}
        \label{Tab:weight_all}
\end{center}
\end{table}

This subsection describes whether the MLF building can support the withstand load
from the detector. Discussion of this topic has already taken place with the JAEA facility section and with ``Nikken-Sekkei", which designed and built the MLF building.

The MLF building tolerance for additional weight loads is determined by two
parameters: the bending moment of the floor concrete and the withstand load of the concrete floor itself.
These parameters are defined by "Nikken-Sekkei" company.

\begin{figure}[htbp]
 \centering
 \includegraphics[width=0.5 \textwidth]{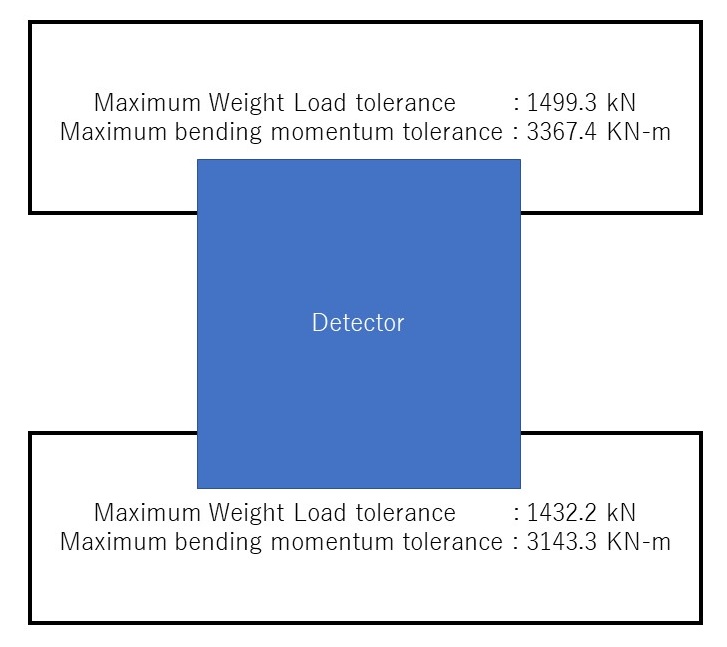}
 \caption{\setlength{\baselineskip}{4mm} 
  The maximum weight load tolerance values on the 3rd floor (after subtraction).
 }
 \label{fig:WLresult}
\end{figure}

Before the comparison between the tolerance and weight for the detector, 
we must investigate the weights and moment which are already used
for the existing materials on the 3rd floor because they must
be subtracted from the tolerance. The existing materials on the 3rd floor are
shown as the shaded regions in Fig.~\ref{FIG:WL} (a).
Then, we subtract them from the tolerance of the building as shown in
Fig.~\ref{fig:WLresult}. The maximum tolerance values of the building
before the subtraction are given by ``Nikken-Sekkei".

As mentioned, the expected total weight of the detector is about 190 tons.
Therefore, the maximum weight load tolerance after subtraction
is much larger than the expected
detector weight (919.5 kN (detector) vs. 1432.2 kN (tolerance)).
The bending moment of the 3rd floor assuming the detector design shown in
Fig.~\ref{FIG:WL} is 3032.7 KN$\cdot$m (the bending moment is determined
by the location of the support points of the detector (weight center)
and weight of the detector). This is within the tolerance of the
MLF building. Note that this calculation includes large safety factors.

After showing the calculation to the JAEA facility section and ``Nikken-Sekkei",
they agree that there are no issues with putting the detector on the MLF 3rd floor.

\subsection{Tanks}

\subsubsection{General structure}
\indent

Figure~\ref{fig:Tank_Overall} shows a schematic drawing of the JSNS$^2$ detector tank system. 
The tank system consists of, from inner to outer, an acrylic vessel, stainless steel (s.s.) tank and an anti oil-leak tank.  
The acrylic vessel contains the Gd-loaded LS which works as the neutrino target. 

Unloaded LS, which vetoes the background signals is contained in the region between the acrylic vessel and the s.s. tank.
193 8-inch PMTs and 48 5-inch PMTs will be installed in the space between the s.s. tank and acrylic vessel. 
The tank system has to be moved at least once per year to make room for maintenance work on the MLF beam line. 

\subsubsection{Stainless Steel Tank and Anti Oil-Leak Tank}
\indent

\begin{figure}[htbp]
\centering
\includegraphics[width=140mm]{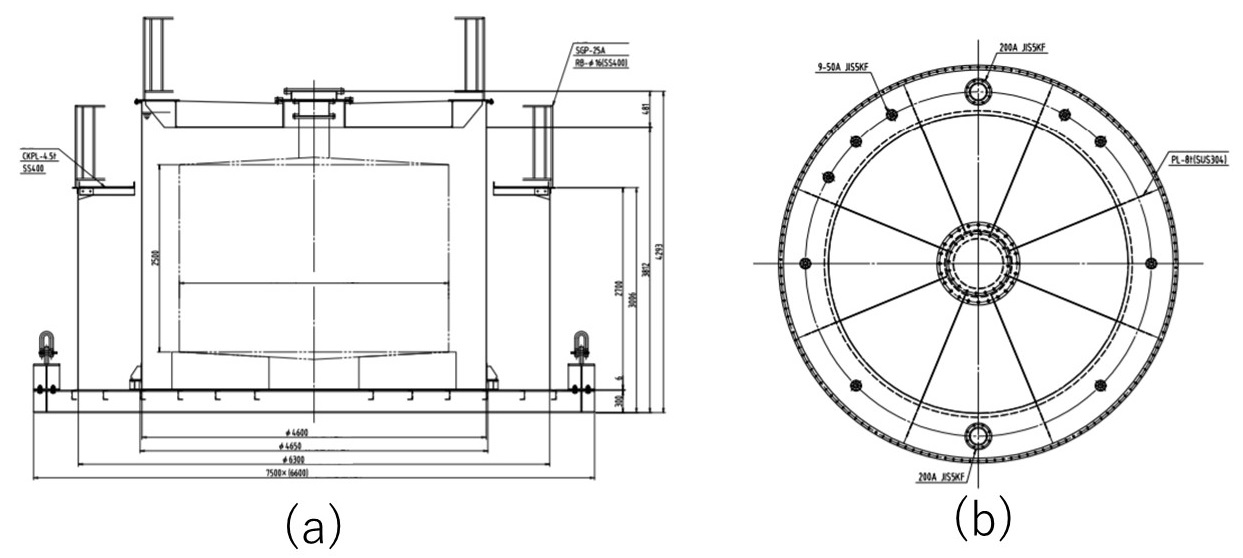}
\caption{\small{Design of the stainless steel tank. ($a$) Side view, ($b$) Top view  }}
\label{fig:Tank_Morimatsu_All}
\end{figure}

Figure~\ref{fig:Tank_Morimatsu_All} shows a drawing of the tank
design. 
The diameter, height and volume of the cylindrical part of the s.s. tank are 4.6~m, 3.5m and 58~m$^3$, respectively. 
The diameter and height of the anti oil-leak tank are 6.6~m and 2.7~m, and the volume is 80~m$^3$, which can contain all the liquid even if there is a leak from the s.s. main tank. 
The thickness of the side wall of the s.s. tank is 5~mm and the bottom plate is 6~mm. 
In order to prevent agitation of the oil surface while the tank is moved by a crane, the lid of the s.s. tank is designed to be below the LS level and the LS can push up the lid.  
The lid is reinforced by 8 welded beams. 
The nominal oil level is 10~cm higher than the s.s. lid.
The diameter of the s.s. tank  chimney is 80~cm.
At the circumference of the stainless steel tank, there is 
30~cm wide LS stabilization region for the veto area. 
This stabilizes the oil level with respect to temperature changes as described in the following sections.
 
\subsubsection{Acrylic Vessel}
\indent

\begin{figure}[htbp]
\centering
\includegraphics[width=130mm]{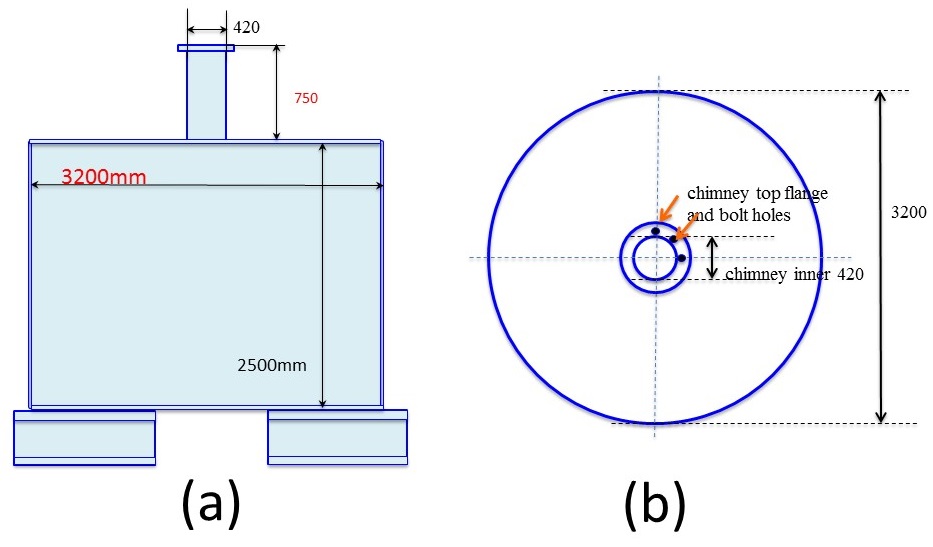}
\caption{\small{$(a)$ Acrylic vessel side view, $(b)$ view of lid.  }}
\label{fig:AcrylicTank_Side_Top}
\end{figure}

Figure~\ref{fig:AcrylicTank_Side_Top} shows side and top views of the acrylic vessel design. 
The diameter, height and volume of the cylindrical part of the acrylic vessel are, 3.2~m and 2.5~m and 19.3~m$^3$, respectively. 
The thickness of the acrylic is 3~cm.
The maximum strength of the acrylic vessel corresponds to 
$\pm 15$~cm head difference.   
Due to the pressure changes, the center of the acrylic lid moves at most $\pm 2$~cm. 
 
\begin{figure}[htbp]
\centering
\includegraphics[width=130mm]{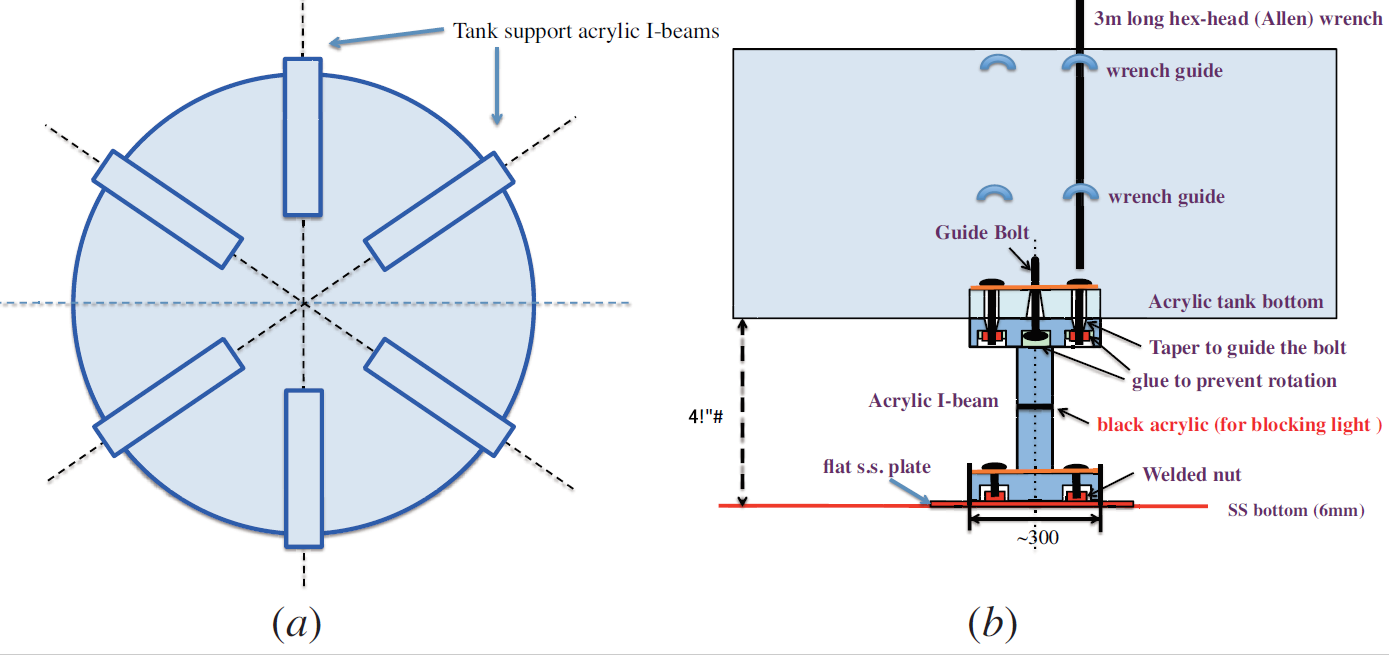}
\caption{\small{$(a)$ Acrylic vessel support acrylic blocks, $(b)$ Details of the support blocks and their installation.  }}
\label{fig:AcrylicTank_Support}
\end{figure}

The total volume of the vessel part of acrylic material is 1.2~m$^3$ and  the total weight is 1.5~ton.
Since the specific gravity of the acrylic is 1.2 and that of the LS is 0.86, the net weight of the acrylic vessel in the oil is 406~kg. 
This weight stabilizes the acrylic vessel in the oil.  
The acrylic vessel stands on the bottom of the stainless steel tank via six acrylic fin blocks as shown in Fig.~\ref{fig:AcrylicTank_Support}.
The support blocks are bolted to the bottom of the s.s. tank before the PMT installation.
After the PMT installation, the acrylic vessel is put in the s.s. tank and bolted to the support blocks from the above using a specially-made long Allen wrench (hex head wrench).
Wrench guide rings are installed on the side wall of the acrylic vessel. 

\begin{figure}[htbp]
\centering
\includegraphics[width=80mm]{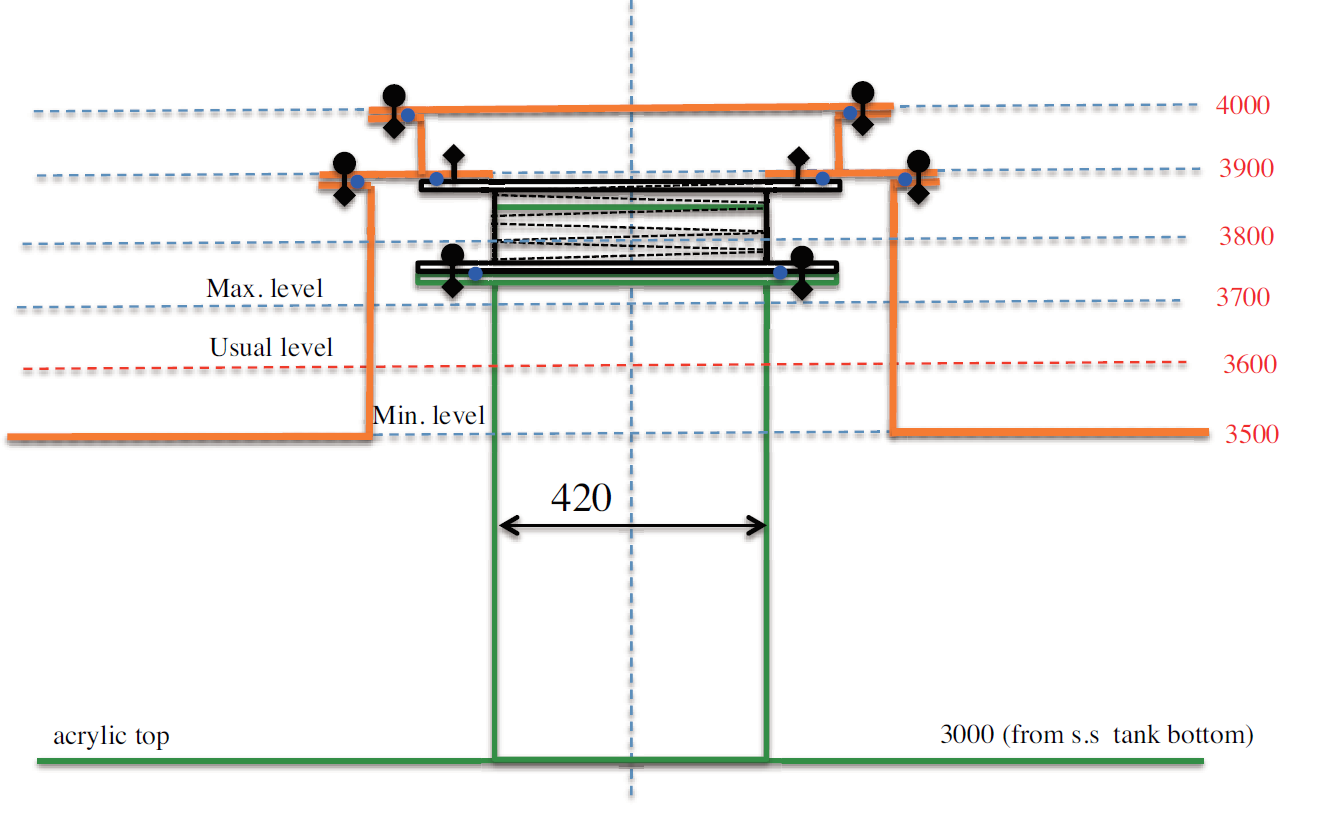}
\caption{\small{Chimney of the JSNS$^2$ detector tanks.}}
\label{fig:Tank_Chimney}
\end{figure}

The acrylic vessel has a 42~cm diameter chimney which is connected to the s.s. tank through a 15~cm long bellows as shown in Fig.~\ref{fig:Tank_Chimney}. 
The bellows have a movable span of $\pm$7.5 cm and 
absorb mechanical changes and possible deformations of the tanks due to internal pressure variations. 

The elastic limit of the acrylic vessel lid deformation due to the internal pressure is $\pm 2$~cm. 

Table~\ref{table:TankLiq_Parameters} summarizes the parameters of the tanks and LS. 
 \begin{table}[htbp]
 \caption{ Parameters of the tanks and liquids.\label{table:TankLiq_Parameters}}
 \begin{center}
 \begin{tabular}{|c||c|c|c|c|} 
 \hline
  Item      & Radius[m]  & Height[m] & Volume [m$^3$] & Weight [ton]    \\ 
\hline
\hline
Acrylic Vessel          & 1.6 & 2.5 & 19.3 & 1.5 \\
\hline
s.s. tank  & 2.3  & 3.5 & 58   & - \\         
 \hline
 S.P.R. & 3.15 & 2.7  & 84 & - \\
 \hline
  Gd-LS     &- & -& 19.3 & 17 \\
 \hline    
  Buffer LS  &- &- & 35 & 30 \\
 \hline 
 \end{tabular}
\end{center}
\label{table:2_HelicitySpinDirection}
\end{table}

%
\subsubsection{Liquid level stabilizing mechanism}
\indent

The thermal expansion coefficient of  linear alkylbenzene is 
$\Delta V /V \sim 9\times 10^{-4}/^o$C~\cite{15_Zhou}. 
On the other hand, the maximum temperature variation of the experimental area is expected to be $\pm$10~degrees and the liquid volume will change as much as  $\pm$0.9\%.
This volume change causes a change in the oil level. 
For example, the change of the Gd-LS level in the acrylic chimney is 14~cm/$^o$C and so a level stabilizing mechanism is necessary. 

\begin{figure}[htbp]
\centering
\includegraphics[width=120mm]{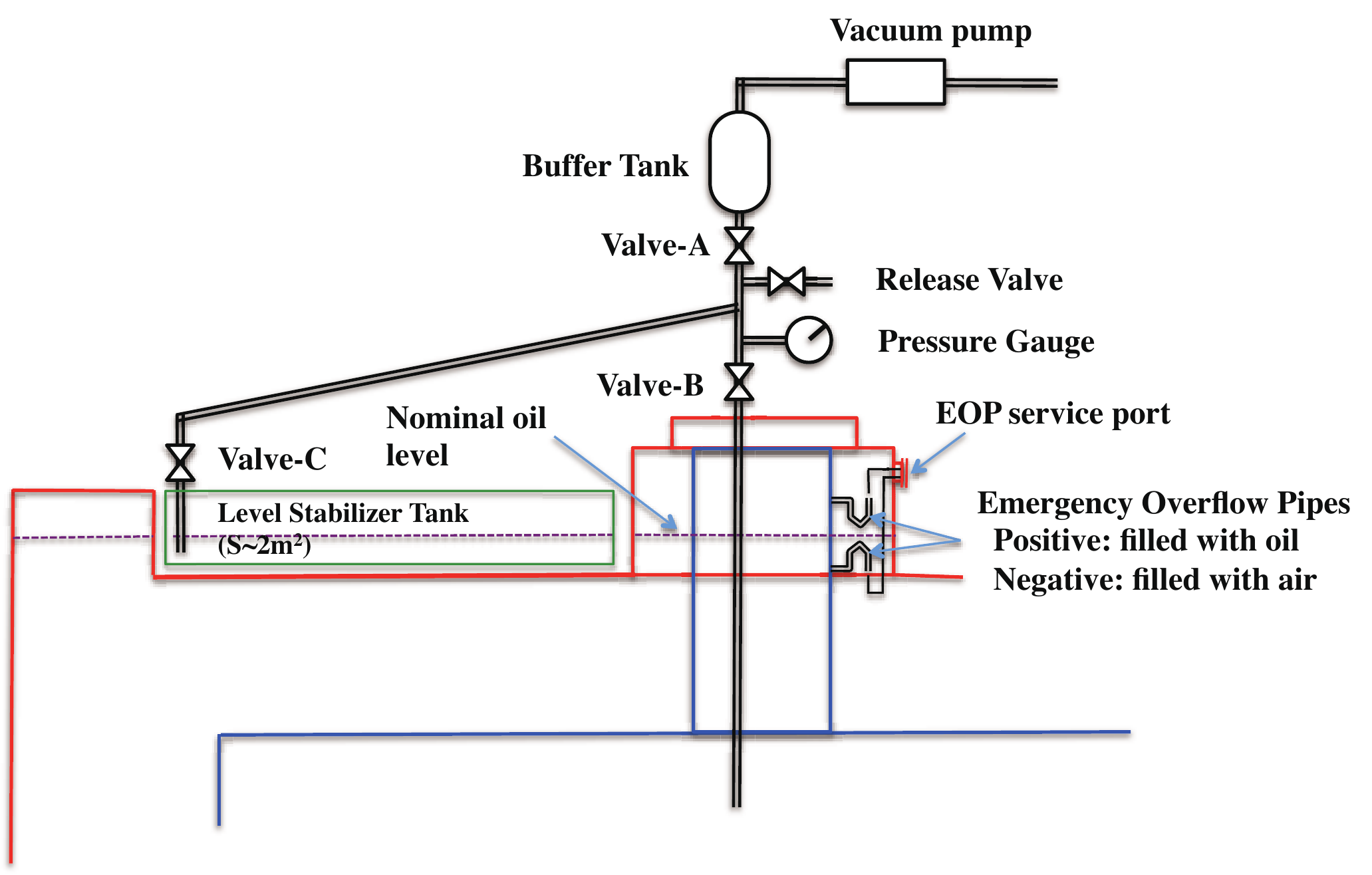}
\caption{\small{Oil level stabilizing mechanism }}
\label{fig:Tank_Stabilizer}
\end{figure}

The oil level to be maintained is shown in Fig.~\ref{fig:Tank_Overall}. 
There is a 30~cm wide stabilization area along the circumference of the stainless tank which will stabilize the veto LS level.
The area of this outer surface ($S_B$) is $\sim 4~{\rm m^2}$ and the change of the oil level caused by $\pm 10^\circ$ temperature change is kept within $\pm 10$~cm. 
As for the Gd-LS liquid, there is an oil level stabilizer tank on the s.s. tank lid as shown in Fig~\ref{fig:Tank_Stabilizer}.
The inner surface of the stabilizer tank is made of plastic in order to prevent potential instability of the Gd-LS. 
The area of the stabilizer tank ($S_T$) is  
$S_T = S_B\frac{V_T}{V_B}\sim 2.5{~\rm m^2}$ to prevent any level difference due to temperature change.

The liquid in the stabilizer tank and the acrylic vessel are connected by a {\it reverse siphon} (RS) system. 
After oil filling, the RS system is established according to the following procedure. 
\begin{enumerate}
\item At first, all the valves are closed. 
\item LS is supplied to the stabilizer tank until the end of the siphon pipe is completely soaked.
\item Valves-A,~B,~C. are opened.
\item The vacuum pump is switched on and LS is sucked from both tanks.
\item When the LS becomes higher than Valve-A, the valve is closed and the RS is established.
\item The pressure gauge keeps monitoring the RS to ensure it is still established.
\end{enumerate}
The measured pressure value also indicates the level of Gd-LS in the tank. 

If, for some reason, the RS is broken, the LS inside the pipe is returned to each tank by opening the Release Valve and the RS establishing process ((1)-(5)) is performed again. 

The reverse siphon method has been utilized effectively in the Double Chooz experiment.

\subsubsection{Emergency overflow pipes}
\indent

The V and $\Lambda$ shaped pipes on the acrylic chimney in Fig.~\ref{fig:Tank_Stabilizer} are emergency overflow pipes. They are only used in the case that
the RS system is somehow broken.
The V-shape pipe is filled with LS using a pipe from the EOP service port and the $\Lambda$-shaped pipe is filled with air to separate the target and veto regions. 
If the Gd-LS level becomes high for some reason and reaches the inlet of the V-shape pipe, Gd-LS starts to flow out to the buffer LS region.
If the Gd-LS level becomes low and reaches to the inlet of the $\Lambda$-shape pipe, the veto LS starts to flow in to the target region. 
In this way, the Gd-LS level is kept between the inlets of the V and $\Lambda$-shape pipes at all times. 

If the overflow case were to happen, the Gd-LS and veto LS would mix. 
However, this would only occur in an emergency case and we expect it will not happen throughout the lifetime of the experiment.

\subsection{Safety Protection System}
\label{SEC:SPS}
\indent

Here we describe the safety protection systems.
First, the additional "anti oil-leak tank" normally required by the
Fire Law (one "anti oil-leak tank") is described.

\begin{figure}[h]
 \centering
 \includegraphics[width=0.7 \textwidth]{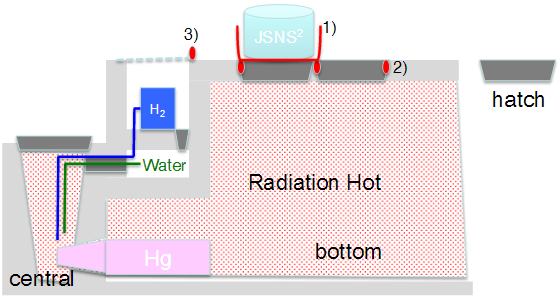}
 \caption{\setlength{\baselineskip}{4mm} 
   The additional oil spill prevention systems on top of the Fire Law requests including
   (1) A second "anti oil-leak tank", (2) Sealant in the space between the
   hatches and concrete, (3) The third oil spill prevention wall to prevent LS from flowing into the mercury target area.
 }
 \label{fig:OP}
 \end{figure}

Figure~\ref{fig:OP} shows the overall oil spill
prevention systems in addition to the
Fire Law requests.

There are three systems as follows;
\begin{enumerate}
\item A second anti oil-leak tank
\item Sealant in the space between the hatches and concrete
\item An anti oil-leak wall which prevents LS from flowing to the mercury target.
\end{enumerate}
The details of the each system are described in subsections below. We also describe the safety slow monitor systems for the detector later.

\subsubsection{Second Anti Oil-Leak Tank}
\indent

\begin{figure}[h]
 \centering
 \includegraphics[width=0.4 \textwidth]{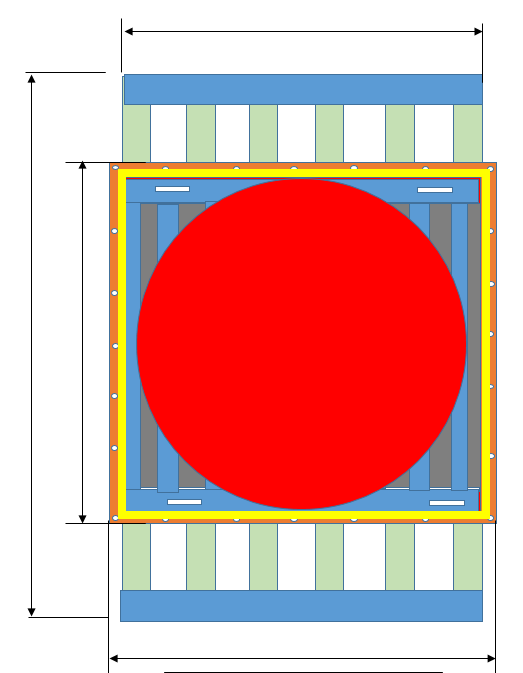}
 \caption{\setlength{\baselineskip}{4mm} 
   The top view of the tank including the first anti oil-leak tank
   (red),
   2nd anti oil-leak tank (yellow) and the tank base.
   The height of 2nd anti oil-leak tank is about 1.4 meter with a thickness of a few mm.
 }
 \label{fig:sOP}
 \end{figure}

As required by the Fire Law, we have the first anti oil-leak tank
around the detector (shown as the red circle in the Fig.~\ref{fig:sOP}).
In addition to the Fire Law requirement,
we will have the second anti oil-leak tank which surrounds the detector
tank and the first anti oil-leak tank (shown as the yellow
box in the figure).

This second tank is made of thin (a few mm) iron and is fixed by bolts and nuts (white circles inside the orange line in the figure).
As explained in the next subsection, sealant can be used around the nuts and bolts to make the second tank even less susceptible to leaks.

\subsubsection{Sealing the Space between Hatches and Concrete}
\indent

\begin{figure}[h]
 \centering
 \includegraphics[width=0.7 \textwidth]{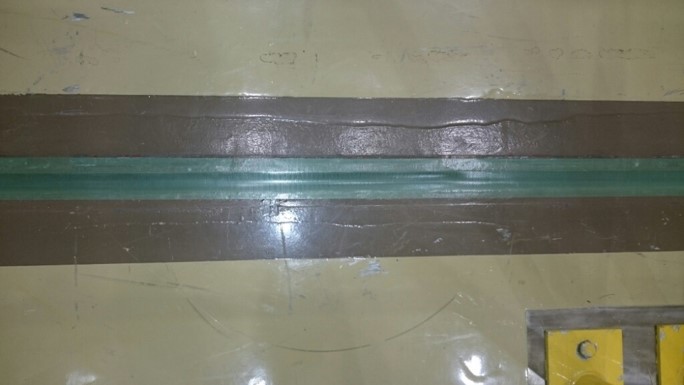}
 \caption{\setlength{\baselineskip}{4mm} 
   The space between hatches and concrete. The green tapes are
   used for the temporary sealing which has weak effects.
 }
 \label{fig:space}
 \end{figure}

Figure~\ref{fig:space} shows the space between the hatches and the concrete which corresponds to several millimeters.

\begin{figure}[h]
 \centering
 \includegraphics[width=0.7 \textwidth]{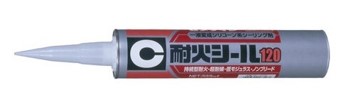}
 \caption{\setlength{\baselineskip}{4mm} 
   The container for the sealing material.
 }
 \label{fig:seal}
 \end{figure}

We propose to fill the space using air tight sealing materials. This sealing material has been successfully used by the MLF muon group~\cite{CITE:MLFMUON}.
Figure~\ref{fig:seal} shows the container for the sealing material.

\begin{figure}[h]
 \centering
 \includegraphics[width=0.4 \textwidth]{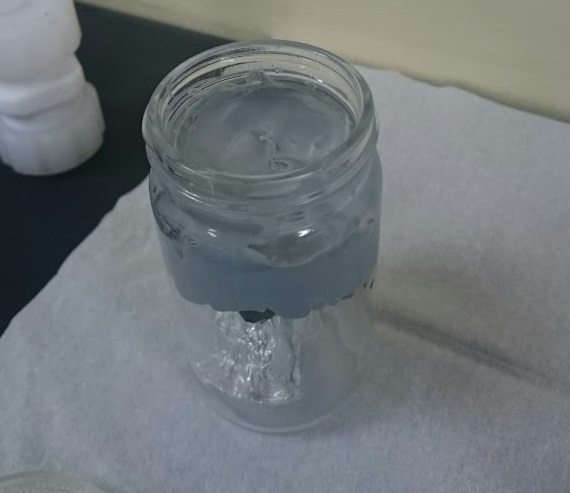}
 \includegraphics[width=0.6 \textwidth]{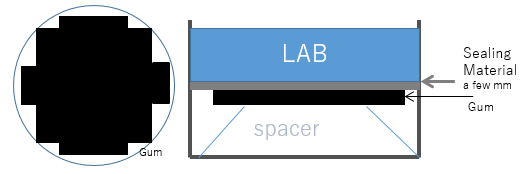}
 \caption{\setlength{\baselineskip}{4mm} 
 The setup for the mockup test to demonstrate the properties of the sealing material.
 Top: picture of the setup. Bottom: Cartoon to show the concept.
 }
 \label{fig:mockup}
\end{figure}

We have created a mockup test to demonstrate the effectiveness of the sealing material against LS.
The setup of the test is shown in Fig.~\ref{fig:mockup}.
The sealing material is filled in the space between container and the gums, and LS is put on the sealing material layer.
With the sealing material, no LS leaks were found during a period of over one month.
In the case of an emergency, the expected duration that the LS is stored on the
MLF third floor is a few days. Therefore, the sealant we have chosen ensures that LS will never fall down into the radiation hot areas, even in an emergency situation.

\begin{figure}[h]
 \centering
 \includegraphics[width=0.7 \textwidth]{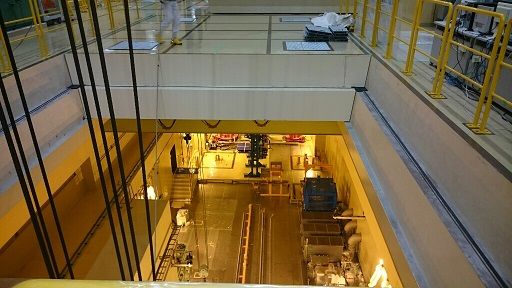}
 \caption{\setlength{\baselineskip}{4mm} 
The gum lines are located under the hatches.
 }
 \label{fig:gum}
 \end{figure}

An additional concern is potential damage to the sealing material due to a large earthquake causing motion between the sealant and concrete. However, gum lines are put under the hatches to prevent motion of the concrete during an earthquake as shown in
Fig.~\ref{fig:gum}. 
We investigated the maximum coefficient of friction for the case that
the gum line is inserted between concretes using various references,
and concluded that it is between 0.7-1.5.
Given that the MLF building can withstand an earthquake up 0.25G, the hatches will not be moved even if a large earthquake occurs.

PTFE or rubber sheets can be put above the hatches and below
the detector in addition to the sealant in order to
avoid oil leakage.

\subsubsection{The Third Anti Oil-Leak Wall}
\indent

We will put a steel L-type angular piece for the third anti oil-leak wall which prevents LS from flowing into the target area. The height of the wall is 5 cm just to prevent LS from moving past it.
This wall will be fixed by bolts and nuts and will also be sealed using the sealing material.


\subsubsection{Slow Monitors}
\indent

\begin{figure}[htbp]
\centering
\includegraphics[width=170mm]{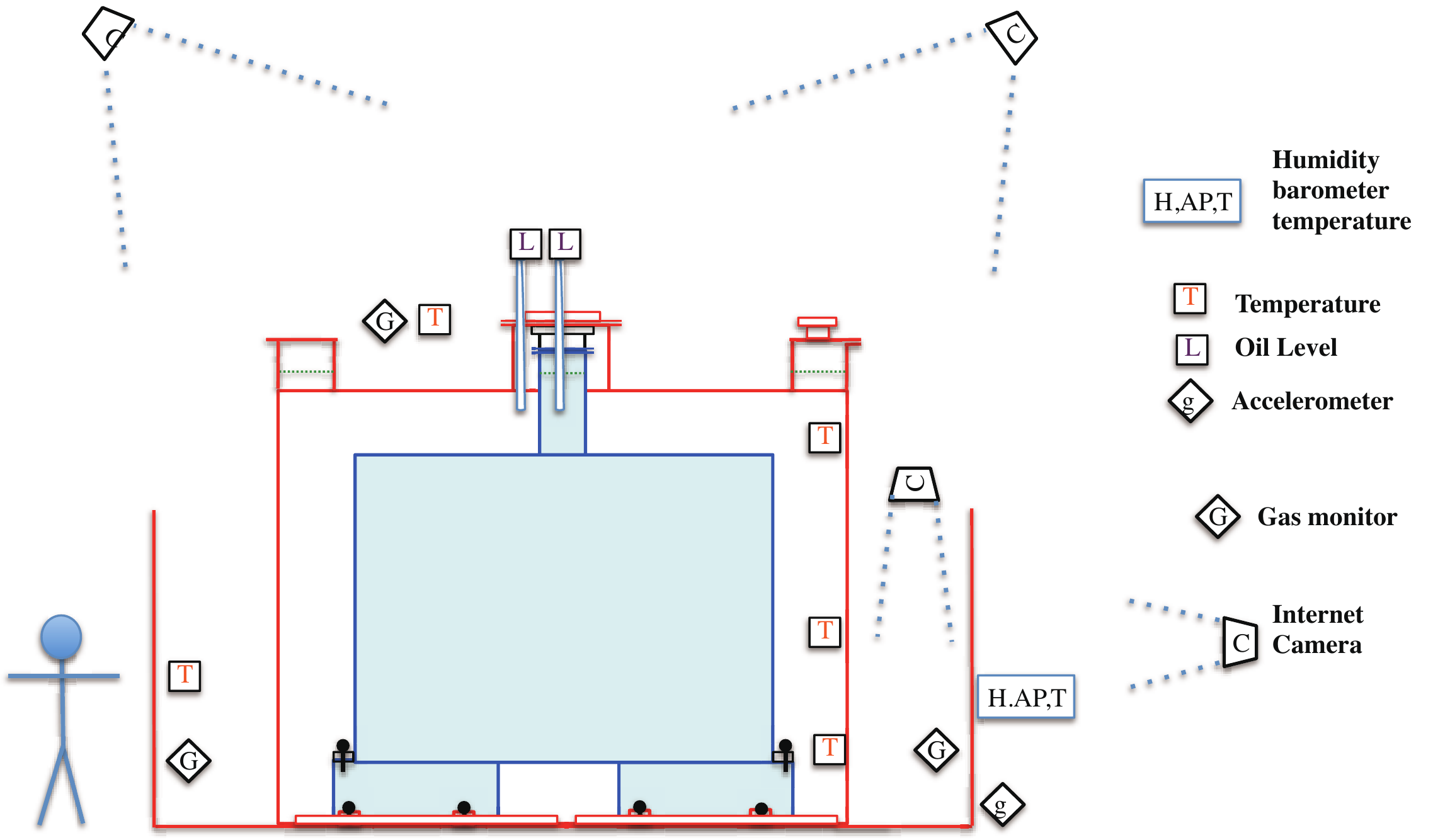}
\caption{\small{Safety and oil monitors.}}
\label{fig:Monitors}
\end{figure}

The JSNS$^2$ detector uses a large amount of organic oil and thus for safety reasons it is necessary to monitor various parameters of the detector at all times. 
Figure~\ref{fig:Monitors} shows a set of possible monitors to accomplish this goal. 
Generic safety monitors include: organic vapor monitors, and thermometers which will be placed around the detector.  
Internet cameras will also be used to view the whole detector, the inside the anti oil-leak tank, and the gauges of the slow monitors.  

In addition, specific tools will be used to monitor the liquid scintillator. Thermometers will be placed in various depths in the stainless steel tanks to make sure the temperature does not change too much. Liquid level monitors will be placed in the chimneys of the tanks which measure the absolute level of the oils and the level difference between the outer and inner tanks. 
There will be also one set of absolute atmospheric pressure, humidity, and environmental temperature monitors. 

An accelerometer will be used to monitor any shocks the detector experiences while it is moved by the crane or when earthquakes happen.

\begin{figure}[htbp]
\centering
\includegraphics[width=150mm]{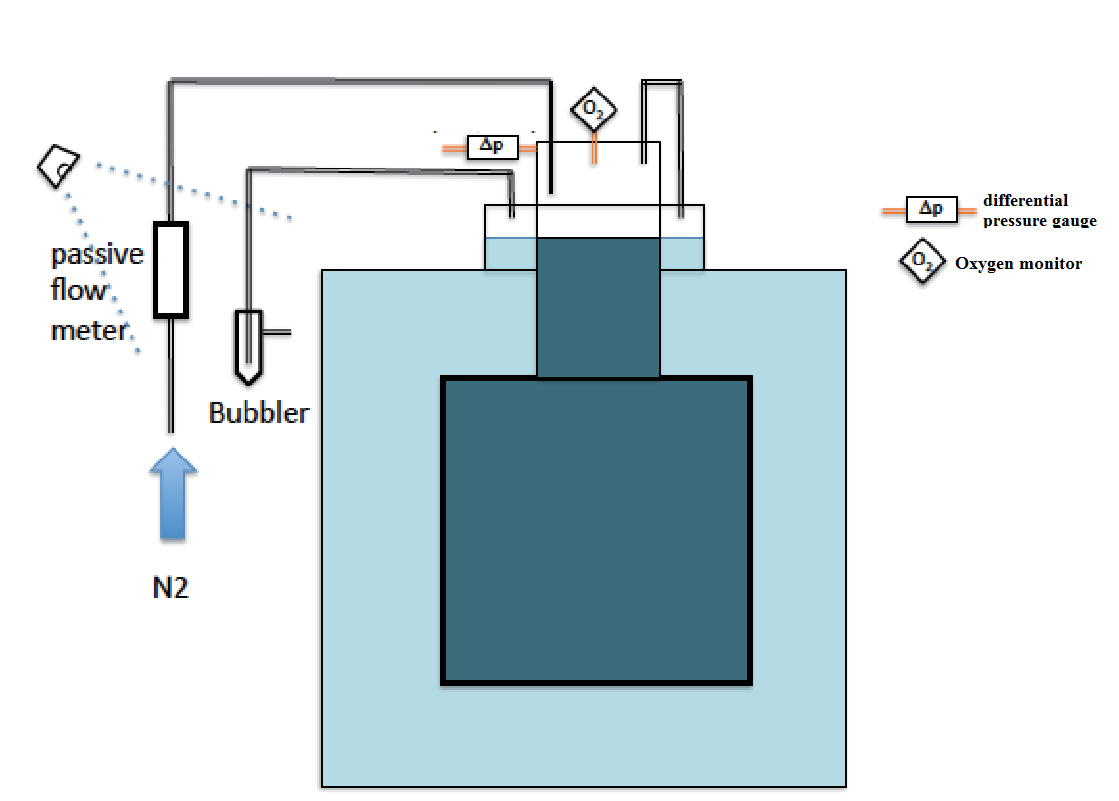}
\caption{\small{Gas monitors}}
\label{fig:Gas_Monitors}
\end{figure}

The gas phase of the tanks will be purged by nitrogen gas continuously. 
Figure~\ref{fig:Gas_Monitors} shows an idea for gas monitors. 
The gas flow will be monitored by internet cameras which view the flow meter and bubblers at the outlet of the nitrogen gas line. 
The relative pressure between the gas phases of the acrylic vessel and atmospheric pressure will be monitored by a differential pressure gauge. 
An oxygen monitors will measure the oxygen concentration in the Gd-LS vapor phase. The gauges will be readout online and checked by the shifters. 



\subsection{PhotoMultiplier Tube and its system }

\subsubsection{Overview of the PMT system\label{sec:PMT_overview}}
\indent

\begin{figure}[htbp]
\begin{center}
\includegraphics[width=0.75 \textwidth]{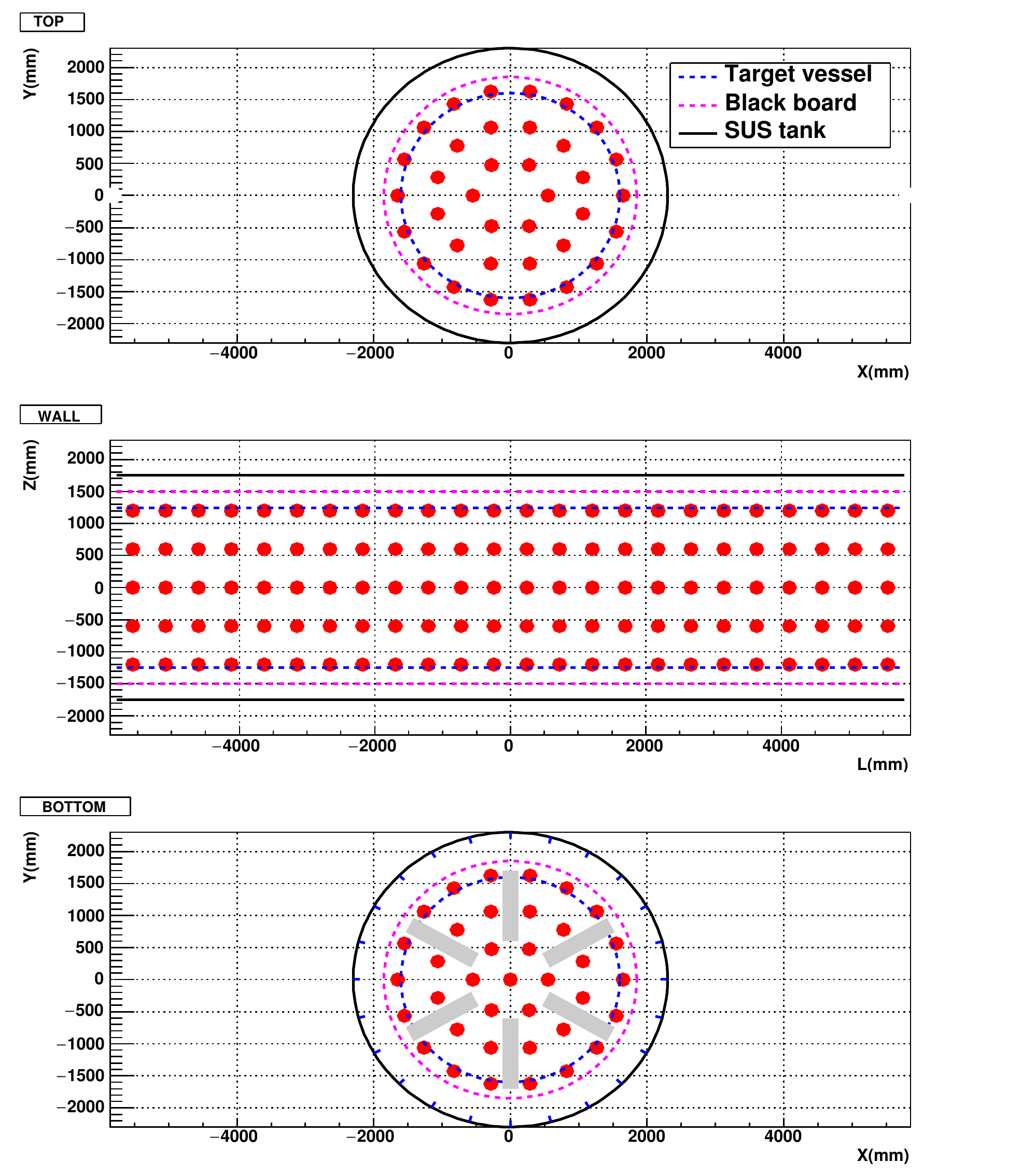}
\end{center}
\caption{\setlength{\baselineskip}{4mm}
The 8-inch PMT map for the inner volume.}
\label{TargetPMTmap2}
\end{figure}

Figure~\ref{TargetPMTmap2} shows the PMT arrangement in the
stainless steel tank. 
The dominant $\bar{\nu}_e$ background from $\mu^{-}$ is separated from the signal using the energy spectrum.
The required energy resolution to perform this separation in JSNS$^2$ is $\Delta E/ E< 15\%/\sqrt{E[{\rm MeV}]}$. 
In order to satisfy the requirements, JSNS$^2$ will use 193 8-inch photomultiplier tubes to detect the scintillation light from the neutrino target.
The photocathode coverage achieved in this design is $\sim$ 11~\%. 
Each PMT is covered by a magnetic shield (FINEMET: see later subsection for more details) to reduce the effect of the Earth's magnetic field. 
The PMTs are totally submerged in LS (organic oil) and the base circuit of the PMT is moulded in an epoxy resin to prevent the contact with the oil. 
This oil-proof treatment has been used in the KamLAND, Double Chooz, DayaBay and RENO experiments. 
The signal and HV are supplied by separated two 25 m Teflon jacket cables.
Positive HV is supplied from the HV supply to the splitter and sent to the PMT after ripple and noise filters in the splitter circuit.

\subsubsection{PMT\label{sec:PMT_PMT}}
\indent

\begin{figure}[htbp]
\centering
\includegraphics[width=80mm]{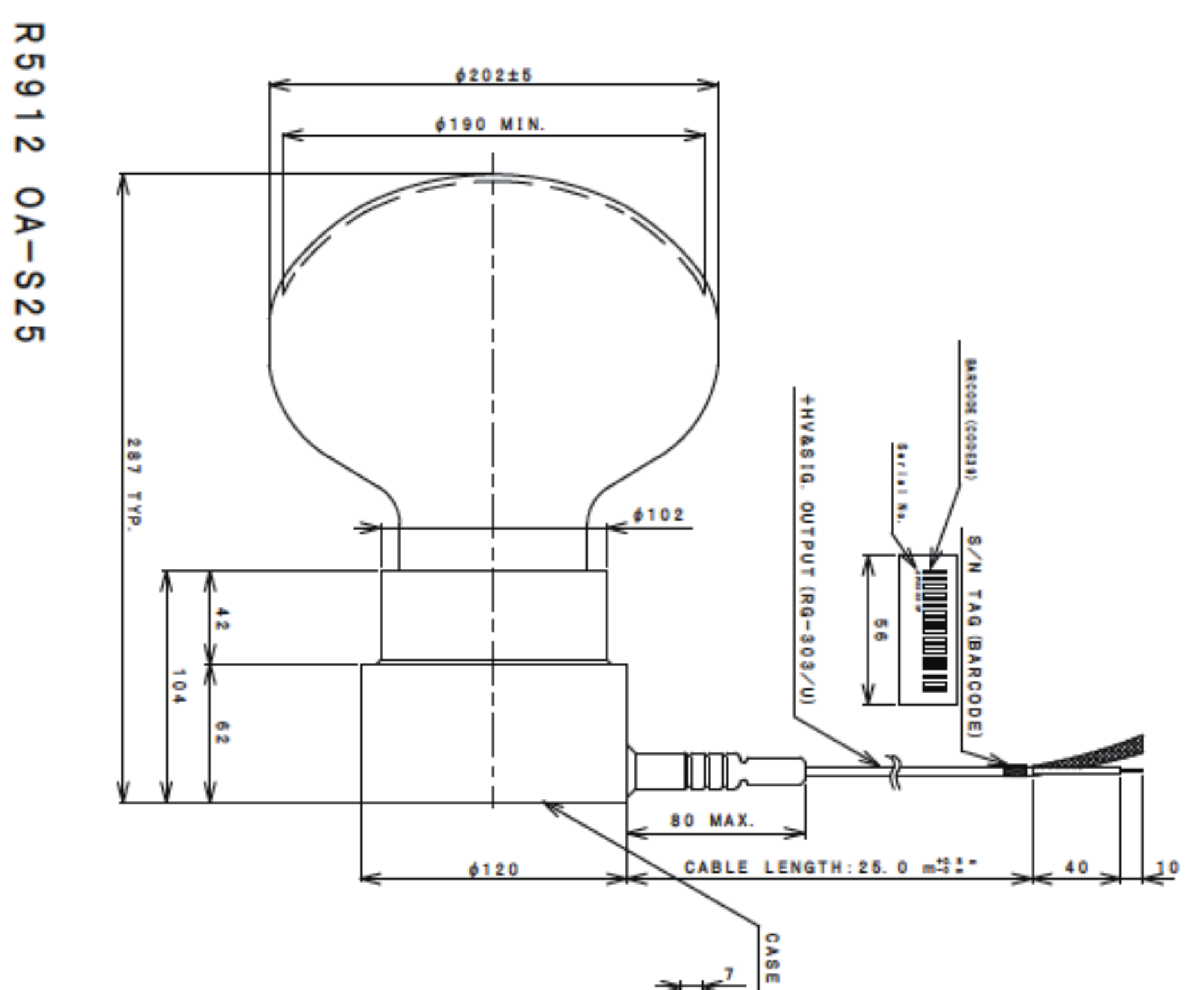}
\caption{\small{PMT structure}}
\label{fig:PMT_shape}
\end{figure}

Figure~\ref{fig:PMT_shape} shows the structure of the 
8-inch R-5912 PMT chosen for this experiment~\cite{HPKCat}. 

The PMTs are 8 inches in diameter with borosilicate glass windows and photocathodes at lease 190 mm in diameter.
The photocathode material is a bialkali with a spectral response range between 300~nm to 650~nm and a peak wavelength of 420~nm.
The typical quantum efficiency at 390~nm is 25~\%.
The typical gain with high voltage 1500~V is $1.0\times 10^{7}$ giving single p.e. pulse heights of a few mV.
The rise time of a pulse is typically 3.6~ns, the transit time (TT) and the transit time spread (TTS) are measured to be 54~ns and 2.4~ns (FWHM), respectively. 
The typical peak to valley ratio for a single photoelectron signal is 2.8 and the typical dark count rate after 15~hours storage in darkness is 4~kHz.
  
\begin{figure}[htbp]
\centering
\includegraphics[width=130mm]{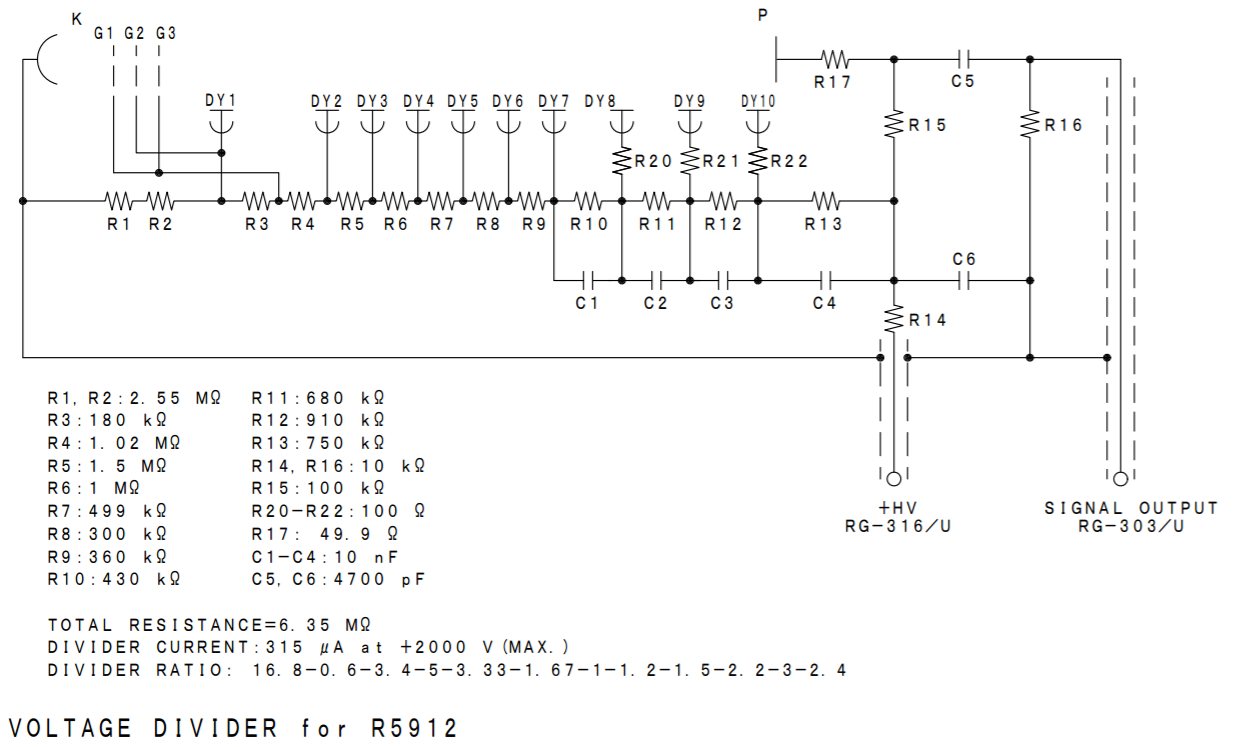}
\caption{\small{Tapered breeder circuit. }}
\label{fig:Breeder_Circuit}
\end{figure}

Figure~\ref{fig:Breeder_Circuit} shows an example of the divider circuit.
Positive HV is supplied and the cathode potential is set to be ground. 
HV is provided by one cable and the signal is readout using a separate one. 
The cable is back-terminated to damp the signal reflections quickly.

\begin{figure}[htbp]
\centering
\includegraphics[width=100mm]{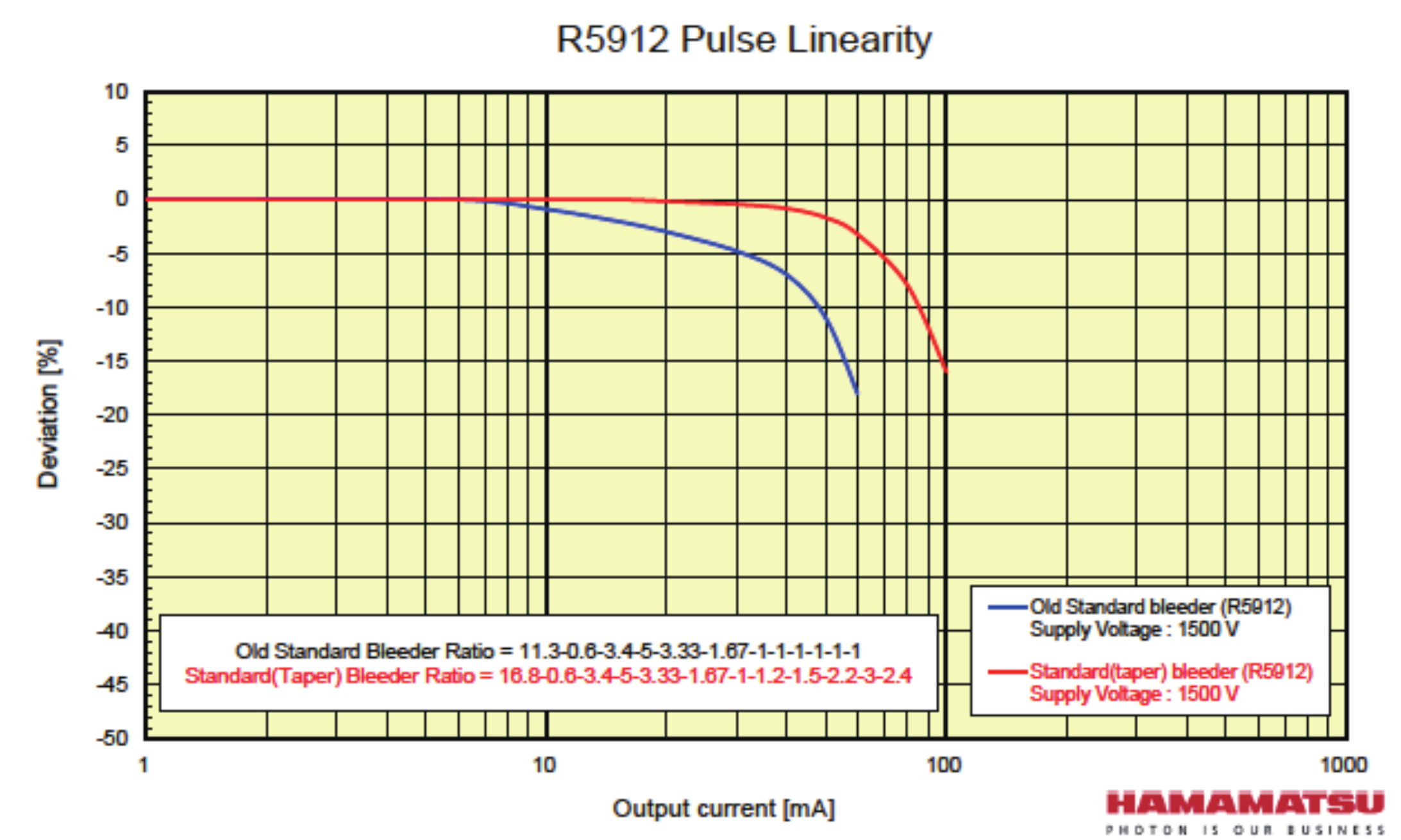}
\caption{\small{Comparison of dynamic ranges for the tapered base circuit and a normal one.}}
\label{fig:PMT_Dynamic_range}
\end{figure}

For a neutrino with the maximum energy of $\sim$50~MeV right next to a PMT, the PMT will receive 2000~p.e. 
On the other hand, the pulse threshold of the PMT is $\sim 0.3$~p.e.
Therefore, the linear dynamic range required for the PMT is 1~p.e. to 2000~p.e. 
To realize such a large dynamic range, a tapered voltage divider will be used. 
Figure~\ref{fig:PMT_Dynamic_range} compares the linear dynamic range for a standard breeder circuit and the tapered breeder circuit. 

\subsubsection{Test, Delivery}
\indent

In this subsection, the PMT delivery scenario, including testing and transportation to the experimental site, is described.
The PMTs will be delivered from the vendor to Kitasato University for the initial acceptance test, and then they will be transported to J-PARC for assembly at the experimental site. 
The PMTs will be attached to the support structure inside the outer stainless tank.
In total, we need 193 PMTs 8-inch PMTs for the construction of a single detector.

\begin{figure}[htb]
\begin{center}
\includegraphics*[width=7.8cm,clip]{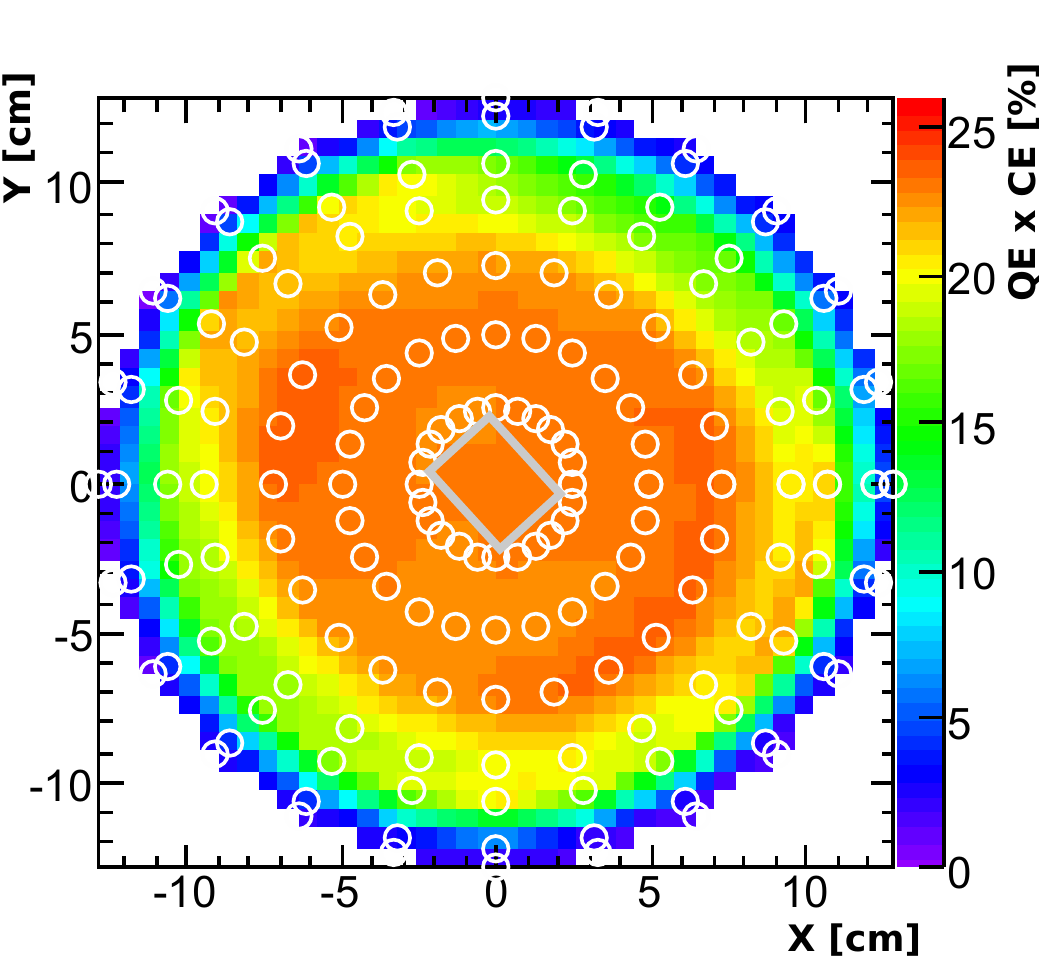}
\end{center}
\caption{\setlength{\baselineskip}{4mm}
QE$\times$CE map for a typical PMT, which is used for the Double Chooz experiment (Hamamatsu R7081, 10-inch). Although we will employ 8-inch PMTs, they have similar shape and dynode structure. Cable outlet is along the x-axis. The box at the center indicates the first dynode. White circles show the light injection points\cite{DCPMT}.}
\label{QECEmap}
\end{figure}

\paragraph{Testing and HV tuning} 
We propose the following plan for testing the PMTs based on experience with the acceptance tests used for PMT delivery in the Double Chooz experiment \cite{DCPMT}.
The items included in the acceptance tests are:
\paragraph{i)} Measure the gain value as function of the supplied HV in order to find the HV values 
which give specific gains ($ 10^6 \sim  10^7$). Compare the results with what is provided by the vendor
and define the nominal HV value for each PMT.
\paragraph{ii)} Measure the peak to valley ratio. This must be greater than 2.0.
Compare the results with what is measured by the vendor.
\paragraph{iii)} Measure the transit time spread which should be around 3 ns(FWHM).
The average value and the deviation are used as input parameters to MC simulation. 
\paragraph{iv)} Measure the QE (quantum efficiency) $\times$ CE (collection efficiency) with a light source.
The average value and the deviation are used as input parameters to MC simulation. 
\paragraph{v)} Measure dark counts at threshold (1/4 of the single photoelectron peak) with the nominal HV value.
\paragraph{vi)} All PMTs are burned-in for more than 24 hours with the nominal HV.
We will require the vendor to perform items i), ii), iii) and v) for all PMTs before delivery in order to compare to our results.
During the acceptance test for the Double Chooz experiment, the position dependence of QE $\times$ CE at the PMT surface was measured in order to check the uniformity of the response. Figure~\ref{QECEmap} shows an example result of the uniformity mapping of QE$\times$CE for a Double Chooz PMT.
We require the vendor to perform the QE $\times$ CE uniformity measurement on the PMT surface for 5 $\sim$ 10 \% of all PMTs delivered so that the typical performance of delivered PMTs are obtained because this is an input to the MC.

\begin{figure}[h]
\begin{center}
\includegraphics*[width=12cm,clip]{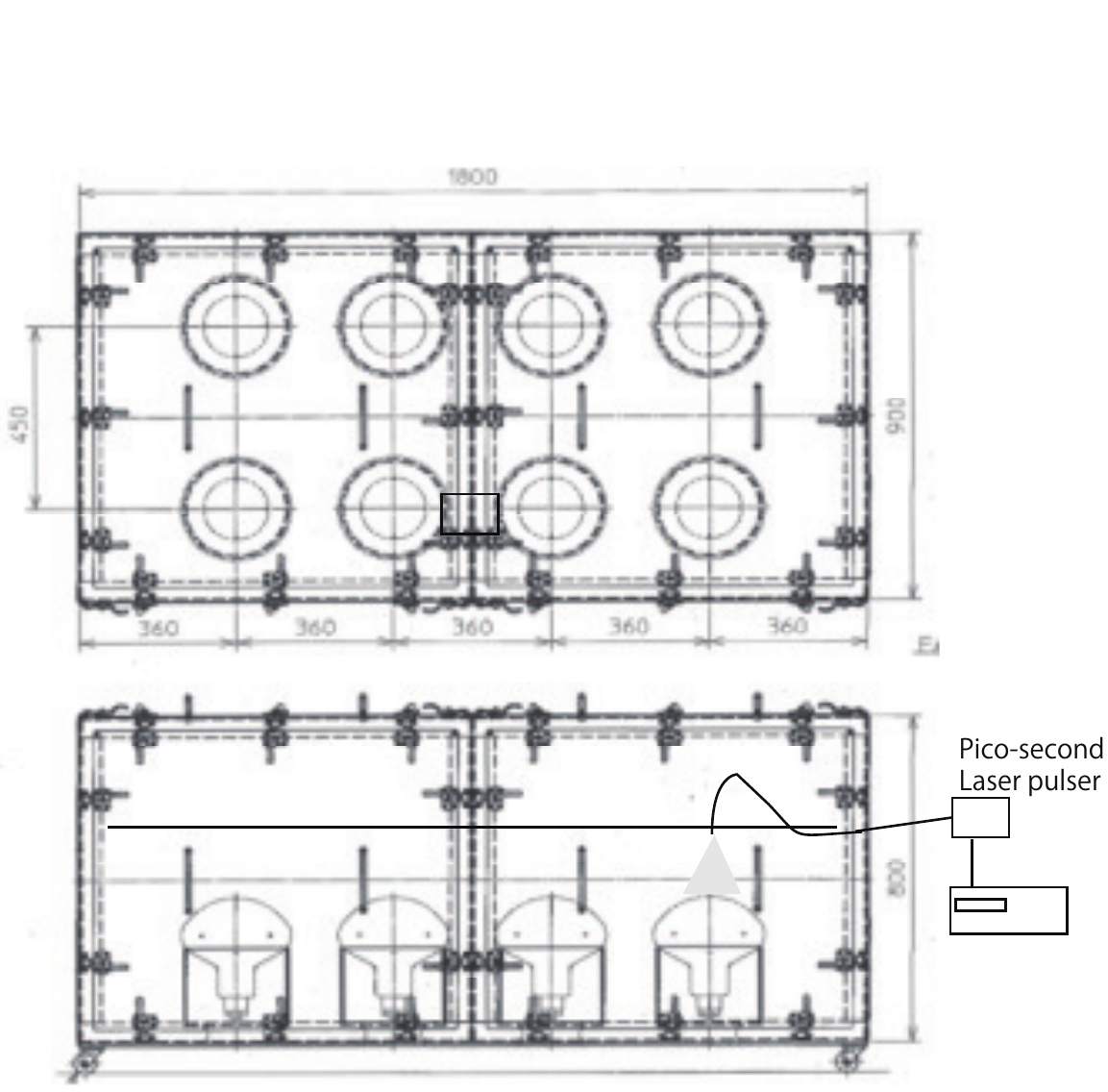}
\end{center}
\caption{\setlength{\baselineskip}{4mm}
PMT test system, which was originally developed for the Double Chooz PMT test. Eight PMTs can be tested at once. We will perform some modifications due to the difference in PMT sizes. Laser light is used to illuminate each PMT.}
\label{fig:8PMTsys}
\end{figure}

\paragraph{Test System}
For the PMT acceptance test, we plan to re-use the PMT test system developed for the Double Chooz experiment. Figure~\ref{fig:8PMTsys} shows the system. It consists of a black box and cylindrical $\mu$-metals to protect PMT from the geomagnetic field. Some modifications are needed due to the difference of the PMT size.

A pico-second laser ($<$ 60ps width) with a wavelength of 440$\pm$3 nm (Advanced Laser Diode System Co,) is used as a light source since a short pulse and precise timing information are necessary to measure the transit time spread. The intensity of the laser pulse is adjusted with the laser driver box and an ND filter.
The light is injected 20 cm away from the PMT surface with the diffuser in order to illuminate whole photocathode.

\paragraph{Test schedule}
The schedule of PMT delivery strongly depends on the vendor's production capability. 
The first delivery of 50 PMTs takes 4.5 months after the order is placed. The second delivery with another 50 PMTs takes 1 more month.
The delivery rate will increase to 100 PMTs/month at most. In total, the delivery of 200 PMTs will take roughly 6.5 months.
The PMTs will be delivered from the vendor to the testing site in a carton box which contains 8 PMTs.
As a part of the acceptance test above, we perform the items from i) to iv) for a single PMT and items v) and vi) for eight PMTs at once. Therefore, we can treat 8 PMTs/day at most.
Every week, we expect that a few cartons of PMTs will be delivered to the testing site and after testing they will be transported to the assembly site within a week. Therefore, a large space to store the PMTs at the testing site is unnecessary.

\begin{figure}[htb]
	\begin{center}
		\includegraphics[width=0.7\columnwidth,clip]{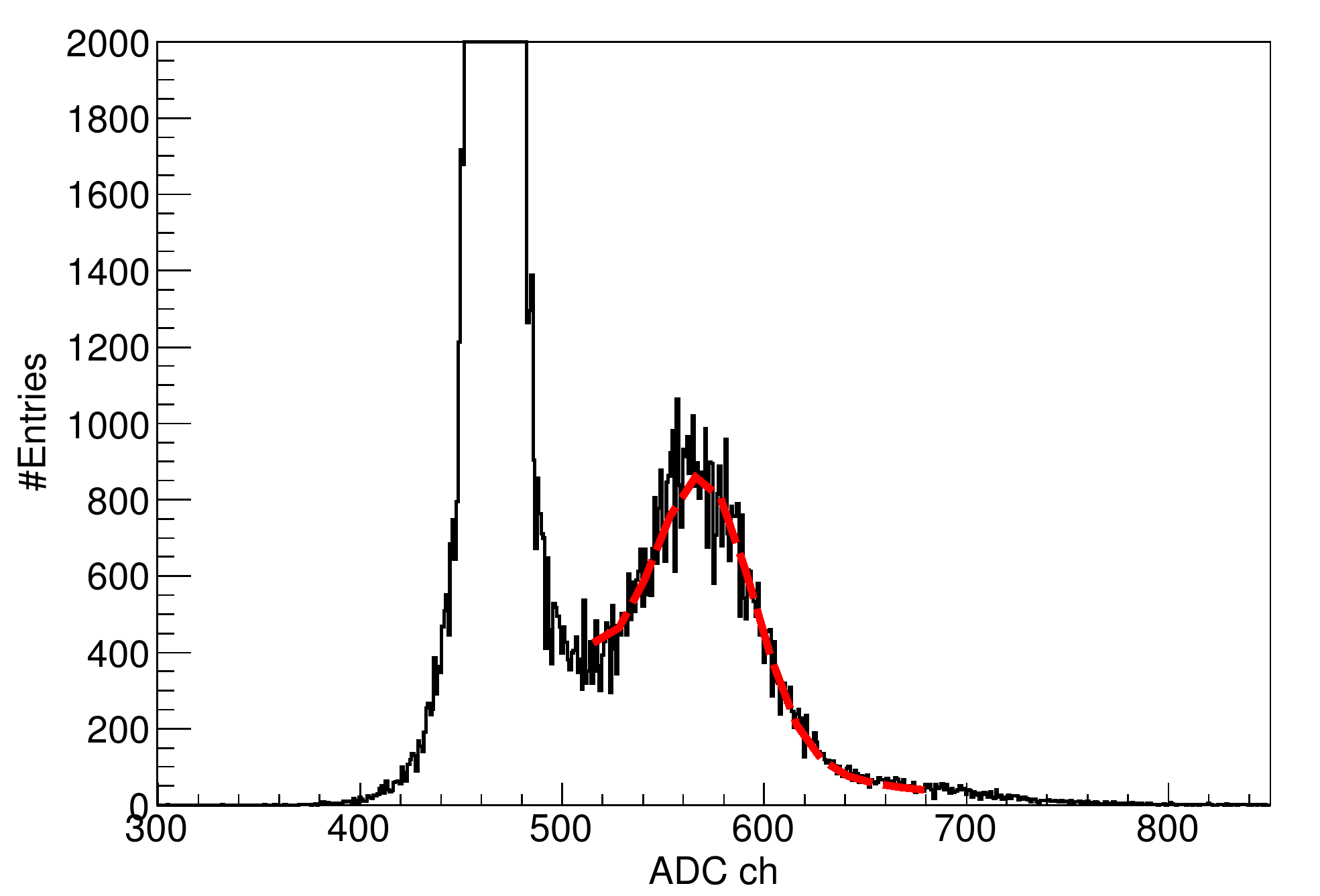}
	\end{center}
	\caption{\setlength{\baselineskip}{4mm}
	Example charge distribution for a Hamamatsu R5912 PMT measured in the acceptance test system.
	}
	\label{fig:ADC_TDC}
\end{figure}

\paragraph{Example results by test system}
Figure~\ref{fig:ADC_TDC} shows an example measurement of the charge distribution of an 8-inch PMT (Hamamatsu R5912) using the acceptance test system.
The supplied high voltage (HV) is $1700\,\mathrm{V}$, which is the typical value to obtain a gain of $10^7$ for the R5912 PMT.
The intensity of injected laser pulses for this test corresponds to about 0.2 photoelectrons (PEs).
In order to match the dynamic range of the ADC, we used a pre-amplifier with a gain of 10.
The 1 PE peak is shown at around 570 ADC count in Fig.~\ref{fig:ADC_TDC}.
The dashed line in Fig.~\ref{fig:ADC_TDC} is the fitting function (exponential + Gaussian) for the 1 PE peak.
From the fit result, the number of entries at the peak and valley were 862.8 and 425.7, respectively, so the peak-to-valley ratio was measured to be approximately 2.0.

The gain of a given PMT is calculated according to the following equation \cite{DCPMT}
\begin{equation}
G = \frac{2 \times ADC_{count} \times G_{ADC}}{G_{Amp} \times e},
\end{equation}
where $ADC_{count}$ is the ADC count of the 1 PE peak after subtracting the pedestal count, $G_{ADC}$ is the calibration factor of charge ADC count, $G_{Amp} = 10$ is amplification gain of pre-amplifier, and $e$ is the electron charge.
The factor of 2 is a correction factor for the back-termination of the base circuit of our 8-inch PMT.
From this expression, we obtained a gain value of approximately $0.8 \times 10^6$ with $1700\,\mathrm{V}$.

We can also measure the related transit time (TT) and its spread (TTS) using the same PMT test system.
The TDC used for this measurement was started with the trigger signal, which generates laser pulse, and was stopped with the signal created by discriminating the PMT signal with a threshold level of 0.25 p.e.
We briefly evaluated TTS for a sample of 8-inch PMTs, and obtained $\sim 1.7\ \mathrm{ns}$, which is consistent with the value $\sim 2.0\ \mathrm{ns}$ obtained by Hamamatsu.

\subsubsection{Noise test at MLF 3rd Floor}
\indent

\begin{figure}[h]
\begin{center}
\includegraphics[scale=0.5]{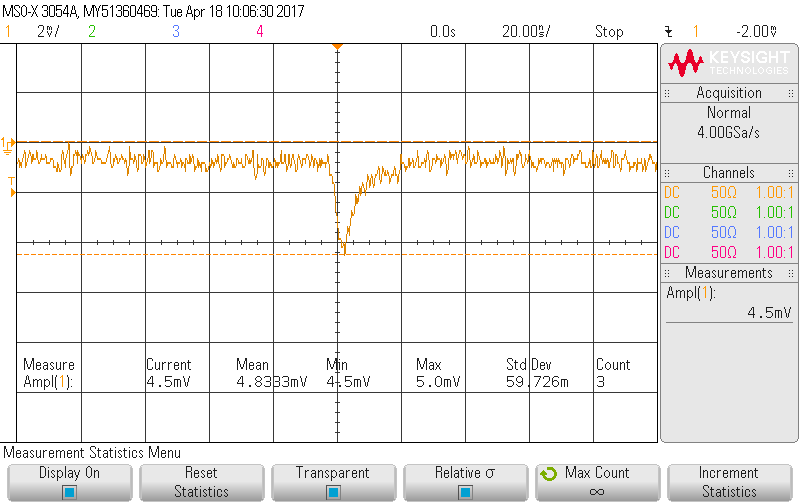}
\end{center}
\caption{\setlength{\baselineskip}{4mm}PMT signal in the oscilloscope. The x-axis bin size is 20 ns and the y-axis bin size is 2 mV. The RMS of the baseline is below 1 mV.} 
\label{fig:noise_test_MLF_jspark}
\end{figure}

We performed a noise test of the 8-inch PMT with a 500 MHz oscilloscope at the MLF 3rd floor while proton beam was on. Positive 1360 V was applied and samples were taken with a 2 mV trigger threshold, as shown in Fig.~\ref{fig:noise_test_MLF_jspark}. The standard deviation of the baseline before the PMT signal region was measured to be $\sim$0.2 mV and the average pulse height of a single photoelectron was $\sim$5 mV. 

\subsubsection{Magnetic Shield}
\indent

\begin{figure}[hbtp]
        \begin{center}
                \includegraphics[scale=0.52]{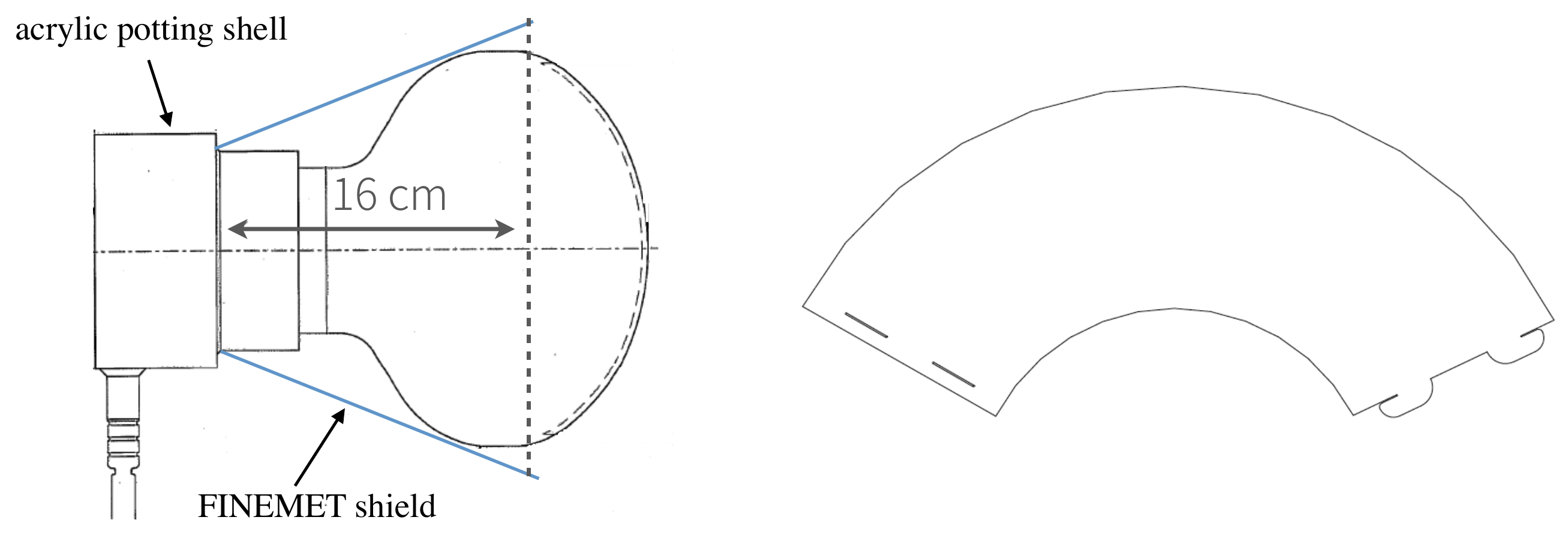}
                \caption{\setlength{\baselineskip}{4mm} 
                Cross section of the FINEMET magnetic shielding around a PMT (left) and an example shielding piece (right) \cite{CITE:DBFINEMET}.}
                \label{finemet}
        \end{center}
\end{figure}

The PMT is surrounded by FINEMET\cite{CITE:FINEMET} which is used as magnetic shielding.
The Daya Bay experiment also uses FINEMET for the Hamamatsu R5912 PMT.
Figure~\ref{finemet} shows a cross section of the FINEMET magnetic shielding around a PMT and an unfurled FINEMET shield \cite{CITE:DBFINEMET}.
The slant height of the magnetic shielding is 16 cm.

\subsubsection{Magnetic Field Measurement on the 3rd Floor of the MLF}
\indent

\begin{figure}[h]
\begin{center}
\includegraphics[scale=0.5]{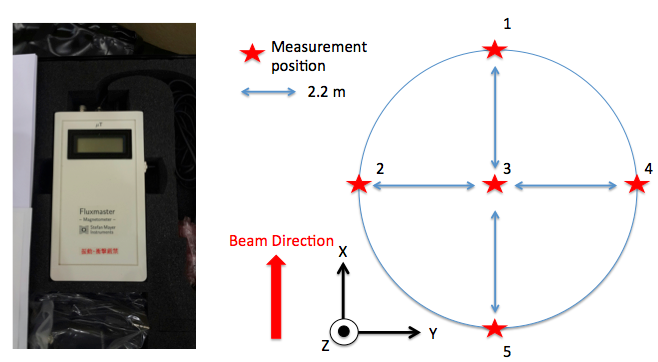}
\end{center}
\caption{\setlength{\baselineskip}{4mm}Left : Fluxmaster magnetometer. Right : Magnetic field measurement points at the proposed detector position. We chose five different points and two different heights (floor level, and 1.3 m above the floor), and measured three perpendicular components of the magnetic field at each point.} 
\label{fig:magnetic_field_point_jspark}
\end{figure}

Each PMT will be surrounded by FINEMET to eliminate interference from the geomagnetic field which will increase PMT performance. The required thickness of FINEMET is proportional to the magnetic filed strength. Therefore, we measured magnetic field strength at the proposed detector position on the 3rd floor of the MLF. We used a Fluxmaster magnetometer, which can measure magnetic fields up to 2000 mG. Three perpendicular direction components were measured at five different points and two different heights (floor level and 1.3 m above the floor). Figure~\ref{fig:magnetic_field_point_jspark} shows the magnetometer and the measurement points along with the definition of the three direction components. Table~\ref{tab:magnetic_field_jspark} summarizes the measurement result and shows that the magnetic field strength at the proposed detector position was $\sim$300-500 mG without FINEMET. 

\begin{table}
\begin{center}
\begin{tabular}{c | c | c | c | c}
\hline
Floor & X (mG) & Y (mG) & Z (mG) & Total (mG) \\ \hline
1 & - 309  &  - 3  & - 233 & 387  \\ \hline
2 &  - 317 &  - 20   &  - 186 & 368 \\ \hline
3 &  - 292  &  - 22   &  - 159 & 333 \\ \hline
4 & - 336 & 18   &  - 140 & 364  \\ \hline
5 & - 378 &  - 5  & - 92  & 389 \\ \hline
\hline
1.3 m height & X (mG) & Y (mG) & Z (mG) & Total (mG) \\ \hline
1 & - 227 &  15  & - 233  & 326 \\ \hline
2 & - 291 &  - 97  & - 82  & 318 \\ \hline
3 & - 71 &  - 36  & - 465  & 472 \\ \hline
4 & - 353 &  - 24  & 225  & 419 \\ \hline
5 & - 478 &  10  & - 40  & 480 \\ \hline
\end{tabular}
\end{center}
\caption{\setlength{\baselineskip}{4mm}Magnetic field strength measured by the magnetometer. Between $\sim$300-500 mG was measured at the proposed detector position.}
\label{tab:magnetic_field_jspark}
\end{table}

\subsection{Veto system}
\indent

\newcommand{\mycolor}{}

\begin{figure}[hbtp]
        \begin{center}
                \includegraphics[scale=1.0]{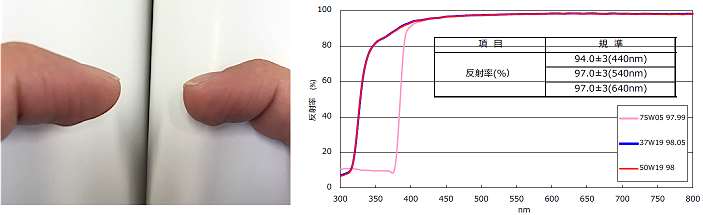}
                \caption{\setlength{\baselineskip}{4mm}
                  Left: Picture of the LUIREMIRROR reflection sheet, a real finger on the left and its mirror image on the right.
                Right: Reflectance as a function of wavelength~\cite{CITE:REIKO}.}
                \label{reflectionSheet}
        \end{center}
\end{figure}

The central target and buffer volumes are surrounded by a veto layer.
The veto layer is used to reject charged and neutral particles coming from outside the detector,
and to detect energy leakage from the central volumes.
The veto layer is {\mycolor 25$\sim$45} cm thick and filled with the same LS filled in the buffer region (LAB-based liquid scintillator without Gd).
The inner and outer surfaces of the veto layer are covered by reflective sheets made of {\mycolor REIKO LUIREMIRROR~\cite{CITE:REIKO}.
This material has good reflectance above 380 nm wavelength and for wavelengths longer than 440 nm the reflectance is more than 94 \% as shown in Fig.~\ref{reflectionSheet}.
}

\begin{figure}[hbtp]
        \begin{center}
                \includegraphics[scale=0.6]{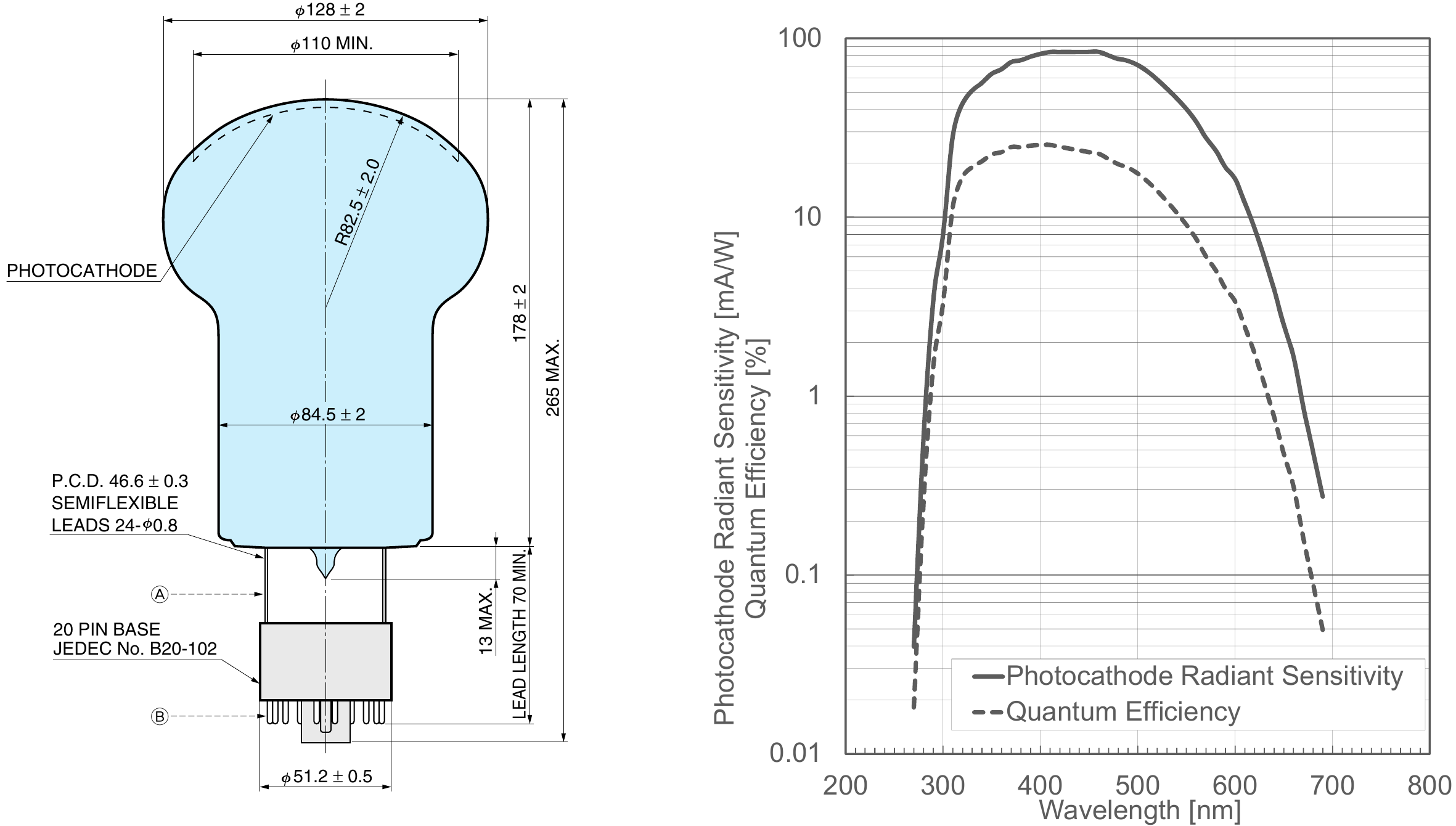}
                \caption{\setlength{\baselineskip}{4mm}
                  Schematic (left) and performance (right) of the Hamamatsu R6594 PMT~\cite{HPKCat}.}
                \label{r6594}
        \end{center}
\end{figure}

\begin{figure}[hbtp]
        \begin{center}
                \includegraphics[scale=0.4]{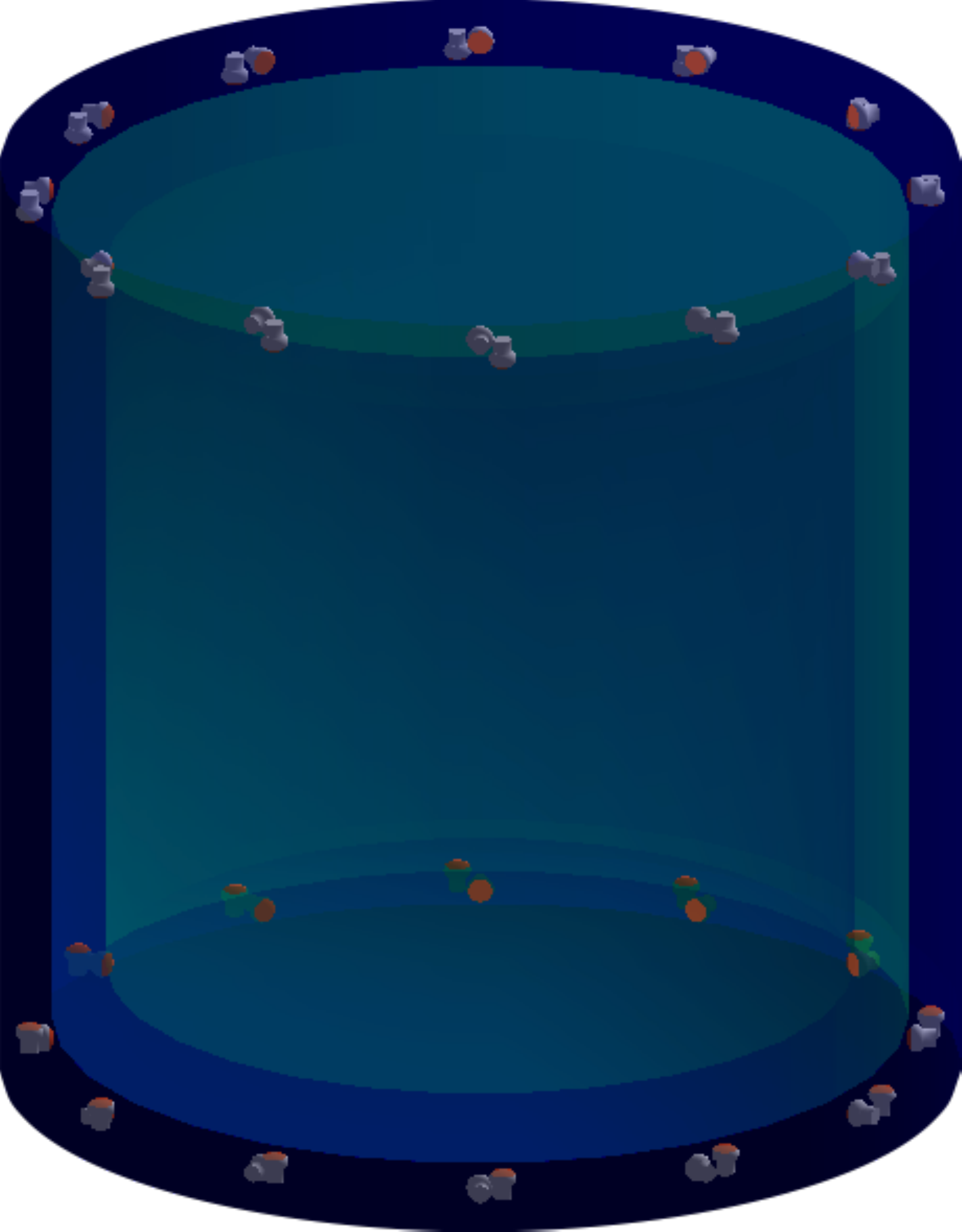}
                \caption{The PMT arrangement in the veto layer.}
                \label{pmtArrange}
        \end{center}
\end{figure}

Scintillation light in the veto layer is converted to the photo-electrons by Hamamatsu R6594 (5") PMTs which are located in the layer (Fig.~\ref{r6594}).
As shown in Fig.~\ref{pmtArrange}, the PMT arrangement has a 12-fold symmetry. Both the top and bottom layers are viewed by 12 PMTs, and the side region is viewed by 12$\times$2 PMTs.
This PMT arrangement was chosen because it provides a sufficient light yield, reasonable light uniformity, and reasonable position resolution with a finite number of PMTs and flexibility in the installation.


\begin{figure}[hbtp]
        \begin{center}
                \includegraphics[scale=0.48]{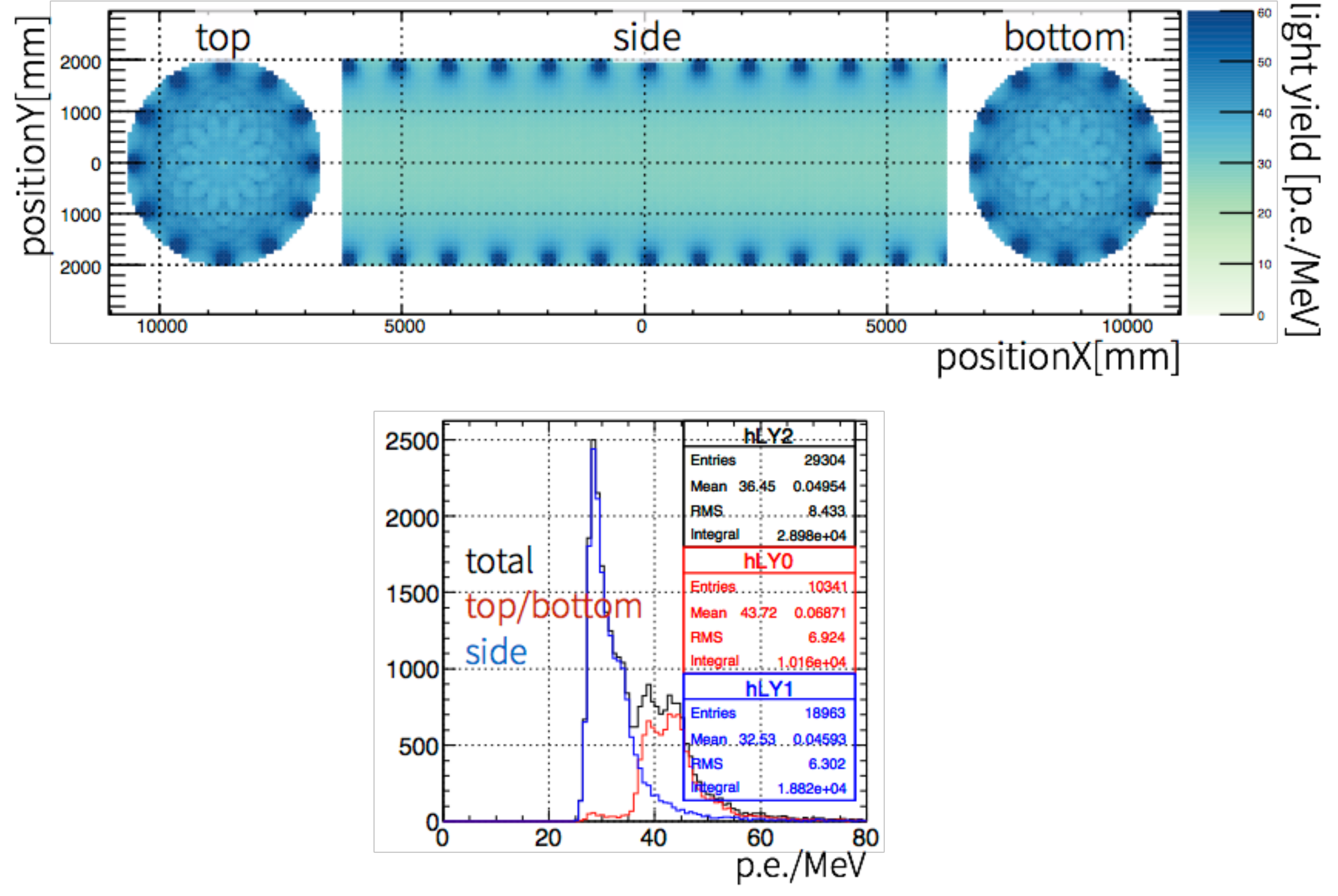}
                \caption{\setlength{\baselineskip}{4mm}
                  Expected light yields as a function of incident
                  positions (top), and its projection (bottom).}
                \label{lightYield}
        \end{center}
\end{figure}

The expected performance was evaluated with a Geant4 based MC simulation.
Here we assumed 90\% reflectance for the reflective sheets.
4 GeV/c muons were vertically and uniformly injected into the veto layer
and stopped at the surface of the buffer volume placed inside of the veto layer.
Fig.~\ref{lightYield} shows the expected light yield distribution.
The absolute light yield is expected to be more than 25 photoelectron/MeV for all incident positions.

\subsection{High Voltage System}
\indent

We plan to set up the high-voltage (HV) system on or close to the detector in order to reduce the required cable lengths and electrical noise.

\begin{figure}
\centering
\includegraphics[width=1.0\textwidth]{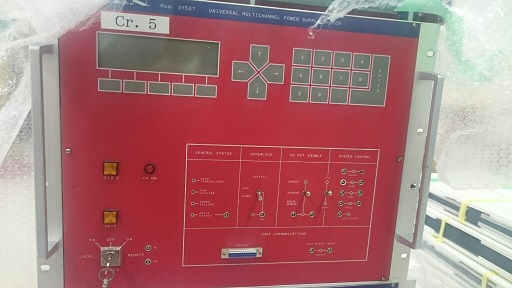}
\caption{\setlength{\baselineskip}{4mm}
Photo of SY527, CAEN mainframe. }
\label{fig:sy527}
\end{figure}

\begin{figure}
\centering
\includegraphics[width=1.0\textwidth]{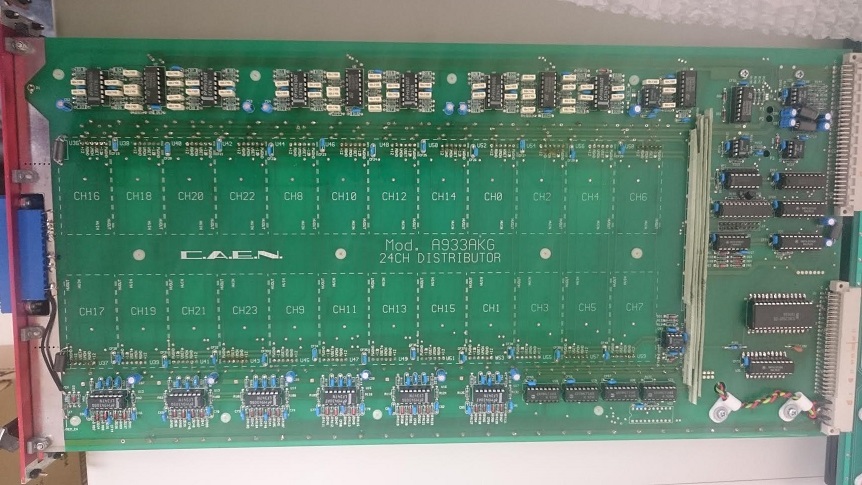}
\caption{\setlength{\baselineskip}{4mm}
Photo of A933KP, positive HV supply and 24-ch distributor module. }
\label{fig:a933kp}
\end{figure}

The HV system consists of a CAEN-SY527 mainframe (Fig.~\ref{fig:sy527}) and A933KP modules (Fig.~\ref{fig:a933kp}). Three SY527 and 29 A933KP have been donated from the Super-Kamiokande experiment, and were reliably working for over 20 years and it was confirmed at J-PARC that all channels are working. Each mainframe can contain ten A933KP and each A933KP has 24 positive-HV output channels. This system is old one, however, it has worked very well for a long time and we have a lot of reserve modules.

Each SY527 has power-supplies for 10 modules, control system circuits and cooling fans. All channels are controlled and monitored by a front panel console and via an RS232c connection to a PC, which provides a TV100-type emulator. It is also possible to use CAENNET to access the registers one-by-one. Through an internet connection to the PC, we can monitor and control the HV modules from outside the experimental area. It is possible to cut off all of the HV outputs using a front-panel lemo-connector interlock.

The A933KP HV module has one power supply and a 24-ch distributor. The power supply generates HV up to +2550 V and a current limit can be selected from 1 to 13mV. On each distributor channel, the voltage can be reduced by anywhere from 50 V to 900 V with a 0.2 V step. This means we can set the HV values for each channel as high as +2500 V. The maximum current for one channel is fixed as 0.5 mV. Within a given module, the HV values for each channel must be within 850 V of one another which can easily be achieved. The ramp up and down speed are also tunable and can be set anywhere between 10 to 500 V/sec. A block-type female connector (75 pins AMP201311-3 type) is used for HV outputs. 

\subsection{Liquid Scintillator}
\indent

There are several requirements for the JSNS$^{2}$ liquid scintillator (LS). First of all, it should be an appropriate chemical substance to satisfy the J-PARC fire safety regulation. It should produce sufficient scintillating light for obtaining good energy resolution that is necessary for observing a spectral modulation due to possible oscillation into sterile neutrinos. The liquid needs to be able to differentiate the antineutrino signal from accidental and neutron backgrounds. It should also be stable to maintain the sterile neutrino search for several years.

LS is a mixture of base solvent, primary fluor, and secondary wavelength shifter (if necessary). Linear Alkyl Benzene (LAB, C$_{n}$H$_{2n+1}$-C$_{6}$H$_{5}$, n = 10 $\sim$ 13) is chosen as the base solvent because it is an environmentally friendly, non-toxic material with a high flash point of 152 $^{\circ}$C. It also provides a long attenuation length of $>$10 m at 430 nm, and produces a large light yield of $\sim$10000 photons per MeV. Isu Chemical Company in South Korea produces high quality LAB and can deliver it in a clean storage container for JSNS$^{2}$ at a reasonable price. 

The density of LAB is $\sim$0.86 g/L. LAB consists of four main molecules with different number of carbon and hydrogen atoms. These different components can be measured by a gas chromatography mass spectrometer (GC-MS). LAB suffers from non-linear energy response due to quenching effects at low energies (i.e. no release of scintillating light even with deposited energy). The RENO experiment has measured the Birk's constant for LAB as $\sim$0.012 cm/MeV using a Ge detector~\cite{cite:RENO_LS_NIM}.  

\subsubsection{Recipe of LS}
\indent

\begin{figure}[h]
\begin{center}
\includegraphics[scale=0.6]{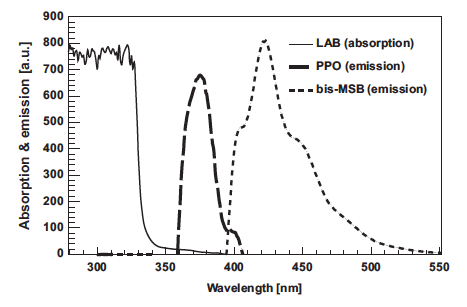}
\end{center}
\caption{\setlength{\baselineskip}{4mm}Emission spectra of LAB, PPO, and bis-MSB. The bis-MSB emission spectrum is well-suited for the high-quantum efficiency of the JSNS$^{2}$ PMTs.} 
\label{fig:fig7_from_nim_jspark}
\end{figure}

Because of LAB's emission spectral peak at $\sim$340 nm, it should be mixed with a wavelength shifter to make the emitted light suitable for PMT detection. We have chosen scintillation-grade PPO (2,5-Diphenyloxazole, C$_{15}$H$_{11}$NO) for a primary fluor and scintillation-grade bis-MSB (1,4-Bis(2-methylstyryl) benzene, (CH$_{3}$C$_{6}$H$_{4}$CH=CH$_{2}$)$_{2}$C$_{6}$H$_{4}$) for a secondary wavelength shifter, as listed in Table~\ref{tab:LS_composition_jspark}. Figure~\ref{fig:fig7_from_nim_jspark} taken from reference \cite{cite:RENO_LS_NIM} shows the emission spectra of LAB, PPO, and bis-MSB. The bis-MSB emission ranges from 390 nm to 500 nm and matches well with the spectral acceptance of the PMTs in JSNS$^{2}$. The LS recipe was shown to give satisfactory performance by the RENO and Daya Bay experiments with several years of stability. The JSNS$^{2}$ experiment will use LS with 3 g/L of PPO and 30 mg/L of bis-MSB.

\begin{table}
\begin{center}
\begin{tabular}{c | c | c | c}
\hline
Material & Company & Quantity & Spec \\ \hline
LAB & Isu Chemical Company   &  54,000 L   & - \\ \hline
PPO &  Alfa Aesar  &  162 kg   &  Scintillation grade \\ \hline
bis-MSB & Alfa Aesar & 1.6 kg   &  Scintillation grade  \\ \hline
\end{tabular}
\end{center}
\caption{\setlength{\baselineskip}{4mm}List of chemical elements to be used for JSNS$^{2}$ LS.}
\label{tab:LS_composition_jspark}
\end{table}

\subsubsection{Purification of LS}
\indent

We will use a water extraction method and membrane filtering to purify the LS. High quality LAB can be contaminated by dust introduced during delivery, but it can be removed using a membrane filter produced by Meissner. A water extraction method will be applied to remove $^{40}$K from the PPO. According to the RENO's measurement\cite{cite:RENO_TDR} the radio-impurity of LS is 13.9 ppt for $^{238}$U, 17.7 ppt for $^{232}$Th, and less than 0.32 ppt for $^{40}$K meaning that additional purification of the LAB is not necessary.

\subsubsection{Mass Production of LS}
\indent

\begin{figure}[h]
\begin{center}
\includegraphics[scale=1.6]{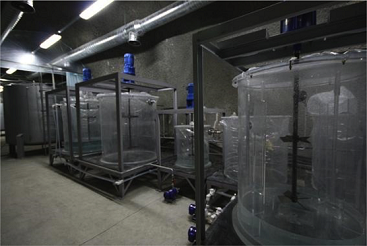}
\end{center}
\caption{\setlength{\baselineskip}{4mm}LS production facility to be used for the RENO experiment. This will be used for making the JSNS$^{2}$ LS after necessary refurbishment and addition.} 
\label{fig:RENO_LS_facility_jspark}
\end{figure}

\begin{figure}[h]
\begin{center}
\includegraphics[scale=0.5]{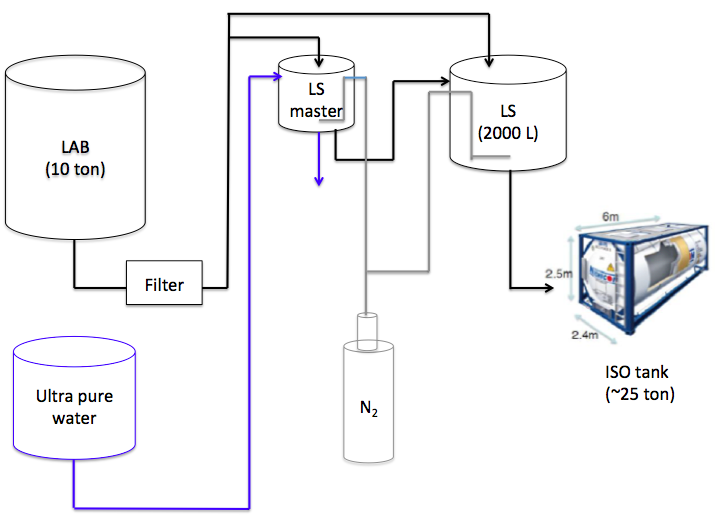}
\end{center}
\caption{\setlength{\baselineskip}{4mm}A schematic diagram of the LS mass production system modified from the one that was used for the RENO experiment. This will be used for making the JSNS$^{2}$ LS after necessary refurbishment and addition. The LAB will go through a micro-filter of 0.5 $\mu$m pore size. The black lines show LAB or LS flow routes, the blue lines show the flow routes of ultra-pure water, and the grey lines show where nitrogen gas will flow.} 
\label{fig:RENO_LS_facility_refurbish_jspark}
\end{figure}

Mass production of the LS is expected to be done using RENO's facility as shown in Fig.~\ref{fig:RENO_LS_facility_jspark}. A refurbishment including cleaning is necessary to reuse the facility and will take place in the period of late 2017 to early 2018. A new pipeline will be added in order to fill an ISO tank with produced LS. The ISO tank will be used for delivery to J-PARC. Figure~\ref{fig:RENO_LS_facility_refurbish_jspark} shows a schematic diagram of the refurbished mass production system for the JSNS$^{2}$ LS. 200 L of 10 times enriched PPO and bis-MSB master solution will be prepared and purified by water extraction using 50 L of ultra-pure water to remove $^{40}$K from the PPO. After the water is drained out, the master solution will go through nitrogen purging to eliminate any moisture. The master solution will be mixed with pure LAB to make LS that will be transported into an ISO tank after additional nitrogen purging. The ISO tank will then be delivered to J-PARC to fill the gamma-catcher and veto layers with the LS. 

\subsubsection{Gadolinium-loaded LS}
\indent

The JSNS$^{2}$ experiment will use 17 tons of Gd-loaded LS as a neutrino target. The IBD reaction is utilized to detect antineutrino appearance from sterile neutrino oscillations. A neutron coming from the IBD reaction is captured by either hydrogen (H) or Gadolinium (Gd). While the 2.2 MeV gamma-ray from the neutron capture on H suffers from a large radioactivity background, the $\sim$8 MeV gamma-rays from the neutron capture on Gd can be well identified to reduce the background events greatly. The mean neutron-capture time on 0.1w\% Gd is $\sim$30 us, more than 6 times shorter than that on H, and can results in significant reduction of accidental backgrounds. 

\subsubsection{Synthesis of Gd-LAB}
\indent

It is necessary to prepare a organometallic complex of Gd using carboxylic acid as an organic ligand. We will make the Gd-carboxylate complex of Gd(RCOO)$_{3}$ from 3,5,5-trimethylhexaonic acid (TMHA) and Gd powder in the form of GdCl$_{3}$. The chemical reaction  for the synthesis consists of the following steps;
\begin{itemize}
\item RCOOH + NH$_{3}$ $\cdot$ H$_{2}$O $\rightarrow$ RCOONH$_{4}$ + H$_{2}$O
\item 3RCOONH$_{4}$ + GdCl$_{3}$ $\rightarrow$ Gd(RCOO)$_{3}$ + 3NH$_{4}$Cl
\end{itemize}

The first step is a neutralization process of TMHA and ammonium hydroxide. The second step is to make the Gd-carboxylate complex which is formed by mixing an aqueous Gd solution with the neutralized TMHA solution. We will apply a liquid-liquid extraction technique to load the Gd-carboxylate compound into LAB (Gd-LAB). The two solutions will be dropped directly into the LAB under strong stirring. The Gd-LAB yield depends on the solution, dropping speed, and stirring power~\cite{cite:RENO_LS_NIM}. The solubility of the organic metal is higher in the organic solvent than in water. Gd-LAB and water can be separated due to the difference in their densities. The separated layer of water can be removed through a drain valve at bottom of the mixing tank.

\subsubsection{Mass production of Gd-loaded LS}
\indent 

\begin{figure}[h]
\begin{center}
\includegraphics[scale=0.5]{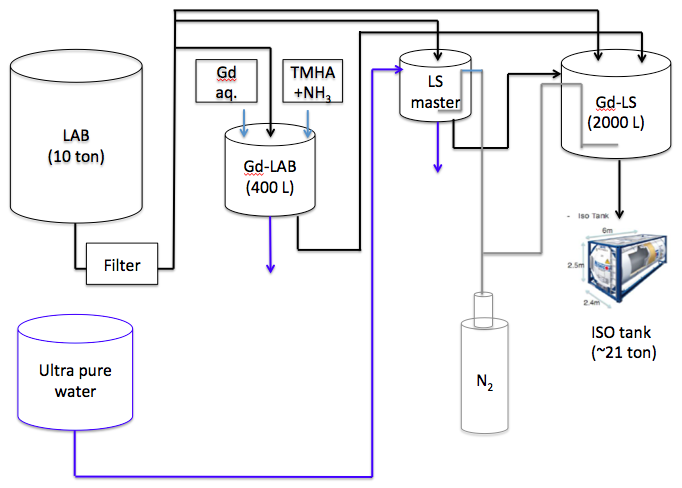}
\end{center}
\caption{\setlength{\baselineskip}{4mm}A schematic diagram of Gd-loaded LS mass production system refurbished from the one used for the RENO experiment. The black lines represent flow routes of LAB, master LS, Gd-LAB, and Gd-loaded LS. The Gd-LAB is produced by mixing a Gd aqueous solution and a TMHA neutralization solution that are dropped from the top of the mixing tank.} 
\label{fig:RENO_GdLS_facility_refurbish_jspark}
\end{figure}

The mass production of Gd-loaded LS is expected to be done using the RENO facility after necessary refurbishment. Fig.~\ref{fig:RENO_GdLS_facility_refurbish_jspark} shows the mass production system of Gd-loaded LS consisting of a 10 ton LAB stainless steel tank, a 600 L Gd-LAB acrylic tank, two 20 L acrylic tanks of Gd aqueous solution and TMHA neutralization solution, a 250 L LS master acrylic tank, and a 2000 L Gd-loaded LS acrylic tank. 

The LAB from the 10-ton storage tank is purified by a 0.5 $\mu$m membrane filter. Gd-loaded LS is produced by mixing Gd-LAB, LS master solution, and pure LAB. Each batch produces 400 L of 0.5w\% enriched Gd-LAB to be diluted with LAB. The LS master solution is prepared as described in the earlier section. The Gd-LS mixing tank produces 2000 L of 0.1w\% Gd-loaded LS based on 400 L Gd-LAB, 200 L LS master, and 1400 L LAB. Additional nitrogen purging is done to remove any remaining moisture before the Gd-loaded LS is transported into an ISO tank.

\subsection{Electronics, DAQ}
\indent

Flash ADCs (CAEN V1730 \cite{CITE:CAENFADC}) are employed as wave-form digitizers for the PMT signals. We have already made liquid scintillator (LS) performance measurements using a V1730 digitizer \cite{CITE:SR_16JAN7}, and have found its 500-MHz sampling with a 14-bit resolution provide good energy resolution and pulse shape discrimination (PSD) performance. 
The Double Chooz experiment \cite{DCPMT} is also taking data successfully with the older version (V1721, 8-bit resolution).
It is expected that the V1730's needed for the experiment will be delivered 3 months after they are ordered.
\par
Table~\ref{tab:V1730} shows the specifications of V1730.
Its size is 1U height and 1-unit width of VME with 16-channels of analog inputs. 
One VME crate can contain all the modules required for a single detector. 
The FADC has 2 V dynamic range (or 0.5~V selective by software) and 14-bit resolution.
The maximum sampling rate is 500~MHz, meaning pulse-height measurements are made every 2~ns. 
The DC offset can be selected anywhere between $\pm$1~V, so measurement between 0~V to -2~V is possible. 
Data buffers are available for each channel and can hold 640~k-samples.
The data buffer can be divided into 1024 event buffers for multi-event data-taking, which reduces readout dead time.
The record length of the wave form is 1.28~$\mu$sec which is enough to utilize PSD techniques, but it can be made shorter in order to limit the data size.
A longer record length is also possible, but would require reducing the number of event buffers.
A daisy-chain with clock-in/out connectors allows synchronization of the FADC chips on all the modules.
The front panel include connectors for many different signals including: trigger-in, reset-in, and busy-out.
\par

\begin{figure}[h]
\centering
\includegraphics[width=0.6 \textwidth]{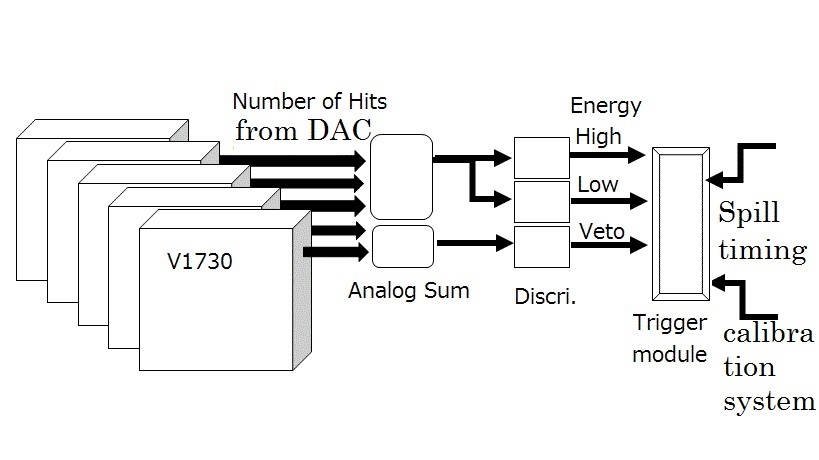}
\caption{\setlength{\baselineskip}{4mm}Conceptual design of the trigger circuit.
Logic signals of energy information, veto information, beam spill timings and calibration triggers are fed into trigger system in order to make gate signals for the FADCs. }
\label{fig:trigger-circuit}
\end{figure}

The trigger is generated using analog outputs corresponding to the number of hit PMTs (Fig.~\ref{fig:trigger-circuit}).
Each module has one digital-to-analog converter (DAC), which provides an output corresponding to how many channels exceed a preset value on the FADC-chip.
These analog signals are fed into an analog sum module then discriminated in order to make logic signals with rough energy information. 
The signal from the veto PMTs is also built in the same way. 
This information is used to make a trigger as well as the spill timing signal. 
\par
Optical link readout is also possible with these modules and a single link can read up to 8 modules at a time.
The average event size is expected to be less than 1 Mb/channel (14~bit, 512~sample) and the number channels to be read is less than 300~ch. 
Thanks to the larger size of the event buffer (1024 event buffers), data can be readout continuously. 
Assuming 4 optical-line readout, the data transfer speed from all FADC module to the readout PC is about 320~Mb/sec, therefore 1~k-event/sec is possible.
We can add more optical links, if additional performance is required. 
Collected data are sent to KEKCC for analysis. 

\begin{table}[h]
\begin{center}
	\begin{tabular}{|l|c|}
        \hline
        Size & VME 6U 1-Unit wide\\
        \hline
	Sampling & 500~MS/s \\
	\hline
        Resolution & 14~bit \\
	\hline
        Number of channels & 16 ch\\
	\hline
        Dynamic range & 0.5 or 2~Vpp \\
	\hline
        DC adjust & $\pm$1~V \\
	\hline
        Data buffer size & 640~kS/ch \\
	\hline
        Maximum data transfer speed & 80~MB/sec by optical link \\
        ~& 200~MB/sec by VME \\
	\hline
        Power requirement & +5~V, 8.2~A and +12~V 840~mA \\
	\hline
	\end{tabular}
	\caption{Specification of CAEN V1730 \cite{CITE:CAENFADC}. } 
        \label{tab:V1730}
\end{center}
\end{table}

\subsection{Calibration System}
\indent

One of the most important calibration sources for the JSNS$^2$ detector is
stopped cosmic ray muons inside the detector because they emit
Michel electrons or positrons, which 
have the same range of the energy as the oscillated signals and dominant
backgrounds. They are easily identified using the veto region of the detector, and 
provide quite useful information. 

In addition to the natural source, we plan to use
two calibration systems with complementarity purposes. One is an embedded LED system. It calibrates gain and timing characteristics for the PMTs and other electronics in a short time whenever we want. It can be performed regularly (e.g. weekly or monthly) to check for stable detector performance and to apply corrections when necessary. It is also useful to check the health of the detector during commissioning or after power loss, for example. Low cost and easy installation can be also expected thanks to simple embedded design. The other calibration system we will use is a laser system with better precision. It measures the PMT gains, PMT charge response functions, PMT time offsets, and timing distributions, as well as the effective speed of light, quantities which are essential in determining the performance of the reconstruction and particle identification algorithms. The timing offsets can be determined with better precision using very short laser pulses (FWHM = 50-70 ps), and the charge response functions can be extracted over a wider range due to the dynamic range of the laser output intensity. In addition, the known positions of the laser diffusers in the JSNS$^2$ detector will provide an in-situ test of the accuracy of the position reconstruction algorithms. Details of those systems are described in following sections. 

\subsubsection{LED System}
\indent

\begin{figure}[htbp]
  \begin{center}
    \includegraphics*[width=10. cm]{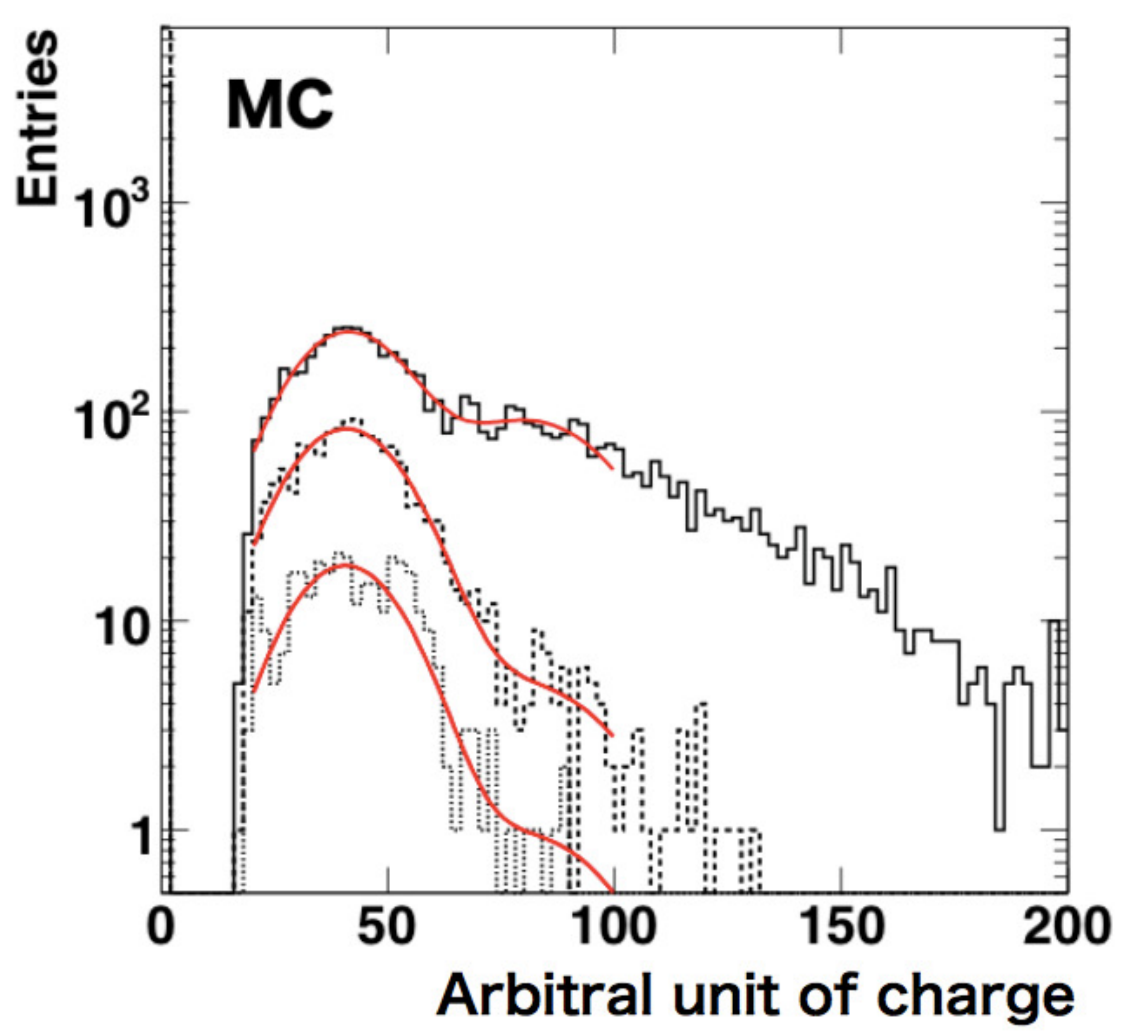}
  \end{center}
  \caption{\setlength{\baselineskip}{4mm}
  MC study of gain calibration for the Double Chooz experiment\,\cite{bib:3}. These charge distributions are obtained from FADC waveforms over a certain threshold. Then the pedestal-subtracted distribution is fitted by multiple Gaussian functions using Poisson statistics. The first peak represents the gain used to convert charge to photoelectrons.}
  \label{fig:LED_gain}
\end{figure}

\begin{figure}[htbp]
  \begin{center}
    \includegraphics*[width=15 cm]{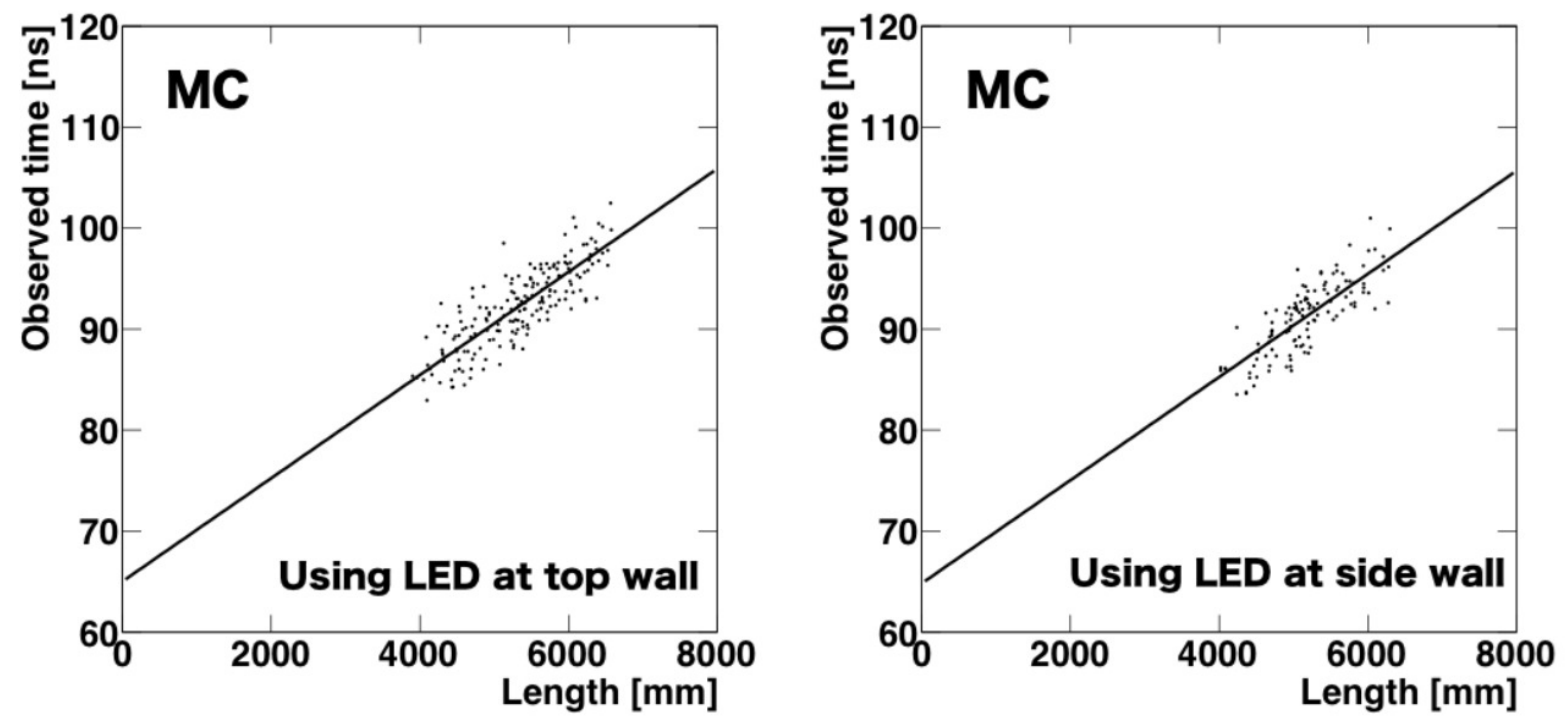}
  \end{center}
  \caption{\setlength{\baselineskip}{4mm}
  MC study of timing calibration for the Double Chooz experiment\,\cite{bib:3}. Timing distribution are extracted from FADC waveforms by finding a peak in the pulse shape. Then mean timing values for various positions of PMTs are plotted. The different TOF for each PMT gives the overall slope, and the variation of an individual PMT from this slop gives the appropriate timing correction.}
  \label{fig:LED_timing}
\end{figure}

The LED system is a conventional system developed and validated by experiments like SNO+\,\cite{bib:1} and Double Chooz\,\cite{bib:2}.
The principle of the calibration by a general LED system is as follows. LED light illuminates PMTs in the detector with variable light intensity. Low intensity light give the charge distribution of the PMT at the single photoelectron level. This allows us to calibrate the PMT gain by fitting the distribution with multiple Gaussian functions as found in Fig.~\ref{fig:LED_gain}. According to MC study for the Double Chooz experiment, we can expect $\sim$2\% precision in the relative gain calibration. On the other hand, high intensity light allows us to calibrate timing. Different distances between light injection points and PMTs provide a slope of observed times as a function of distance. Using variations in the timing measurements from the fitted slope, we can obtain relative timing offsets for PMTs as found in Fig.~\ref{fig:LED_timing}. In the case of the MC study for the Double Chooz experiment, $\sim$0.3\ ns precision of the relative timing offset was expected. As the detector size and structure are similar, we can expect similar calibration precision for the JSNS$^2$ experiment. 

\begin{figure}[htbp]
  \begin{center}
    \includegraphics*[width=9 cm]{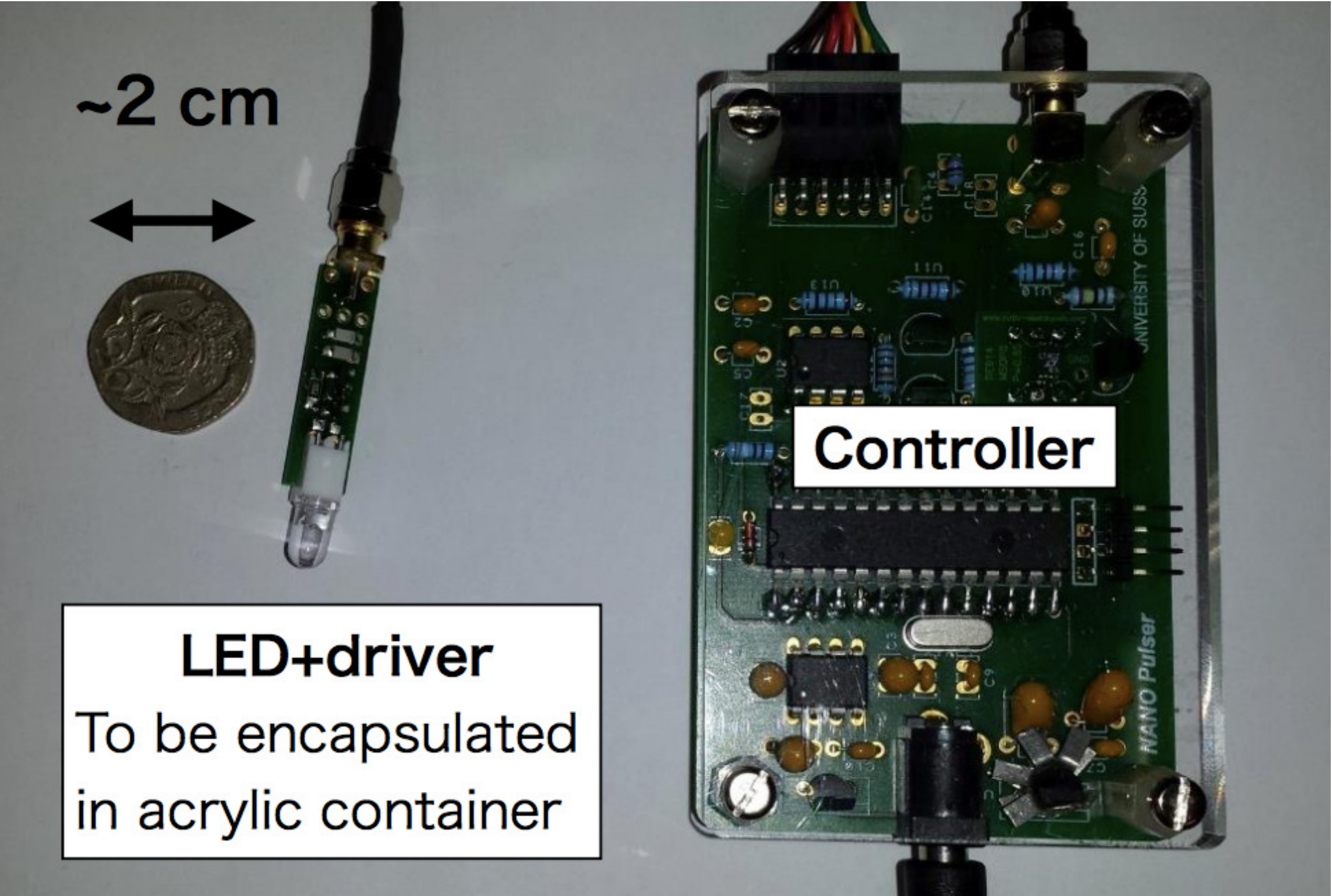}
  \end{center}
  \caption{\setlength{\baselineskip}{4mm}
  A picture of Nanopulser prototype. The left board has an LED and its driver and to be mounted in the detector after being encapsulated in an acrylic container. A twenty pence coin is placed as a size reference. The right board is a controller for the system and connects to the LED with an electrical cable.}
  \label{fig:LED_picture}
\end{figure}

We will employ a ``nanopulser system" developed by Sussex University shown in Fig.~\ref{fig:LED_picture}. One difference between this system and conventional LED systems is the use of direct LED light in the detector without an optical fiber. This gives us the capability to provide higher light intensity and no bias on the timing properties as function of LED emission angle. Therefore, better calibration accuracy is expected than the traditional technique. Furthermore, the device is suitable for the JSNS$^{2}$ detector because optical fibers are not compatible with LAB. In the case of a conventional system, the LED is located outside detector and an optical fiber is used to guide LED light into the detector. The Nanopulser system, on the other hand, can embed an LED with small driver board in the detector. The LED and drive board are encapsulated in a material compatible container made of acrylic. The specifications of the Nanopulser system are summarized in Table~\ref{fig:LED_specifications}. 

\begin{table}[htbp]
  \centering
  \begin{tabular}{ll} \hline
    Item &  Performances \\ \hline
    Wavelength & 355 and 420\,nm \\
    Timing profile of light pulse & 355\,nm: 0.40\,nsec (rise), 0.75\,nsec (fall), 0.79\,nsec (width) \\
                                                      & 420\,nm: 0.40\,nsec (rise), 0.63\,nsec (fall), 0.65\,nsec (width) \\
    Opening angle of LED light & 355\,nm: 30.2 $\pm$ 3.5\,degree\\
                                                      & 420\,nm: 26.9 $\pm$ 1.8\,degree\\
    Light intensity &  Capability to provide single to thousands of photoelectrons \\
    Flushing rate & Up to 100\,kHz \\ 
    Trigger & Capability to produce/accept a TTL trigger \\ \hline
  \end{tabular}
  \caption{Major specification of the Nanopulser system}
  \label{fig:LED_specifications}
\end{table}

\begin{figure}[htbp] 
  \begin{center}
    \includegraphics*[width=10 cm]{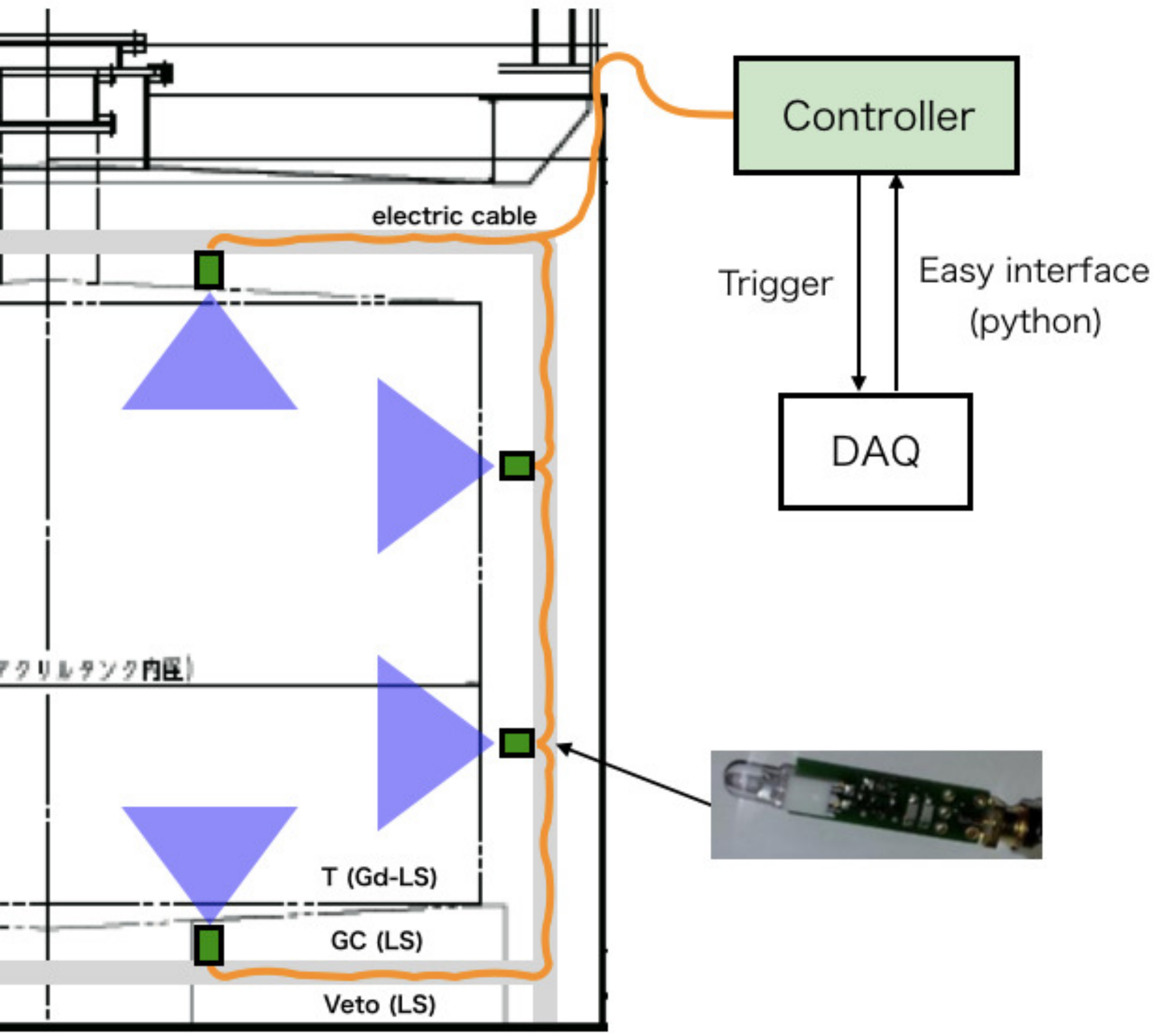}
  \end{center}
  \caption{\setlength{\baselineskip}{4mm}
  Schematic drawing of the Nanopulser system with multiple channels in the detector.}
  \label{fig:LED_design}
\end{figure}

A schematic drawing of the system with multiple channels in the detector is shown in Fig.~\ref{fig:LED_design}. The LED and driver boards are encapsulated in an acrylic container and mounted in the support structure. A single electric cable from the board is placed along the support and taken from the inner detector. It is connected to the controller, which provides a trigger to the DAQ and receives a command through an easy interface with python. The light intensity and flushing rate can be controlled within the specifications of the system. 

To achieve full coverage for gain and timing calibration, 12 LEDs with 420 nm wavelength are located in the detector; 2 positions at the top and bottom of the detector, and 4 positions for 2 rings on the side wall. In addition 2 LEDs are mounted with 355 nm for possible calibration of the attenuation length of the liquid scintillator and angular dependences of the PMT response. Therefore, a total 14 LEDs will be mounted with two different wavelengths. 

\subsubsection{Laser System}
\indent

The system consists of a short-pulsed PicoQuant PDL-800B laser driver and two
laser heads, LDH-P-C-375B and LDH-P-C-470 of wavelengths 375 and 470 nm,
respectively.
The system can be operated with a repetition rate up to 40 MHz, which is well
above the rate envisaged for the experiment.
Typical calibration runs are expected to be taken at rates of ${\cal O}(100)$ Hz,
which means that reasonably high statistics can be achieved in a relatively
short time.
The system can be driven by either an external pulser or internal trigger, and
will deliver a reference timing signal and trigger bit to the JSNS$^2$ DAQ.
The intensity of the laser light can be adjusted through the PDL-800B driver
and through additional neutral density attenuation filters.

The UV laser light is used to extract the PMT gains -- when operated at very
low intensities, as well as the charge and time likelihoods -- when operated
over a large range of intensities.
The charge likelihoods, ${\cal L}_q(q;\mu)$, characterize the probability for
any given PMT to measure a charge $q$ for a predicted charge $\mu$, while the
time likelihoods, ${\cal L}_t(t;t^{pred},\mu)$, characterize the probability
to measure a time $t$ for a predicted time $t^{pred}$ and charge $\mu$.
The UV laser light is delivered to a diffuser through a multi-mode optical
fibre, as additional widening of the original pulse is irrelevant compared to
the long time constants of the scintillator.
The $\Lambda=470\,\mbox{nm}$ laser light does not excite the scintillator, and
therefore is ideally suited to extract the PMT time offsets.
This light has to be delivered to a different diffuser through a single-mode
optical fiber, in order to preserve the short-pulsed characteristics of the
laser driver.
In Double Chooz such a system was shown to be able to measure the time offsets
with a precision of about $0.1\,\mbox{ns}$.

Although light isotropy is not a requirement for the laser light diffusers,
their design will attempt to maximize the isotropy and the diffusers will be
fully characterized before installation.
A known diffuser light output profile will allow additional quantities to be
extracted, such as optical medium or PMT characteristics (e.g., effective
attenuation length, relative quantum efficiencies, etc.).

\subsection{Installation works inside Tank}
\indent

In this subsection, the procedure for the installation of the PMTs, optical
separators between different detector regions, reflection sheets, and FINEMET magnetic shields is described.

First, each PMT is covered by a thin black PET sheet under
the FINEMET sheet to establish
light separation between the inner and veto volumes, and to prevent light noise
in the base circuit of the PMT as shown in Fig.~\ref{WallPMTBox}.
Furthermore, the PMT support structure shown in Fig.~\ref{WallPMTBox}
is attached to each PMT. 

\begin{figure}[htbp]
\begin{center}
\includegraphics[width=0.85 \textwidth]{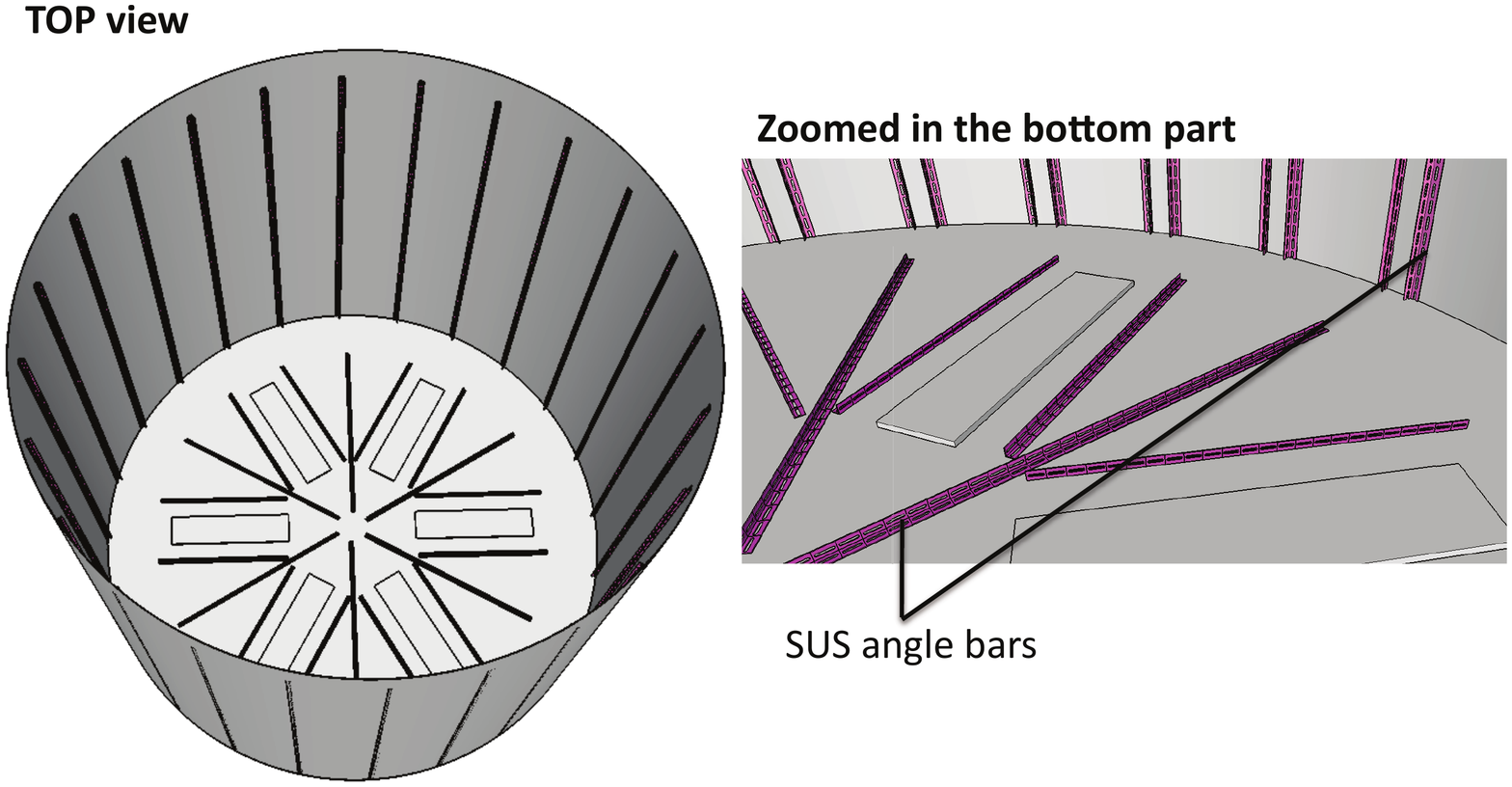}
\end{center}
\caption{\setlength{\baselineskip}{4mm}
Inside of the stainless tank before the PMT installation. The left figure shows the top view, and right figure shows a zoomed-in view of the bottom part.}
\label{WeldedAnglePlate}
\end{figure}

\begin{figure}[htbp]
\begin{center}
\includegraphics[width=0.85 \textwidth]{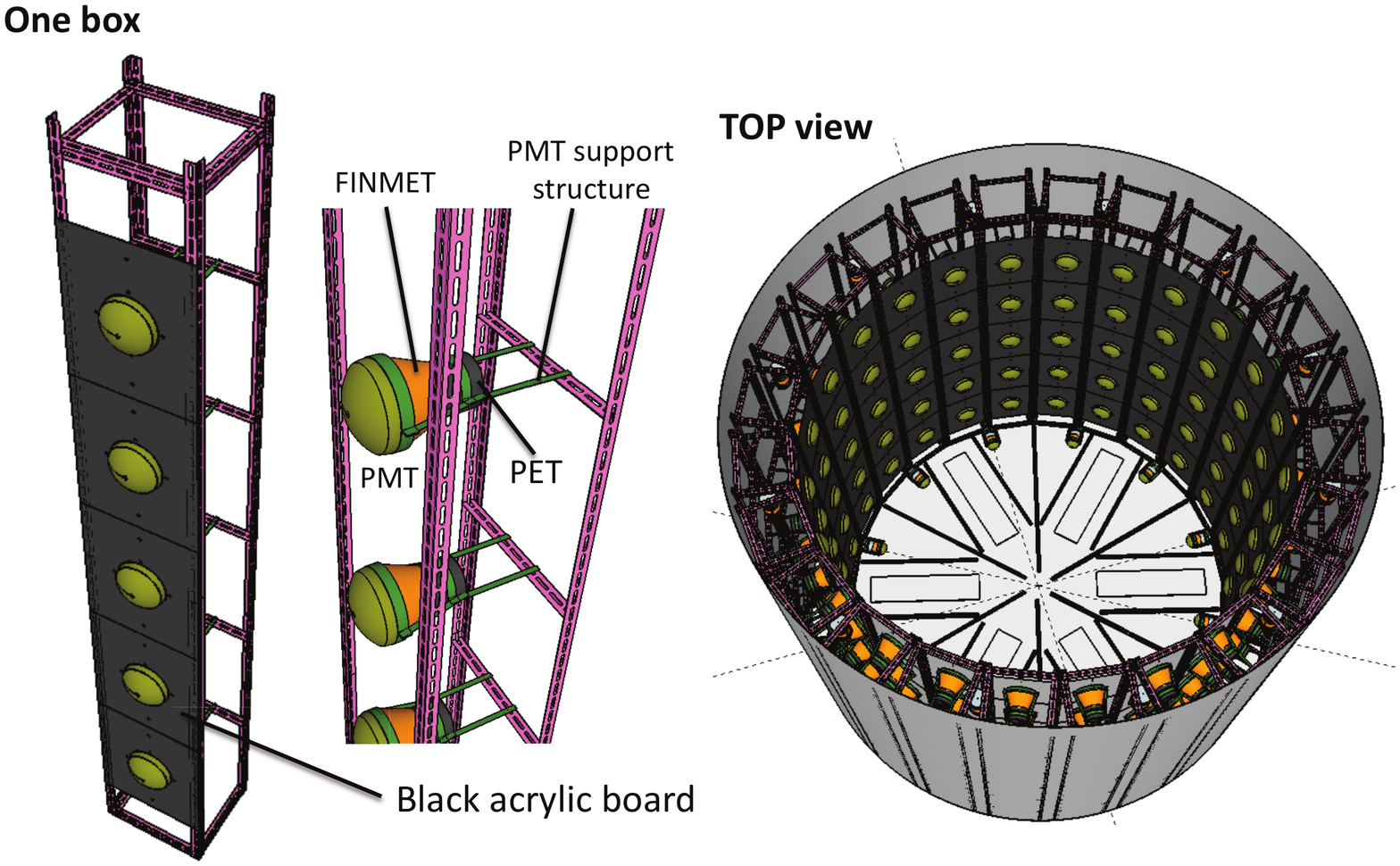}
\end{center}
\caption{\setlength{\baselineskip}{4mm}
Wall PMT installation. The left figure shows one unit of wall PMT support structure. The right figure shows the inside of the stainless tank after the installation of the box units on the walls.}
\label{WallPMTBox}
\end{figure}

In parallel to the installation of the optical separators for each of the PMTs, 
stainless L-type angle bars to fix the PMTs are welded to the
stainless tank (Fig.~\ref{WeldedAnglePlate}),
and the inner surface of the stainless tank is covered by reflective sheets made from
REIKO LUIREMIRROR (see veto sections).
To fix the sheets, welded L-type angle bars are placed in advance.

Next, the PMTs are attached to the stainless tank wall.
The PMT support structure is attached to a box unit assembled with stainless
L-type angle bars (40 mm width) including the welded bars. Then,
a total of five PMTs are
attached to the box unit as shown in the left image of Fig.~\ref{WallPMTBox}.
Black acrylic boards are attached to the box unit for further light
separation from the veto region.
A total of 24 box units are attached along
the stainless tank wall. The right image in Fig.~\ref{WallPMTBox} shows an overview
inside the stainless tank after the installation of the box units.
After the box units are in place, the PMT cables are fixed
with cable ties along the welded angle bars on the wall.

\begin{figure}[htbp]
\begin{center}
\includegraphics[width=0.85 \textwidth]{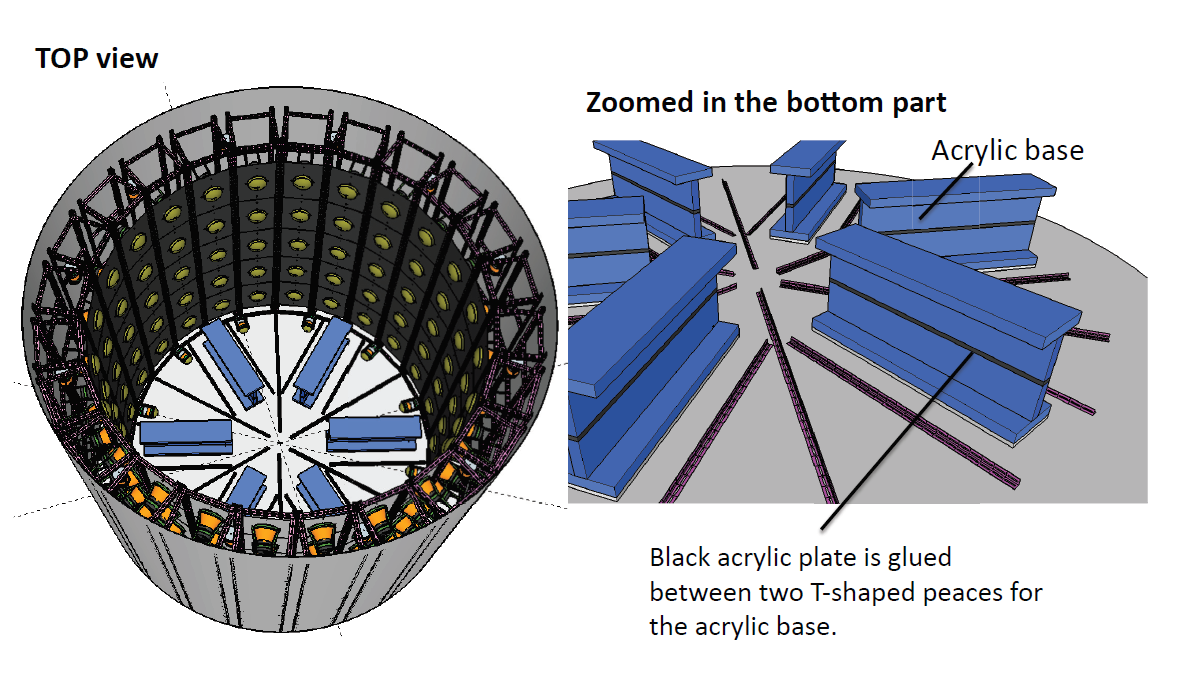}
\end{center}
\caption{\setlength{\baselineskip}{4mm}
Installation of the acrylic base. The left figure shows a top view and the right figure shows a zoomed-in view of the bottom part of the detector.}
\label{AcrylicBase}
\end{figure}

\begin{figure}[htbp]
\begin{center}
\includegraphics[width=0.85 \textwidth]{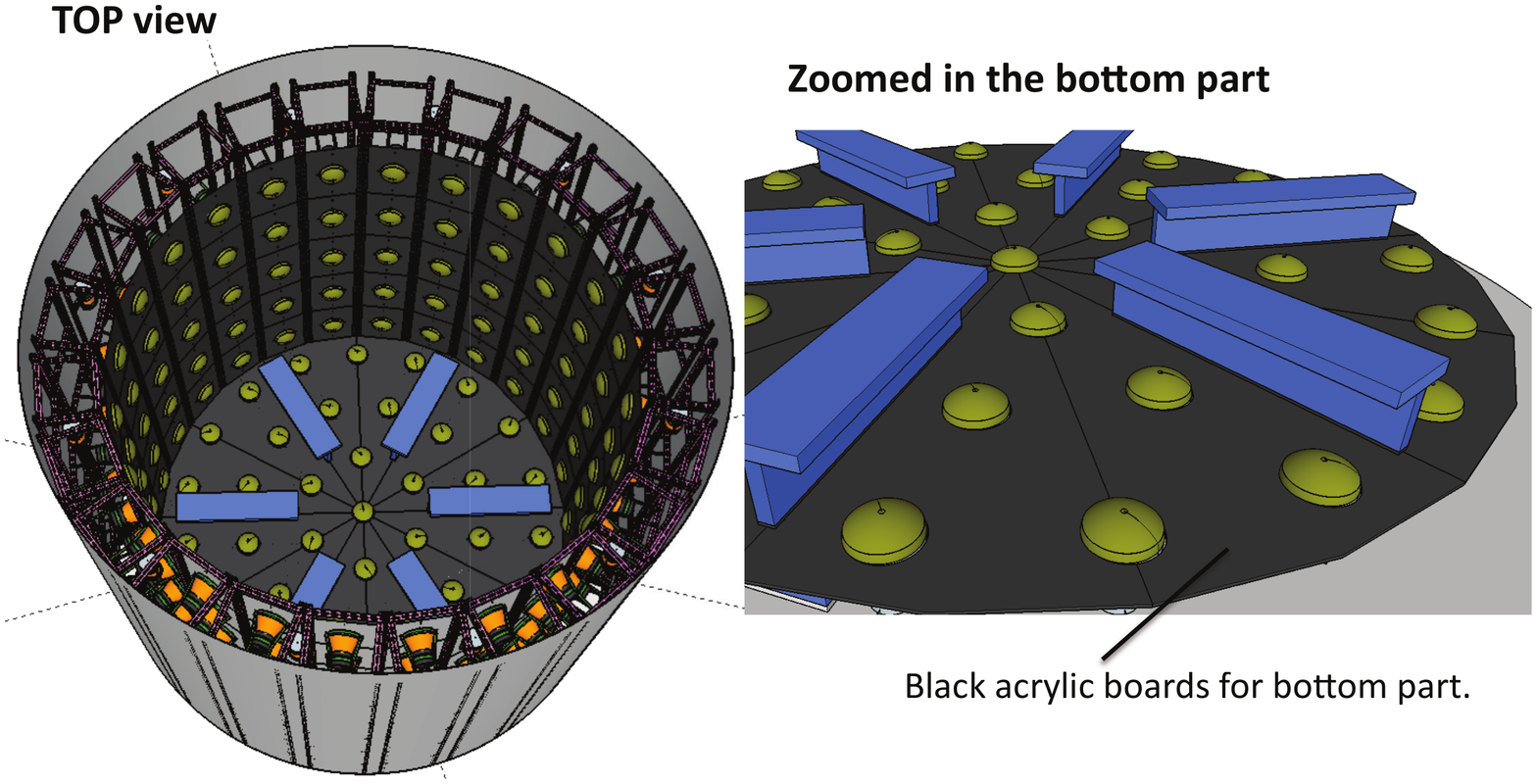}
\end{center}
\caption{\setlength{\baselineskip}{4mm}
Installation of the bottom PMTs and optical separators. The left figure shows a top view and right figure shows a zoomed-in view of the bottom part of the detector.}
\label{TargetPMTBottom}
\end{figure}

Six acrylic bases (supports) are installed on the bottom part of the
stainless tank as shown in Fig.~\ref{AcrylicBase}. 
After that, the PMTs and the blackboards in the bottom part of the detector are installed 
using stainless L-type angle bars (30 mm width) welded on the bottom
part of the stainless tank.
Figure~\ref{TargetPMTBottom} shows an overview of the inside of the stainless tank after
the installation of the acrylic supports and the bottom PMTs.

\begin{figure}[htbp]
\begin{center}
\includegraphics[width=0.85 \textwidth]{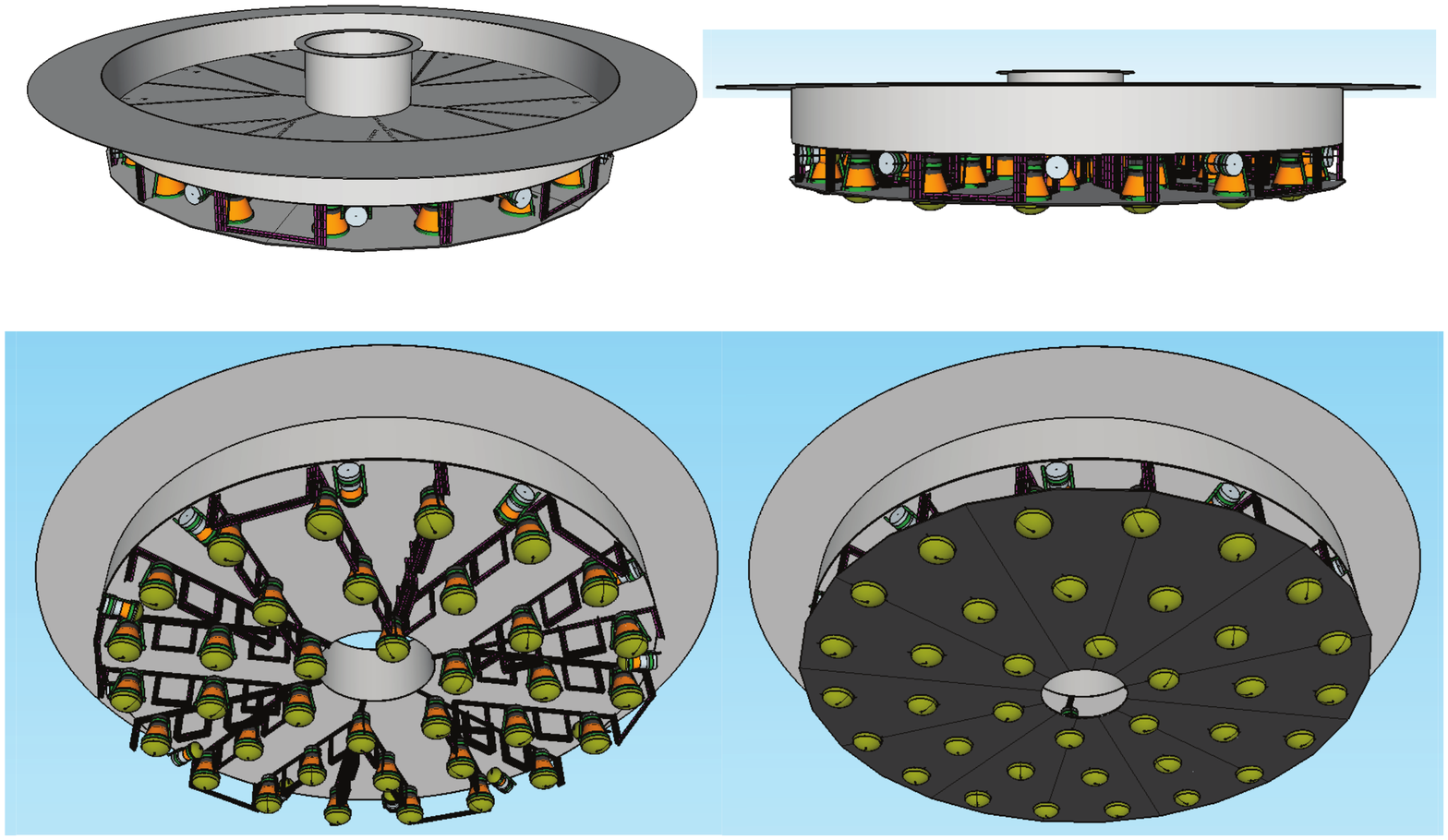}
\end{center}
\caption{\setlength{\baselineskip}{4mm}
  The stainless tank lid after attaching PMTs and black boards.
  The upper left and right figures show the top and side views, respectively.
  The lower left figure shows the bottom of the lid after attaching the PMTs and
  the stainless L-type angle bars, and the right figure shows the bottom
  after covering the lid with black acrylic boards.}
\label{TopPMTLid}
\end{figure}

While the PMT installation on the walls and bottom part of the detector is performed, 
the PMTs and the black acrylic boards for the top part are attached
to the stainless tank lid using the same method as the bottom part.
The installation of the PMTs and the optical separators in the stainless steel tank lid are done in parallel to the other installation work.
Figure~\ref{TopPMTLid} shows the lid part after attaching
the PMTs and the black acrylic boards. 

\begin{figure}[htbp]
\begin{center}
\includegraphics[width=0.85 \textwidth]{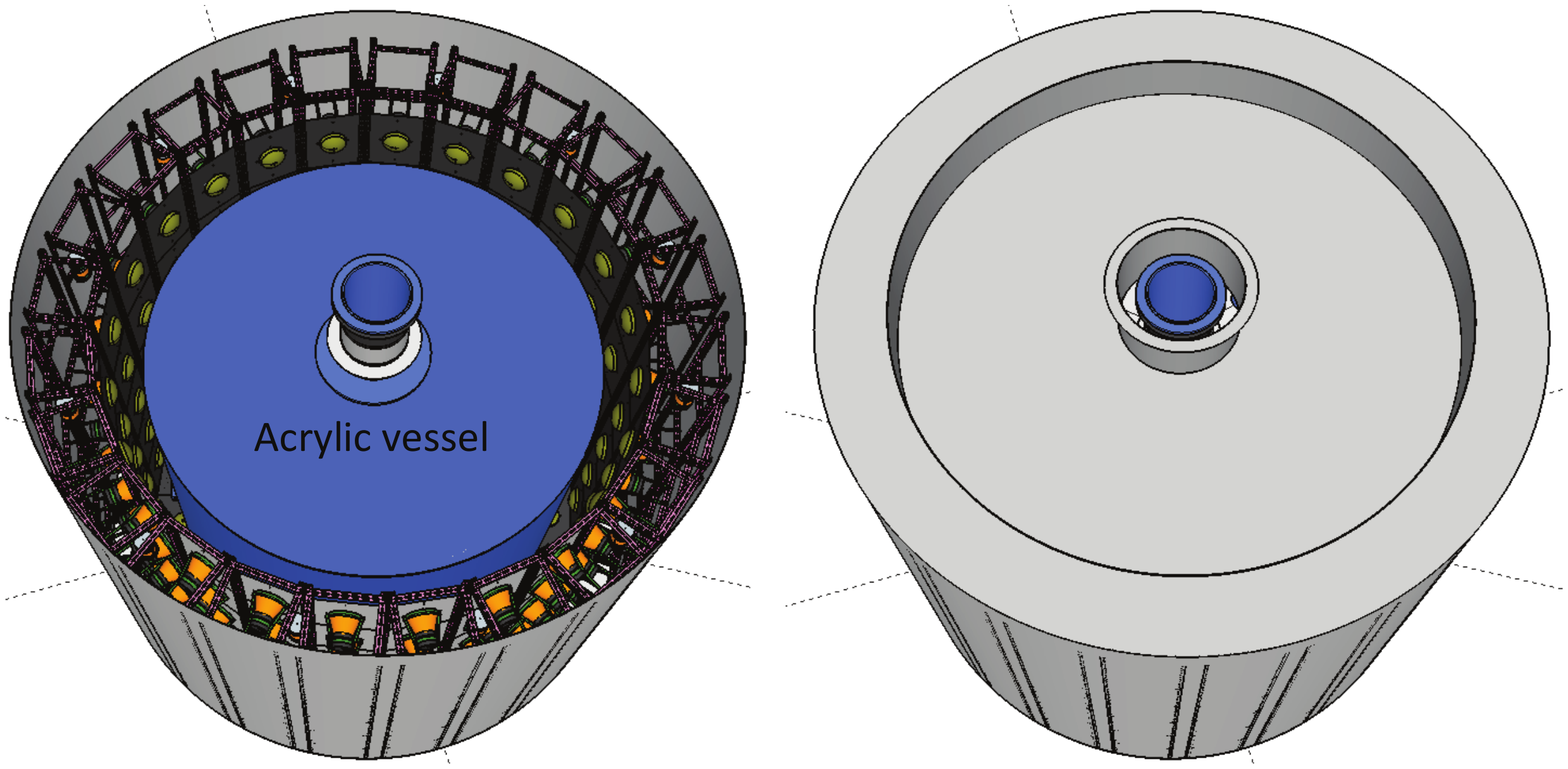}
\end{center}
\caption{\setlength{\baselineskip}{4mm}
  Inside of the stainless tank after the acrylic vessel is inserted (left), and after the lid is on (right).}
\label{InstallAcrylicVessel}
\end{figure}

After the installation of the wall and bottom PMTs, the acrylic target
vessel is lifted down by the crane and is fixed to the acrylic
bases with screws. Finally, the lid is placed on the stainless tank.
(Fig.~\ref{InstallAcrylicVessel}). 

\subsection{Expected Experimental Operation}

\subsubsection{Normal Operation}
\indent

\begin{figure}[h]
 \centering
 \includegraphics[width=1.0 \textwidth]{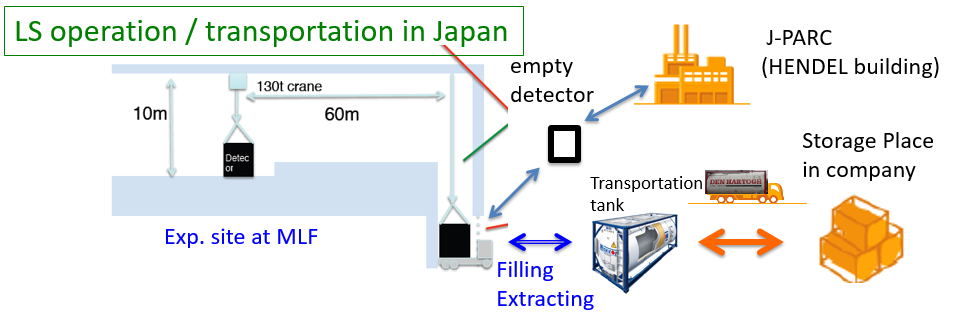}
 \caption{\setlength{\baselineskip}{4mm} 
   The detector and LS operation during the experiment through a given year. Details are given in the main text.
 }
 \label{fig:LSOP}
\end{figure}

The expected detector operation through a given year is illustrated
in Fig.~\ref{fig:LSOP}.
The following are the phases of operation for the detector:
\begin{itemize}
\item In October, we fill the detector with LS at the entrance of the
  MLF using ISO tanks (used to store the LS).
  The ISO tanks, pumps and pipes are prepared by experimental group.
\item After filling, the detector will be moved from the entrance to the
  3rd floor (``Large Component Handling Room'') using a 130-ton crane.
\item Physics data is taken from October until the following June. (5000 total hours)
\item In June, the detector is moved from the MLF 3rd floor to the MLF entrance
  area again. The LS is extracted from the detector and put
  into the ISO tanks.
\item The ISO tanks are stored in Kasawaki by a privae company. They can regularly inject cool nitrogen gas into the detector to preserve the integrity of the scintillator and have experience with this procedure.
\item The weight of the empty detector is small ($\sim$20 tons) and the
  Fire Law is not applicable after the extraction of the LS, therefore the
  detector can be stored in a normal area (i.e.an area not registered under the Fire Law). We plan to put the detector in the HENDEL building at J-PARC during
  that period (no beam).
\end{itemize}

We have already contacted two companies to arrange purchase of ISO tanks,
and the procedure above has been verified to be realistic.

\subsubsection*{ISO Tank and Filling/Extracting System}
\indent

An ISO tank is a safe container made based on an international standard (ISO standard). 
Because ISO tanks satisfy international shipping standards, it is possible to ship them not only in Japan, but also overseas.
Since each ISO tank is equipped with the relevant parts for filling, extracting, and maintaining the liquid, the LS can be safely handled. The ISO tank has satisfactory performance as a dangerous goods container, and thus can be used as a transportation container and as a storage container for the LS with out concerns over degraded quality.

\begin{figure}[h]
\centering
\includegraphics[width=0.5 \textwidth]{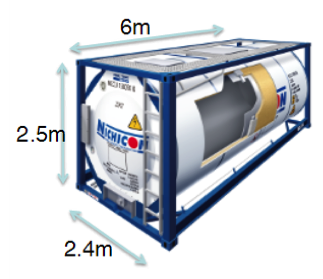}
 \caption{\setlength{\baselineskip}{4mm} 
   Cut model of an ISO tank with the dimensions labeled.
   Figure is reprinted from http://n-concept.co.jp/service/tank/
    }
\label{fig:ISO_tank}
\end{figure}

 The ISO tank has a capacity of about 25,000 L, though other smaller sizes exist as well. 
 When transporting ISO tanks, it is required that the tank only be filled to between 80 to 95  \% of the total volume.  
 In the JSNS$^2$ detector, the volume of Gd-LS is about 19,300 L, and the volume of the standard LS is 35,000 L, so it is possible to store all the liquid in a total of 3 tanks.
ISO tanks have heat insulation to cover the inner tank part of stainless steel, and the exterior is covered in FRP. 
This structure reduces heat intrusion. A metal frame surrounds the container and makes loading the tank easier. 
Figure \ref{fig:ISO_tank} shows construction and size of an ISO tank.

Any metal contact with Gd-LS could accelerate the aging effect of the Gd-LS, resulting in diminished performance. We decided to teflon coat the inside of ISO tank and outlet valve that will be used for storage of the Gd-LS. Two standard stainless-steel ISO tanks will be used to store the veto LS.

\begin{figure}[h]
\centering
\includegraphics[width=0.8 \textwidth]{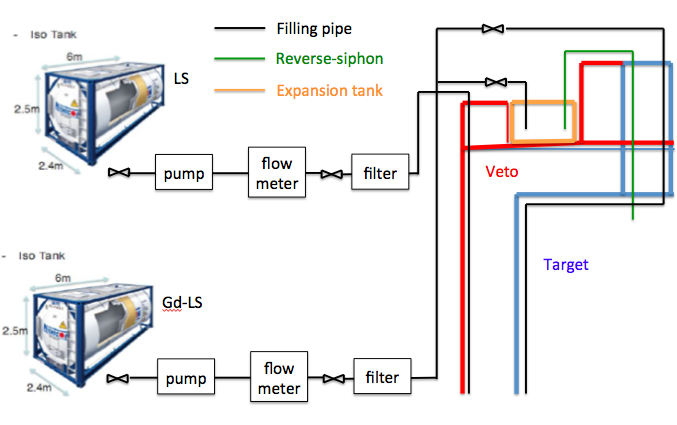}
\caption{\setlength{\baselineskip}{4mm} Schematic diagram of the liquid handling system while filling. The thick blue line is acrylic vessel,  the thick red line is stainless-steel tank.  The green line shows the revrese-siphon system, and the black line is the installed filling pipe. At the target, there are two filling pipes to help protect the fragile chimney region.}
\label{fig:LS_handling_filling_jspark}
\end{figure}

We need to consider several things in designing the liquid handling system. First of all, filling and extracting should be done within a few days because if an emergency happens, the detector must be moved to HENDEL within one week. In addition, the flow path of Gd-LS should be made with non-metallic materials. 1 inch teflon pipe will be used to construct the liquid handling system. Because filling and extracting will be done around the equipment entrance of MLF building, 1 inch teflon pipe will be wrapped with black sheet to reduce exposure to sunshine. We will use a MEGA 960 pump from TREBOR, which is pneumatically driven with 95 LPM of max flow rate and whole flow path is coated with either PFA or PTFE. The max suction lift is 5.5 m, which is larger than height of the JSNS$^{2}$ detector. Even if the quality of the Gd-LS and the LS is good enough, contamination can occur with dust particles while filling or extracting. We will install a 0.1 $\mu$m pore size membrane filter made by Meissner before the detector while filling and before the ISO tank while extracting. A flow meter will be used to check the accumulated amount of liquid and to control the spontaneous flow rate to prevent a difference of liquid level at target, gamma-catcher, and veto layers. Figure~\ref{fig:LS_handling_filling_jspark} shows a schematic design of the handling system for the filling case. 

\begin{figure}[h]
\centering
\includegraphics[width=0.8 \textwidth]{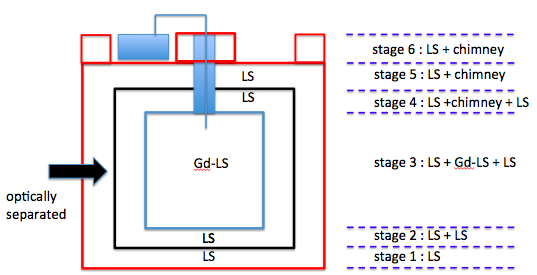}
\caption{\setlength{\baselineskip}{4mm}Definition of each stage for safely filling and extracting.}
\label{fig:filling_stage_jspark}
\end{figure}

Filling and extracting will be performed according to six stages. These stages are important to ensuring that the liquid level in the different detector layers is kept the same. Figure~\ref{fig:filling_stage_jspark} shows each stage for safely filling / extracting and Table~\ref{tab:filling_speed_jspark} summarizes detailed information on each stage. The max flow rate is assumed to be 30 LPM in reality because of resistance in the liquid handling system. Extracting will be performed in reverse order. 

\begin{table}
\begin{center}
\begin{tabular}{c | c | c | c | c | c | c}
\hline
  & Target & Catcher+Veto & Height & Target & Catcher+Veto & Time \\ 
  & [LPM] & [LPM] & [cm] & [L] & [L] & [h] \\ \hline
stage 6 & 0.1 & 1  & 12 & 4 + expansion & expansion & few \\ \hline
stage 5 & 0.1 & 30  & 25 & 8 & 4000 & 2.2 \\ \hline
stage 4 & 0.1 & 30  & 25 & 8 & 4000 & 2.2 \\ \hline
stage 3 & 30 & 29.5  & 250 & 19350 & 19000 & 10.8\\ \hline
stage 2 & 0 & 30 & 25 & 0 & 4000 & 2.2 \\ \hline
stage 1 & 0 & 30 & 25 & 0 & 4000 & 2.2 \\ \hline
Total  &  &   &  & 19370 & 35000 & 19.6 \\ \hline
\end{tabular}
\end{center}
\caption{\setlength{\baselineskip}{4mm}Flow rate, height, volume, and expected time of each stage.}
\label{tab:filling_speed_jspark}
\end{table}

Because the chimney is much narrower than the detector vessel, a small difference in the amount of liquid between the target and veto layers can generate a liquid height difference of more than 15 cm, especially at target chimney. We will install two filling pipes for the target to avoid such a situation. Near the target chimney, the valve of the detector filling pipe will be closed and valve of expansion tank filling pipe will be opened. Then, the Gd-LS expansion tank will be moved automatically to target chimney through reverse-siphon system. In the veto case, the detector itself contains a donut-shaped, region which prevents liquid level changes in response to thermal variations (i.e. we do not need to change filling method at the chimney part for the veto). Extraction will be performed in reverse order.

\subsubsection{Operation in the Emergency}
\indent

If the MLF faces an emergency, the JSNS$^2$ detector should be removed from
the MLF building as soon as possible. We expect that the detector can be
moved within a few days to $\sim$1 week using the 130-ton crane.
The MLF usually needs more than a week to exchange the target so that the target area can become radiation cool.
Therefore no interference in the case of an emergency is expected.

In the case of a detector emergency such as an LS leak, all of the liquid will be contained in the first and second oil-leak tanks. On the off chance that LS gets onto the floor of the MLF, it is not possible for the LS to get into the target area due to the sealant and oil leak barriers. In this case, the LS will be removed from the floor immediately.

\subsection{Offline Software}
\indent

\subsubsection{Event Reconstruction and Particle Identification}
\indent

The event reconstruction developed for JSNS$^2$ is based on a maximum
likelihood algorithm that makes use of all information available in any
given event.
An event is fully characterized by the four-vertex $(x,y,z,t)$ in the
coordinate system of the detector, direction $(\phi,\theta)$, and energy $(E)$.
Thus, for any given event defined by the set of parameters $\vec\alpha$,
\[ \vec\alpha=(x,y,z,t,\phi,\theta,E), \]
the likelihood for measuring a set of PMT charges $(q_i)$ and times $(t_i)$ in
the JSNS$^2$ detector is the product over the individual charge and time
likelihoods at the PMTs:
\[ {\cal L}_{event} = \prod_{i=1}^{N_{pmts}} {\cal L}_q(q_i;\vec\alpha)
                                             {\cal L}_t(t_i;\vec\alpha). \]
Reversing the meaning of the likelihood function, ${\cal L}_{event}$ is the
likelihood that the event is characterized by the set $\vec\alpha$ given the
set of measured charges $(q_i)$ and times $(t_i)$.
Maximizing the event likelihood ${\cal L}_{event}$ (or equivalently minimizing
$- \ln {\cal L}_{event}$) with respect to $\vec\alpha$ determines the optimal
set of event parameters.

Given the neutrino energies and detector size in JSNS$^2$, all events can be
assumed to be point-like.
Neutron/proton events can be well approximated to produce only isotropic
scintillation light, where the flux $\Phi$ (photons per steradian) is
proportional to the event energy $E$.
The average number of photoelectrons (PEs), $\mu_i$, expected at a PMT of
quantum efficiency $\varepsilon_i$, at a distance $r_i$ from the event vertex,
and subtending a solid angle $\Omega_i$ is given by
\[ \mu_i = \varepsilon_i \, \Omega_i\,\Phi\,\exp(-r_i/\lambda_s), \]
assuming that light attenuation is only due to extinction.
Although the scintillation attenuation length $\lambda_s$ and the individual
quantum efficiencies of the PMTs are wave-length dependent, only average,
effective values are used in this approach.
All reconstruction parameters (attenuation lengths, solid angles and quantum
efficiencies) are determined self-consistently, from control data samples.
Furthermore, since any constant can be easily absorbed in the definition of the
light flux $\Phi$, only the relative quantum efficiencies are relevant in this
approach.

The charge likelihood ${\cal L}_q(q;\vec\alpha)$ for any given PMT is directly
obtained from the (normalized) probability of measuring a charge $q$ for
a predicted value $\mu$, ${\cal P}(q;\mu)$, since $\mu$ itself depends on the
set of event parameters $\vec\alpha$.
The negative charge log-likelihood look-up tables will be obtained from the
laser calibration data.

The time likelihood for any {\em hit} PMT is a function of both the corrected
time, $t_{corr}^{(i)}$:
\[ t_{corr}^{(i)} = t_i - t - \frac{r_i}{c_n}, \]
and also the predicted scintillation and Cherenkov charge at that particular
PMT, $\mu_s^{(i)}$ and $\mu_c^{(i)}$, respectively, namely
\[ {\cal L}_t = w_s\,T_s(t_{corr}^{(i)},\mu_s^{(i)})
              + w_c\,T_c(t_{corr}^{(i)},\mu_c^{(i)}), \]
with the weights
\[ w_s = \frac{\mu_s^{(i)}}{\mu_s^{(i)}+\mu_c^{(i)}} \qquad\mbox{and}\qquad
   w_c = \frac{\mu_c^{(i)}}{\mu_s^{(i)}+\mu_c^{(i)}}. \]
The underlying scintillation and Cherenkov time likelihoods,
$T_s(t_{corr},\mu_s)$ and $T_c(t_{corr},\mu_c)$, respectively, will be
extracted from laser calibration data, and verified self-consistently from the
regular data.

Particle identification (PID) can be based on the standard pulse shape
discrimination, which exploits the difference in the scintillation timing
profile between positrons and neutrons (protons).
In particular, the fraction of late light calculated under a single hypothesis
event reconstruction can yield a PID which yields sufficient separation between
signal and background events.
Alternatively, a stronger separation can be achieved using the event likelihood
ratio (or log-likelihood difference), where each event is reconstructed under
two hypotheses, an electron hypothesis -- which yields ${\cal L}_e$, and
a neutron hypothesis -- which yields ${\cal L}_n$.
The resulting log-likelihood difference, $\ln ({\cal L}_e/{\cal L}_n)$, fully
utilizes all differences between electrons and neutrons, both in timing and in
spatial charge distribution.

\section{\setlength{\baselineskip}{6mm} Understandings of the Background and Expected Detector Performance}

\subsection{Expected Detector Performance}
\subsubsection{Energy Resolution}
\indent

\begin{figure}[htb]
\begin{center}
\includegraphics[width=0.85 \textwidth]{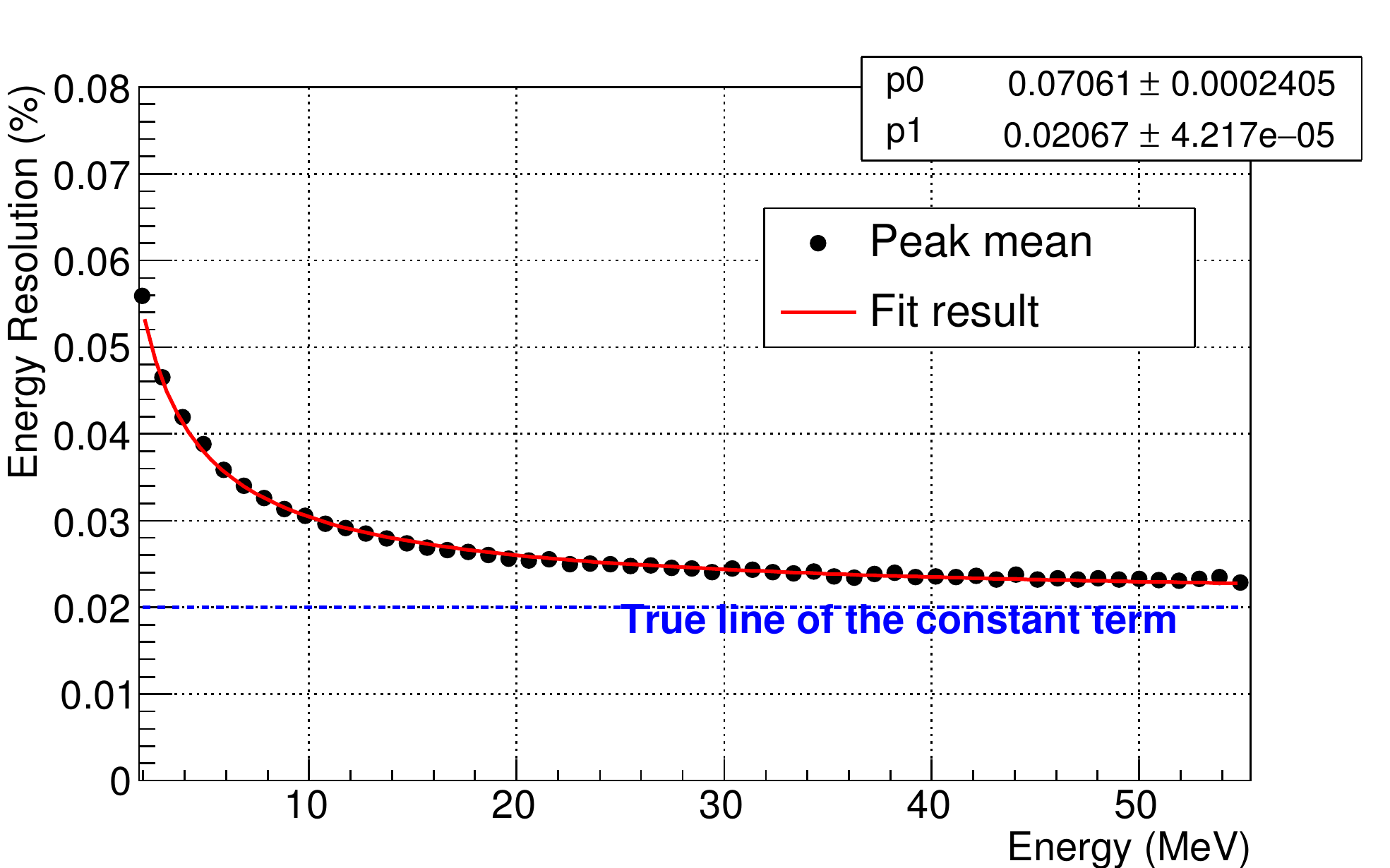}
\end{center}
\caption{\setlength{\baselineskip}{4mm}
Correlation between energy resolution and energy. Black points show the MC result, red line shows fit result to the black points with an approximate formula.}
\label{ERESOLUTION}
\end{figure}

The light yield of the 0.1w\%Gd-LS is 10000 photons/MeV, and the attenuation length is more than 10 m at a wavelength of 430 nm. The acceptance of the 193 8-inch PMTs is $\sim$11\% at the detector center, and considering the QE x CE for the PMTs, the total number of photoelectrons detected for all 8-inch PMTs is $\sim$200/MeV. Figure~\ref{ERESOLUTION} shows the visible energy resolution as a function of the energy. The energy resolution can be assumed to be approximately given by $\sqrt{\frac{p0^2}{E}+p1^2}$, where, $E$ is energy, $p0$ is contribution of the total number of photoelectrons, and $p1$ is a constant term related to hardware. The constant term is assumed to be 2\% in Fig.~\ref{ERESOLUTION}, consistent with the choice in reactor neutrino experiments \cite{ReactorExp}. The black point at each energy in Fig.~\ref{ERESOLUTION} shows the energy resolution of the MC result including the light yield, attenuation length of the Gd-LS and uniformity of the energy resolution due to vertex position dependence. The red line shows the fit result to the black points with the approximate formula from above. The overall energy resolution (p0) from the fit result is calculated as 7\%/$\sqrt{MeV}$.  
In the higher energy range of several tens MeV, the effect of the constant term on the energy resolution is dominant in this detector design. 

\subsubsection{Pulse Shape Discrimination (PSD) capability}
\indent

Because the JSNS$^{2}$ detector is set above ground, it is necessary to reduce the cosmic-induced fast neutron background dramatically compared to an equivalent underground experiment. It is then indispensable for the JSNS$^{2}$ detector to have strong PID capabilities to be able to reject signals from cosmic-induced fast neutrons.
Cosmic-induced fast neutrons mimic the IBD coincidence. Recoil protons from the fast neutron mimic the prompt IBD ionization signal and the neutron eventually thermalizes and captures on Gd. This completely mimics the IBD process.    
Compared to the positron from the IBD signal, the waveform generated by a recoild proton is wider due to differences in the dE/dx between the two particles. By exploiting this difference, the IBD signals can be distinguished from the cosmic-induced fast neutrons analytically. In this TDR, a ratio of the integrated charge in the tail of the prompt waveform to the charge in the full waveform is used as a PSD variable for the analysis (Tail Q/Total Q).  

\begin{figure}[htb]
\begin{center}
\includegraphics[width=1.1 \textwidth]{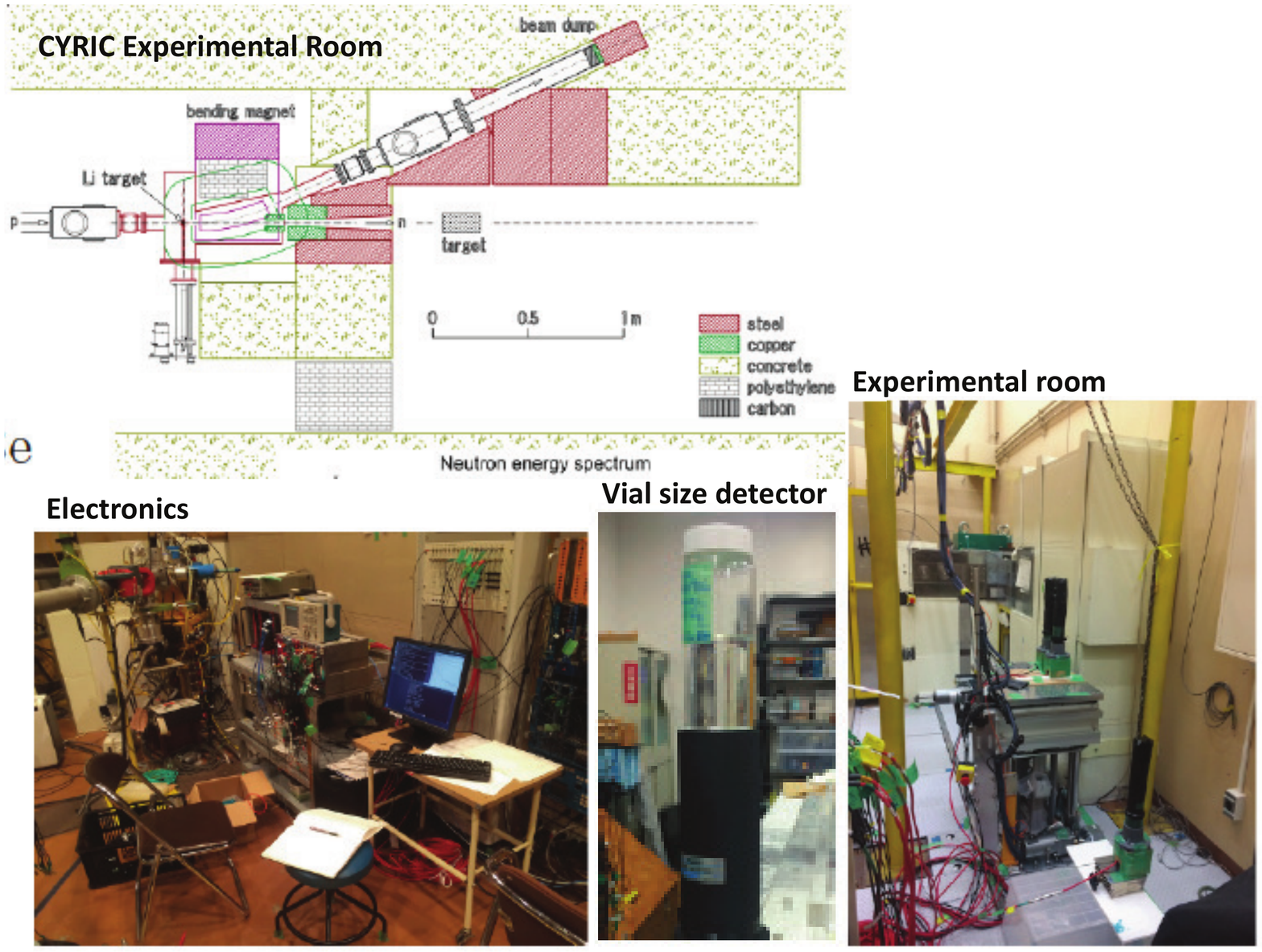}
\end{center}
\caption{\setlength{\baselineskip}{4mm}
Figure showing the CYRIC experimental room and pictures of the experimental setup.}
\label{CYRIC_ROOM}
\end{figure}

In order to evaluate the PSD capability of the JSNS$^{2}$ detector, the PSD performance of the RENO type Gd-LS was measured using a 100 mL vial exposed to a 70 MeV neutron beam produced by CYRIC \cite{CYRIC} at Tohoku University in Japan in November, 2015. The data was taken using a CAEN V1730 (500MS/s, 14bits) digitizer. 
The top image in Fig.~\ref{CYRIC_ROOM} shows the experimental room at CYRIC. The proton beam comes from the left side, and hits $^{7}$Li target, which causes the emission of neutrons. The neutrons pass through a collimator and interact with hydrogen and carbon in the Gd-LS. The scintillation light is viewed by a 2-inch PMT. The bottom pictures in Fig.~\ref{CYRIC_ROOM} show the experimental setup.

\begin{figure}[htb]
\begin{center}
\includegraphics[width=0.85 \textwidth]{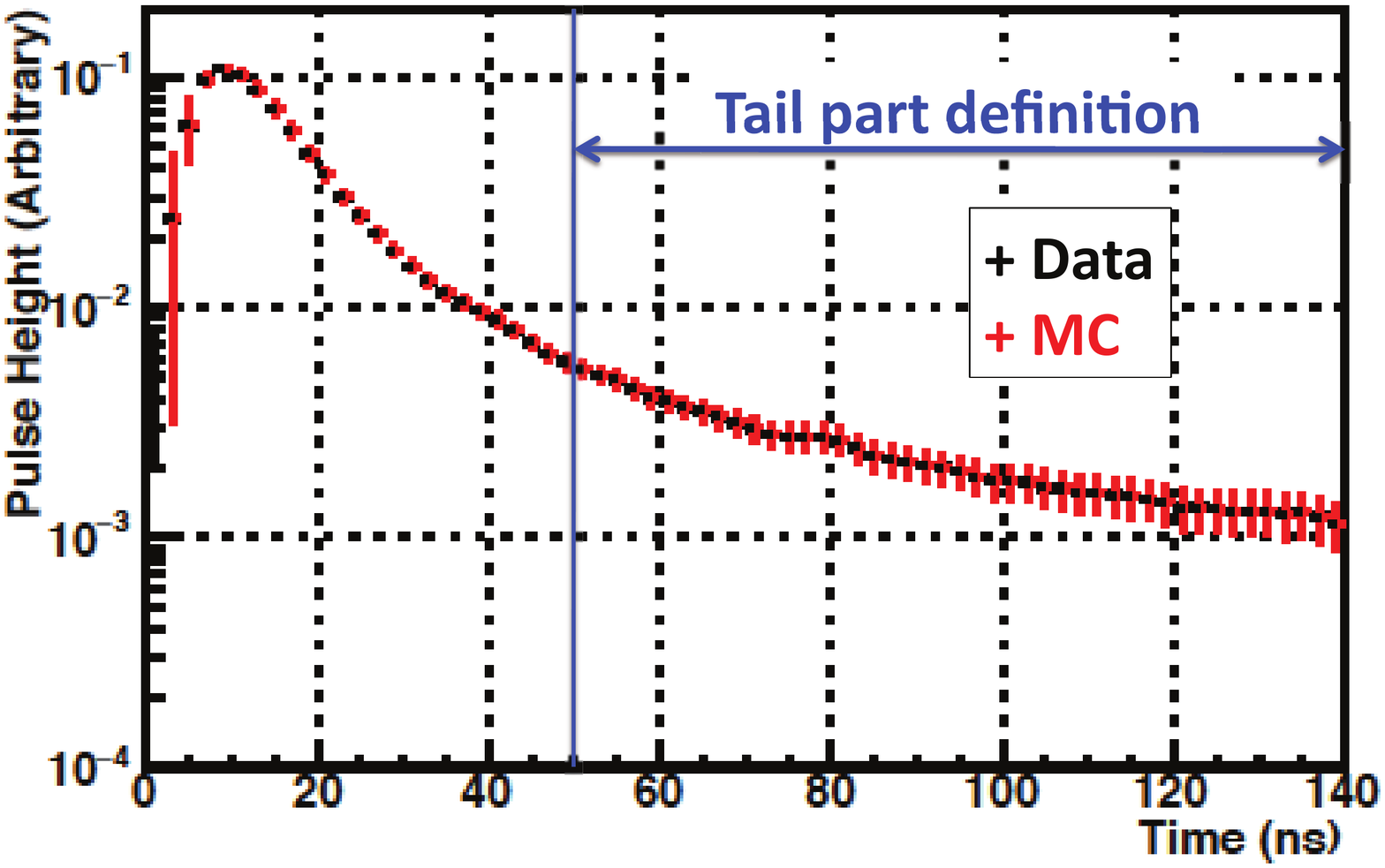}
\end{center}
\caption{\setlength{\baselineskip}{4mm}
Mean waveforms of the recoil protons of data and the MC simulation. The blue arrow shows the definition of the tail region for computing Tail Q/Total Q.}
\label{PSDDef}
\end{figure}

\begin{figure}[htb]
\begin{center}
\includegraphics[width=0.85 \textwidth]{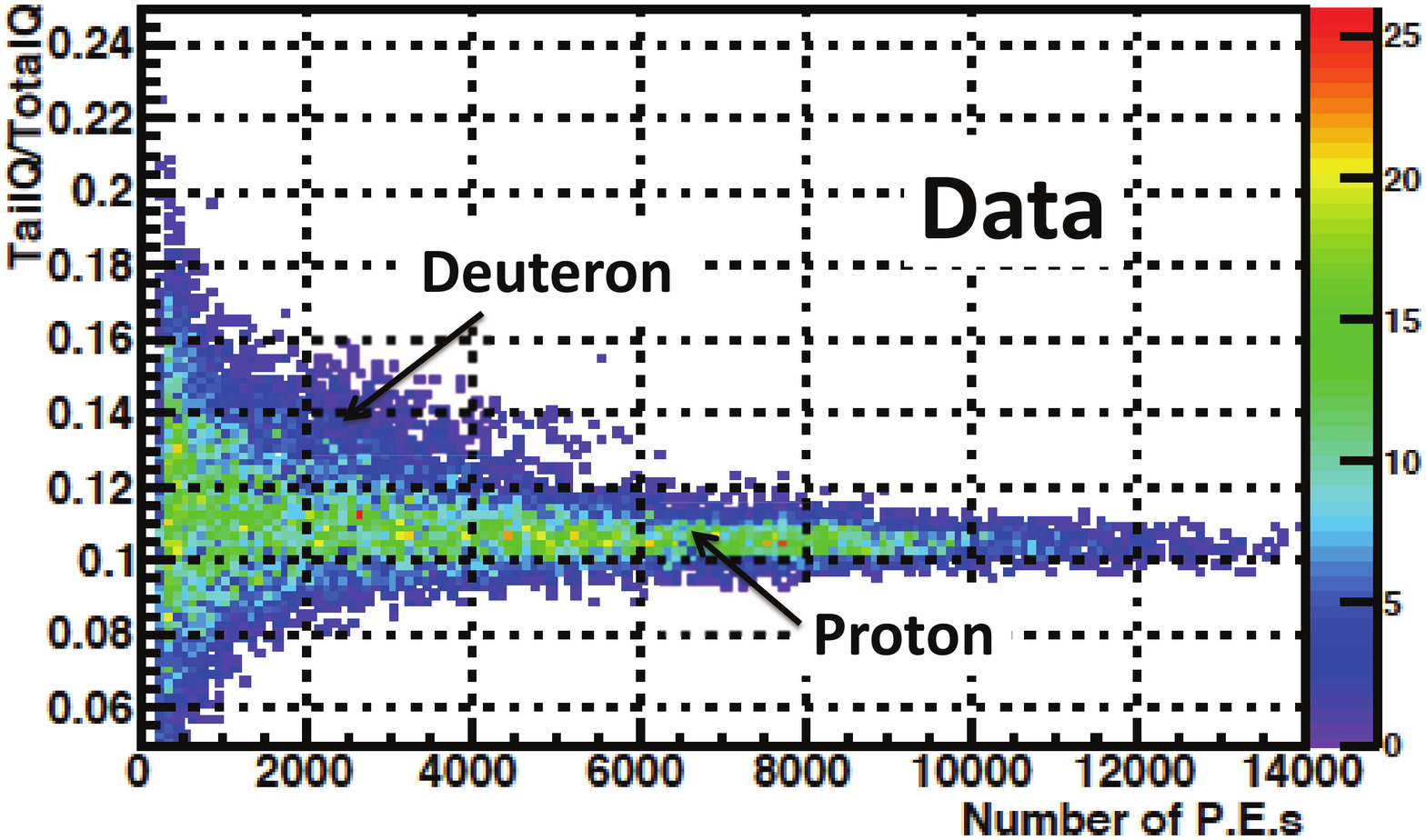}
\end{center}
\caption{\setlength{\baselineskip}{4mm}
Correlation between the Tail Q/Total Q and the visible energy of measured data.}
\label{PSDEnergy_Data}
\end{figure}

\begin{figure}[htb]
\begin{center}
\includegraphics[width=0.85 \textwidth]{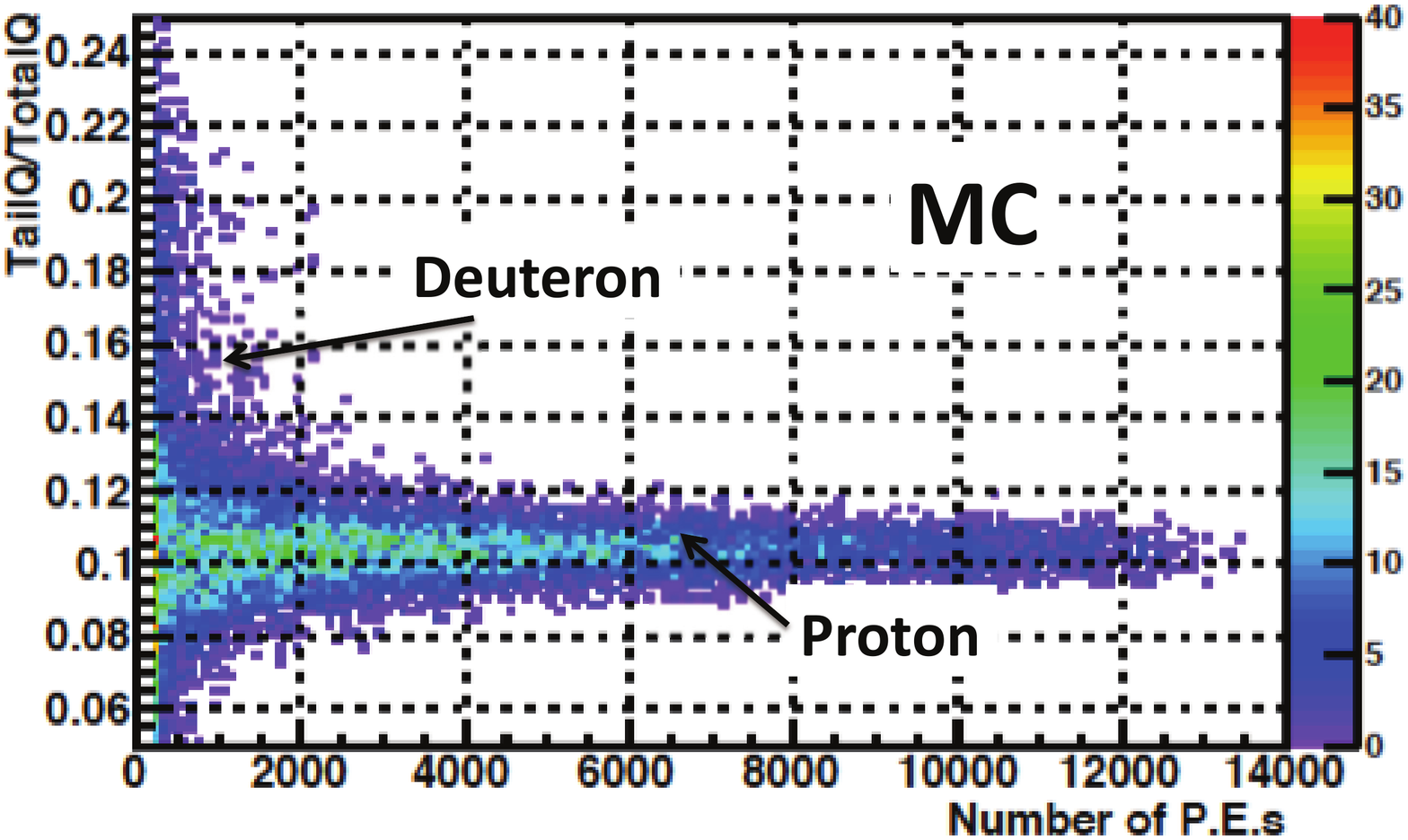}
\end{center}
\caption{\setlength{\baselineskip}{4mm}
Correlation between the Tail Q/Total Q and the visible energy in the MC samples. The recoil proton events are reproduced especially well.}
\label{PSDEnergy_MC}
\end{figure}

\begin{figure}[htb]
\begin{center}
\includegraphics[width=0.85 \textwidth]{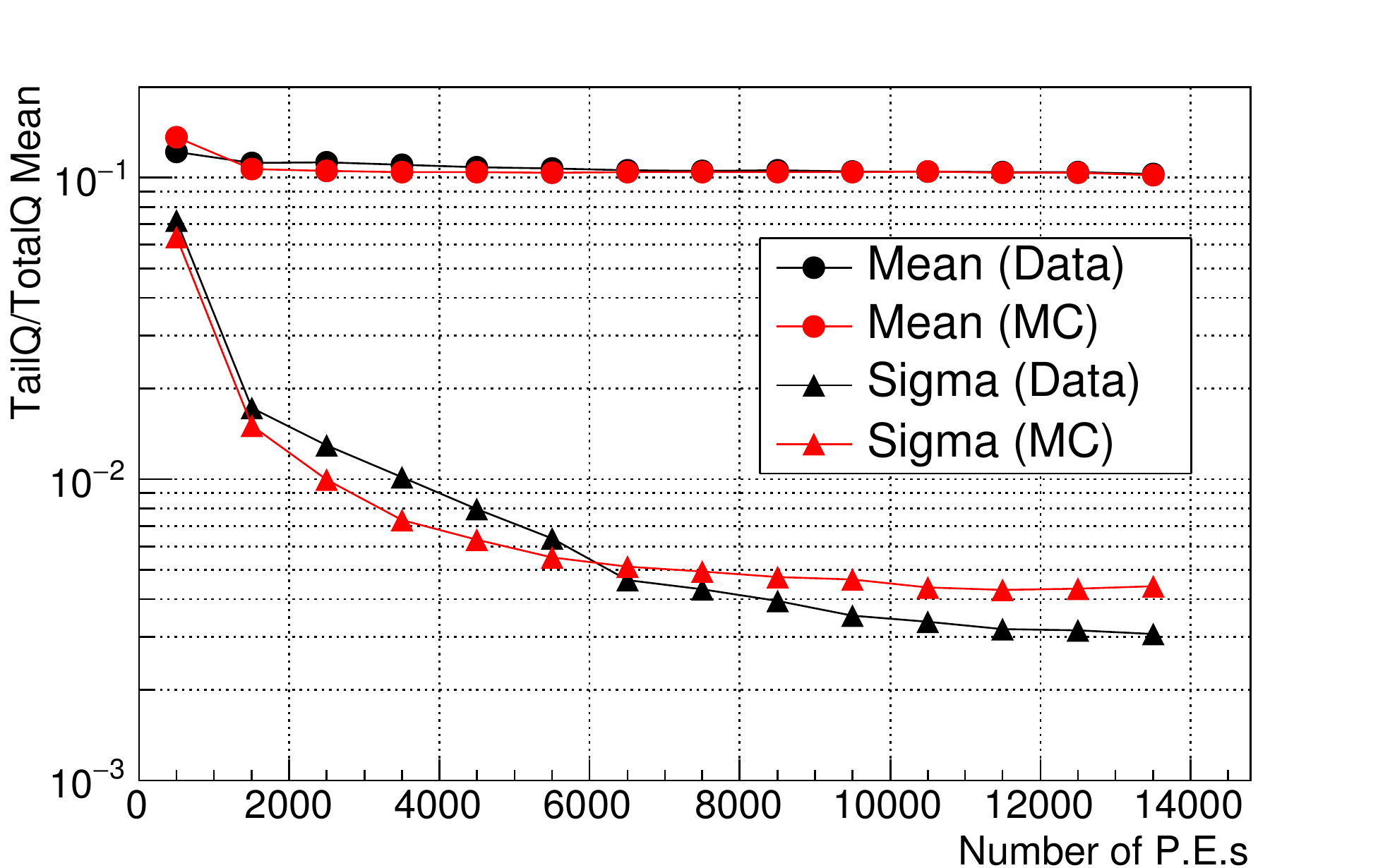}
\end{center}
\caption{\setlength{\baselineskip}{4mm}
Correlations between mean (RMS) of the TailQ/TotalQ and the visible energy for the measured data and the MC samples.}
\label{PSDMEANSIG}
\end{figure}

The measured data is compared to the MC samples produced by Geant4 which include the optical process and reconstruction of the waveform including electronics effects. The MC simulation was already tuned with radioactive sources of $^{60}$Co and $^{241}$Am$^{9}$Be before the test-beam. Figure~\ref{PSDDef} shows mean waveforms of the recoiled protons of measured data and MC, the measured waveform is reproduced well in MC.
Figures.~\ref{PSDEnergy_Data} and \ref{PSDEnergy_MC} show correlations between the TailQ/TotalQ and the visible energy of measured data and the MC samples, respectively. The TailQ/TotalQ definition is shown in Fig.~\ref{PSDDef}. 
Figure~\ref{PSDMEANSIG} shows correlations between the mean (RMS) of the TailQ/TotalQ distributions and the visible energy for the measured data and the MC samples. Though there is a small discrepancy between the RMS curves and the mean curves, the MC samples reproduce the measured data reasonably. The PSD performance of the scintillator when exposed to fast neutrons with several tens MeV in the Gd-LS is well-understood using the MC simulation.    

\begin{figure}[htb]
\begin{center}
\includegraphics[width=0.85 \textwidth]{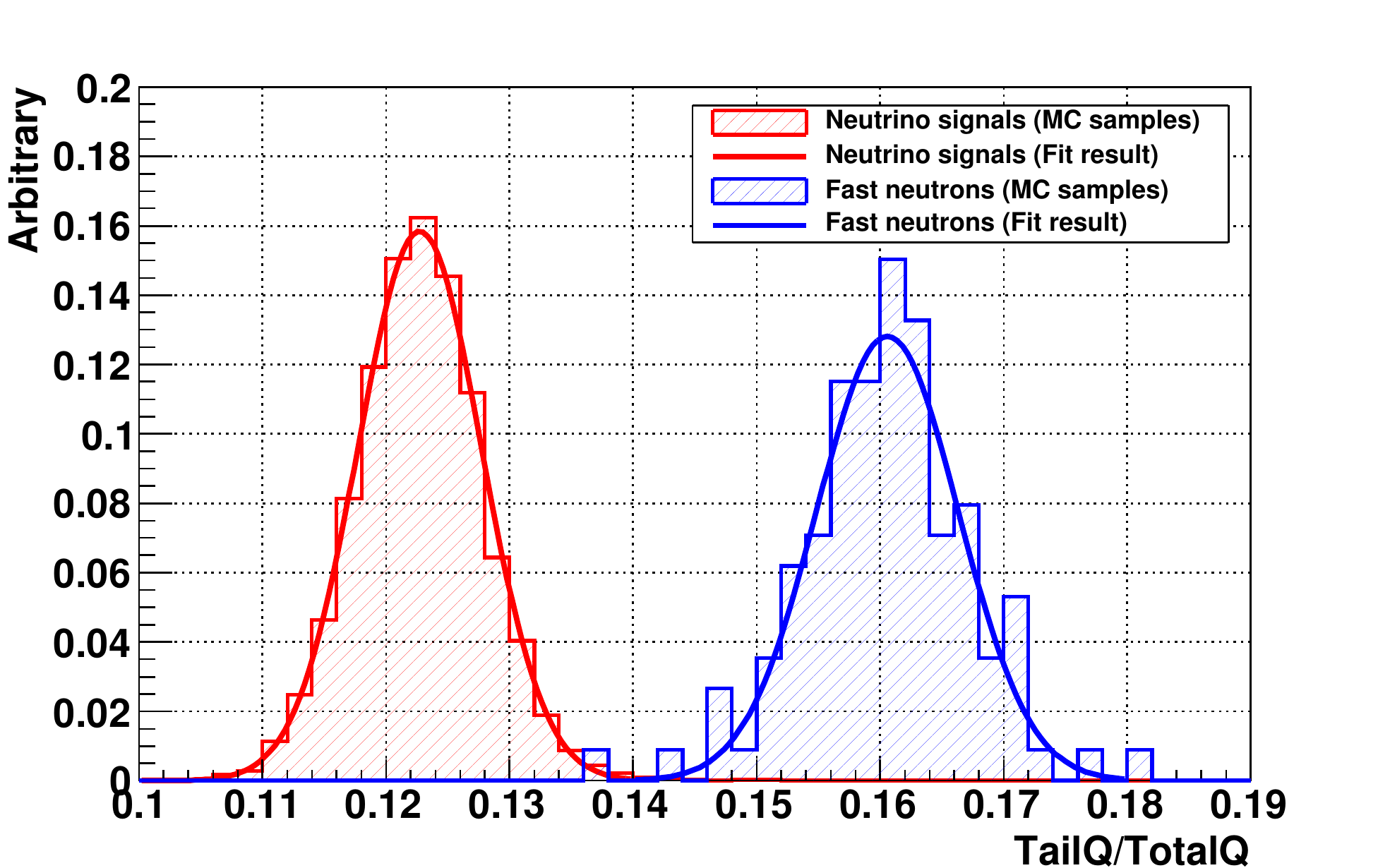}
\end{center}
\caption{\setlength{\baselineskip}{4mm}
The Tail Q/Total Q distributions for the neutrino signals (red) and the cosmic-induced fast neutrons (blue) after applying the usual neutrino selection criteria.}
\label{PSD1D}
\end{figure}

Finally, using the MC simulation, the PSD capabilities of the real JSNS$^{2}$ detector are checked. 
Figure~\ref{PSD1D} shows the Tail Q/Total Q distributions for the neutrino signals and the cosmic-induced fast neutrons after applying the usual neutrino selection criteria. The neutrino signals and the fast neutrons are generated with same methods as the previous status report \cite{CITE:SR_14NOV1}. The result indicates that the JSNS$^{2}$ detector has good enough PSD capability to distinguish the neutrino signals from the cosmic-induced fast neutrons, and most of fast neutron events can be rejected without reduction of the signal efficiency. For example,
if we cut the events by 0.135 in the horizontal axis in Fig.~\ref{PSD1D}, the
detection efficiency for the signal is 99$\pm 0.1 \%$.
Usually, the PSD capability of a large detector such as JSNS$^{2}$ detector becomes worse than the vial size test sample because the waveform containing the timing information from the scintillation emission is distorted by vertex reconstruction bias. In that case, the discriminating power offered by the Tail Q/Total Q distribution becomes worse and the PSD performance suffers. Such and effect is considered in this result.

\subsubsection{Selection criteria and the signal efficiency}
Table~\ref{TAB:SigEff} shows selection criteria for the neutrino signals and the signal efficiencies. A detailed explanation is given in \cite{CITE:SR_14NOV1}. The total signal efficiency is 38\%.

\begin{table}[ht]
\begin{center}
\begin{tabular}{|c|c|}\hline
Cut condition & Efficiency\\\hline
$1\le \Delta t _{prompt}\le 10 \mu s$ & 74\% \\\hline
$7\le E _{delayed}\le 12 MeV$ & 71\% \\\hline
$20\le E _{prompt}\le 60 MeV$ & 92\% \\\hline
$\Delta t _{delayed}\le 100 \mu s$ & 93\% \\\hline
$\Delta VTX _{prompt-delayed}\le 60cm$ & 96\% \\\hline
$\Delta VTX _{OB-delayed}\ge 110cm$ & 98\% \\\hline
Life Time $\le 11$ & 91\% \\\hline
PSD cut & $\sim$ 99\% \\\hline\hline
Total& 38\% \\\hline

\end{tabular}
\caption{\setlength{\baselineskip}{4mm}
Selection criteria for the neutrino signals and the signal efficiencies.}
\label{TAB:SigEff}
\end{center}
\end{table}

\subsection{Summary of 2014 Measurement}
\label{SEC:BKG2014}
\indent

To investigate the feasibility of the experiment, we measured the
background at the detector location using 500 kg of plastic scintillator in 2014.
The results were summarized in the status report and a PTEP publication ~\cite{CITE:SR_14NOV1,CITE:SR_14NOV2}, and therefore only the relevant parts are written in the TDR.

There are four categories for the backgrounds:
\begin{itemize}
\item For the IBD prompt region (20 $<$ E$_p$ $<$ 60 MeV, 1 $<$ T$_p$ $<$ 10 $\mu$s)
  \begin{itemize}
  \item Beam neutrons and gammas  
  \item Gamma rays and neutrons induced by cosmic rays.
  \end{itemize}
\item For the IBD delayed region (7 $<$ E$_d$ $<$ 12 MeV, T$_p < $T$_d$ $<$ 100 $\mu$s)
  \begin{itemize}
  \item Gammas and neutrons induced by beam
  \item Gammas and neutrons induced by cosmic rays
  \end{itemize}  
\end{itemize}  

\begin{figure}[h]
 \centering
 \includegraphics[width=0.8 \textwidth]{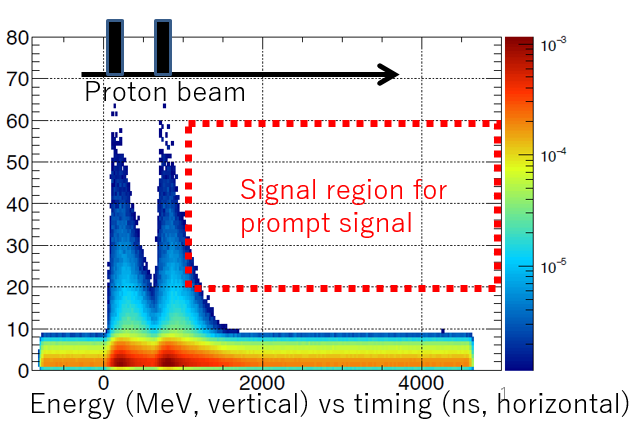}
 \caption{\setlength{\baselineskip}{4mm} 
   The measured energy vs timing w.r.t. the proton beam timing of neutral
   particles using 500 kg plastic
   scintillator~\cite{CITE:SR_14NOV1,CITE:SR_14NOV2}. }
 \label{fig:ET}
\end{figure}

The dominant background for the IBD prompt region comes from gammas induced by cosmic rays, not from the beam.
Figure~\ref{fig:ET} shows the energy (MeV) vs timing (ns) of the neutral particles (neutrons or gammas) around
the proton beam timing. As seen in the plot, there are no backgrounds
in the prompt (positron) signal region.

We also measured the neutral particle backgrounds induced by cosmic rays. 
As shown in the Fig. 11 of ~\cite{CITE:SR_14NOV1}, the MC simulation
reproduces the data well and the estimated background is shown in the table.

The delayed backgrounds are due to beam, especially, gammas from the
floor concrete under the detector induced by beam.
You can see the neutral particle
activity with less than 10 MeV in Fig.~\ref{fig:ET} as well.
To terminate the gammas, we plan to put 30 cm thick
iron plates (equivalent to 12.5cm thick lead blocks)
under the JSNS$^2$ detector which makes this background manageable.
The effect of the lead with thickness up to 10 cm was measured and is shown
in Appendix C of ~\cite{CITE:SR_14NOV1}. 
For this background, we also created a MC model for the gammas, and the energy
and relative rate as a function of lead thickness is well reproduced by the MC.

Thermal neutrons are not observed in the plastic scintillator, and the 
rejection power for the thermal neutrons coming from outside of the detector
within the veto region of the JSNS$^2$ detector is quite high
as shown in Fig. 17 in ~\cite{CITE:SR_14NOV1}. The rejection power is more than 10$^5$,
and therefore the effects of thermal neutrons are negligible.
The delayed background from neutrons
comes from beam neutrons which thermalize inside the detector.
The neutrons in the beam timing window can be tagged using LS
light, therefore the thermal neutron backgrounds from beam are significantly reduced
using the proton on-bunch timing activity inside the LS.

In summary, all dominant backgrounds were measured by a 500 kg detector and
were well reproduced by MC simulations which we implemented. The
simulations were used to estimate the backgrounds in the real JSNS$^2$
detector.  
As a result of the simulation, we found that all these backgrounds can be
controlled by the detector design presented here allowing us to perform the experiment effectively, and therefore the PAC recommended
stage-1 status after seeing the results in 2014.

\subsubsection{Revised Numbers from the Reference}
\label{sec:LATACC}
\indent

After submitting ~\cite{CITE:SR_14NOV1} to PAC, we found that the
gammas induced by cosmic rays  (Fig. 12 in \cite{CITE:SR_14NOV1})
can also be a prominent background in the IBD delayed region.
We discarded this due to the time window difference unintentionally.

The estimated number of gammas / spill / 100$\mu$s is
4.4$\times$10$^{-3}$/spill/100$\mu$s. This gives a larger accidental background
3.8 times higher than the rate given in ~\cite{CITE:SR_14NOV1}.

We use the latest estimation of the accidental background
in this TDR.


\clearpage

\subsection{Summary of Backgrounds}
\indent

Table~\ref{TAB:grandsum} shows a summary of the number of
background and signal events in the JSNS$^2$ experiment.
The baseline of the detector from the target is 24~meters and the
operation period is 5000h$\times$3 years for one 17-ton detector,
which includes the latest accidental background estimation described in section
~\ref{sec:LATACC}.
and it also shows
the 5000h$\times$5 years case with the 50~tons used in ~\cite{CITE:SR_14NOV1}.

The dominant background is $\overline{\nu}_{e}$ from $\mu^{-}$, which is 43,
while the signal is 62 in case of the best fit values of LSND, with a 17-tons detector and a 3-year measurement period (5000~h$\times$3~years).

\begin{table}[ht]
\begin{center}
\begin{tabular}{|c|c|c|c|}\hline
& Contents & this TDR & Reference\\
& & 1 detector & 50tons\\
& & 5000h$\times$3y & 5000h$\times$5y\\ \hline
&&& \\
&$sin^22\theta=3.0\times10^{-3}$& 87 &480\\
&$\Delta m^2=2.5eV^{2}$ & & \\ 
Signal&(Best fit values of MLF)&&\\ \cline{2-4}
&$sin^22\theta=3.0\times10^{-3}$&&\\
&$\Delta m^2=1.2eV^{2}$& 62 & 342\\
&(Best fit values of LSND)&&\\\hline\hline
&$\overline{\nu}_{e}$ from $\mu^{-}$&43&237\\\cline{2-4}
&$^{12}C(\nu_{e},e^{-})^{12}N_{g.s.}$& 3 &16\\\cline{2-4}
background&beam-associated fast n & $\le$2 & $\le$13 \\\cline{2-4}
&Cosmic-induced fast n& negligible &37\\\cline{2-4}
&Total accidental events & 20 & 32\\\hline
\end{tabular}
\caption{\setlength{\baselineskip}{4mm}
Summary of the event rate for 5000h$\times$3years for one 17-ton detector (middle) and the same for 5000h$\times$5years for the 50~ton case\cite{CITE:SR_14NOV1}(right). Note that the event rate for 5000h$\times$3years for one 17-ton detector (middle) includes the latest accidental background estimation described in section~\ref{sec:LATACC}
}
\label{TAB:grandsum}
\end{center}
\end{table}

\section{Sensitivity for the Sterile Neutrino Search}

\subsection{Fit method}
\indent

The binned maximum likelihood method is used for the analysis. The 
method fully utilizes the energy spectrum of the background and 
signal components, and thus the amount of the signal can be 
estimated efficiently.

\begin{figure}
\centering
\includegraphics[width=0.85 \textwidth]{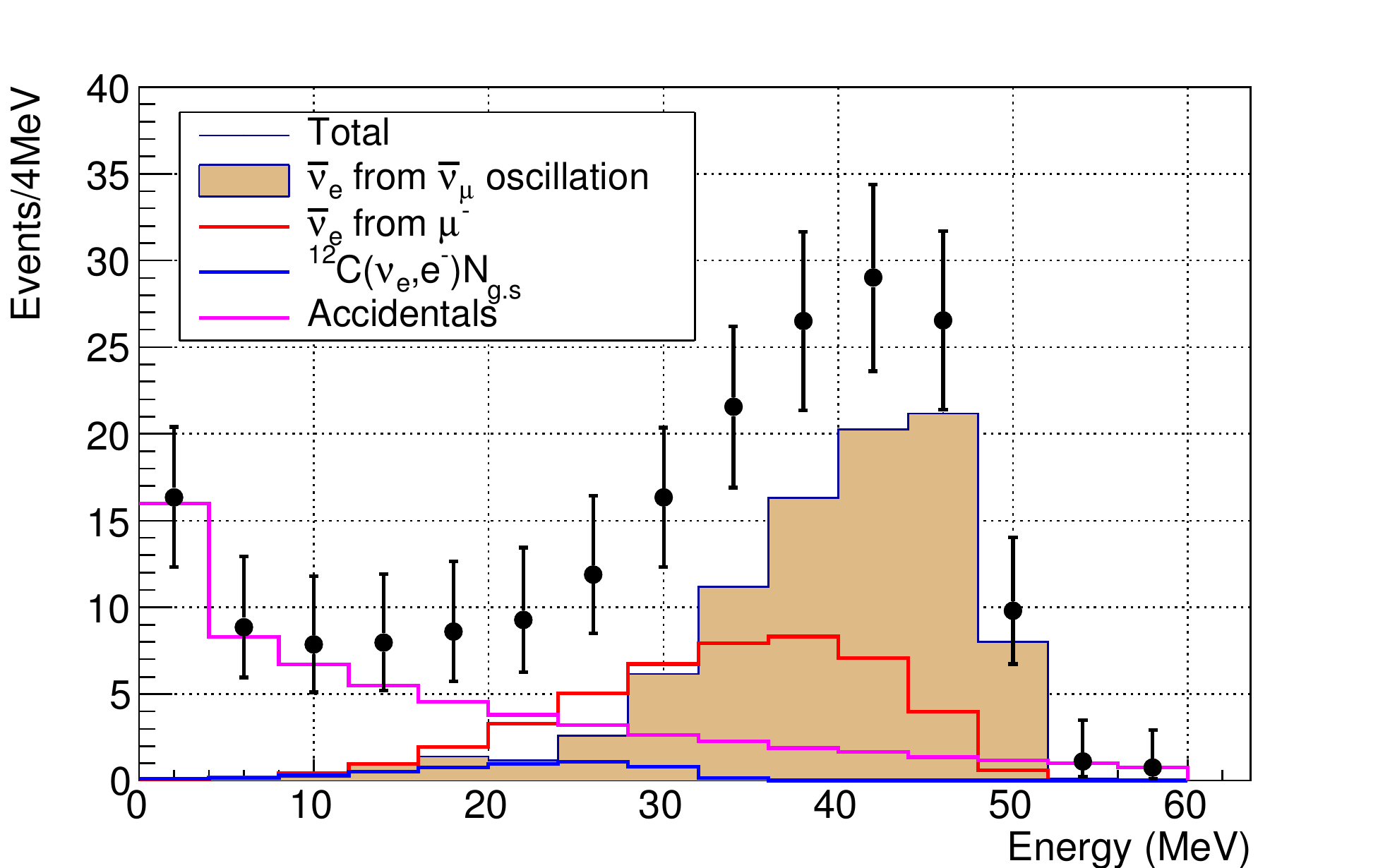}
\caption{\setlength{\baselineskip}{4mm}
Energy spectra of $\bar{\nu}_{e}$ from $\mu^{-}$ (red),$^{12}{C}(\nu_{e},e^{-})N$ (blue), accidentals (pink), and the oscillated signal with ($\Delta m^{2}$, $\sin^2 2\theta$) = (2.5, 0.003) (brown shaded; best $\Delta m^{2}$ for the MLF experiment) are shown. Black points with error bars correspond to the sum of the all components. All spectra of the neutrino signals and all backgrounds except for the beam neutrons include the effect of the energy resolution shown in Fig.\ref{ERESOLUTION}. 
}
\label{fig:tenergy1}
\end{figure}

\begin{figure}
\centering
\includegraphics[width=0.85 \textwidth]{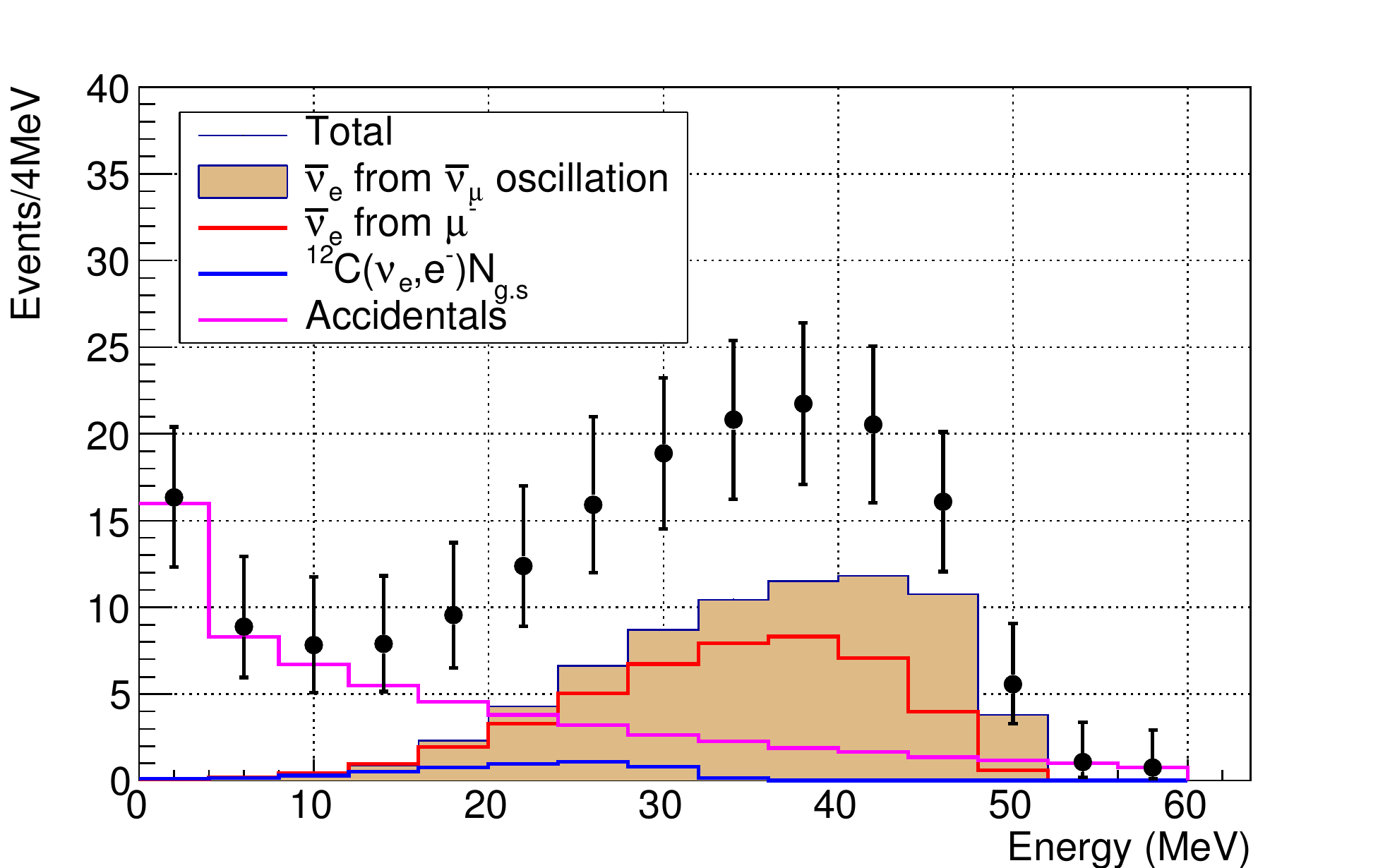}
\caption{\setlength{\baselineskip}{4mm}
Energy spectra of $\bar{\nu}_{e}$ from $\mu^{-}$ (red),$^{12}{C}(\nu_{e},e^{-})N$ (blue), accidentals (pink), and the oscillated signal with ($\Delta m^{2}$, $\sin^2 2\theta$) = (1.2, 0.003) (brown shaded; LSND best $\Delta m^{2}$ for the MLF experiment) are shown.
}
\label{fig:tenergy2}
\end{figure}

The typical energy spectrum from $\mu^{-}$ decay (blue), the oscillated signal with ($\Delta m^{2}$, $\sin^2 2\theta$) = (2.5, 0.003) (brown shaded; best $\Delta m^{2}$ case) and (1.2, 0.003) (LSND best fit case),
are shown in Figs.~\ref{fig:tenergy1} and \ref{fig:tenergy2}, respectively. 
Here we assume the fiducial mass of the detector is 17 tons,
1 MW beam power is available at the MLF, the detector is 24 m from the target, and three years of operation with 5000 hours 
of exposure each year is achieved. 
The signal detection efficiency is assumed to be 38$\%$ in Table~\ref{TAB:SigEff}. 
The number of events in each energy bin is
statistically small, therefore we use maximum likelihood instead of the usual 
minimum $\chi^{2}$ method. The fitter estimates 
the oscillation parameters by varying the size and shape of the brown histogram
to best reproduce the energy distribution of the black points. 

For this analysis, the following equation is used to compute the likelihoods:
\begin{eqnarray}
\label{Eq:likelihood}
L &=& \displaystyle \Pi_{i} P(N_{exp} | N_{obs})_{i} \\
P(N_{exp} | N_{obs}) &=& \frac{e^{-N_{exp}} \cdot (N_{exp})^{N_{obs}}}{N_{obs}!}
\end{eqnarray}
where, $i$ corresponds to $i$-th energy bin, $N_{exp}$ is expected 
number of events in $i$-th bin, $N_{obs}$ is number of observed
events in $i$-th bin. $i$ starts from 20 MeV and ends at 60 MeV
because the energy cut above 20 MeV is applied for the primary signal
as explained before. Note that $N_{exp} =  N_{sig}(\Delta m^{2}, \sin^{2}2\theta) + \displaystyle\sum N_{bkg}$, and $N_{sig}(\Delta m^{2}, \sin^{2}2\theta)$ is 
calculated 
using the two flavor neutrino oscillation equation shown before, 
$P(\bar{\nu_{\mu}} \rightarrow \bar{\nu_{e}}) = \sin^{2}2\theta \sin^{2}(\frac{1.27 \cdot \Delta m^{2} (eV^{2}) \cdot L (m)}{E_{\nu} (MeV)})$.

The maximum likelihood point gives the best fit parameters,
and 2$\Delta lnL$ provides 
the uncertainty of the fit parameters. As shown in the 
PDG~\cite{pdg}, we have
to use the 2$\Delta lnL$ for 2 parameter fits to determine the uncertainties
from the fit.   
 
\subsection{Systematic uncertainties}
\indent

Equation~\ref{Eq:likelihood} takes only statistical 
uncertainty into account, therefore the systematic uncertainties
should be incorporated in the likelihood.  
Fortunately, the energy spectrum of the oscillated signal and background
components are well known, thus the error (covariance) matrix of 
energy is not needed.
In this case, uncertainties on the overall normalization
of each component have to be taken into account, and this assumption
is a good approximation at this stage. 

In order to incorporate the systematic uncertainties, constraint terms
should be
added to Equation~\ref{Eq:likelihood} and the equation is changed as follows:

\begin{eqnarray}
L &=& [\Pi_{i} P( N_{exp}^{'} | N_{obs})_{i}] \times e^{- \frac{(1-f_{1})^{2}}{2 \Delta \sigma_{1}^{2}}} \times e^{- \frac{(1-f_{2})^{2}}{2 \Delta \sigma_{2}^{2}}} 
\label{Eq:likelihood2}
\end{eqnarray}\
\noindent
where $f_{j}$ are nuisance parameters to give the constraint term on the overall
normalization factors. $N_{exp}^{'} = f_{1} \cdot N_{sig} (\Delta m^{2}, \sin^{2}2\theta) + f_{2} \cdot N_{bkg}$. 
$\Delta \sigma_{i}$ gives the uncertainties on the normalization factors of
each component.

In this TDR, the profiling fitting method is used to treat the systematic 
uncertainties. This method is widely known as the correct fitting method.
The profiling method fits all nuisance parameters as well as 
oscillation parameters. 

As mentioned above, 
the flux of the $\bar{\nu}_{e}$ from $\mu^{-}$ decays around 
the mercury target has very poor constraints from external information.
For this situation, the uncertainty of this background
component is set to be 50$\%$ and the uncertainty of number of $^{12}{C}(\nu_{e},e^{-})N$ events is set to be 10$\%$. We neglect the contributions from the beam and the cosmic-induced neutrons in this study.

\subsection{Sensitivity for $\bar{\nu_{\mu}} \to \bar{\nu_{e}}$ oscillation}
\indent

\begin{figure}[h]
 \centering
 \includegraphics[width=0.9 \textwidth]{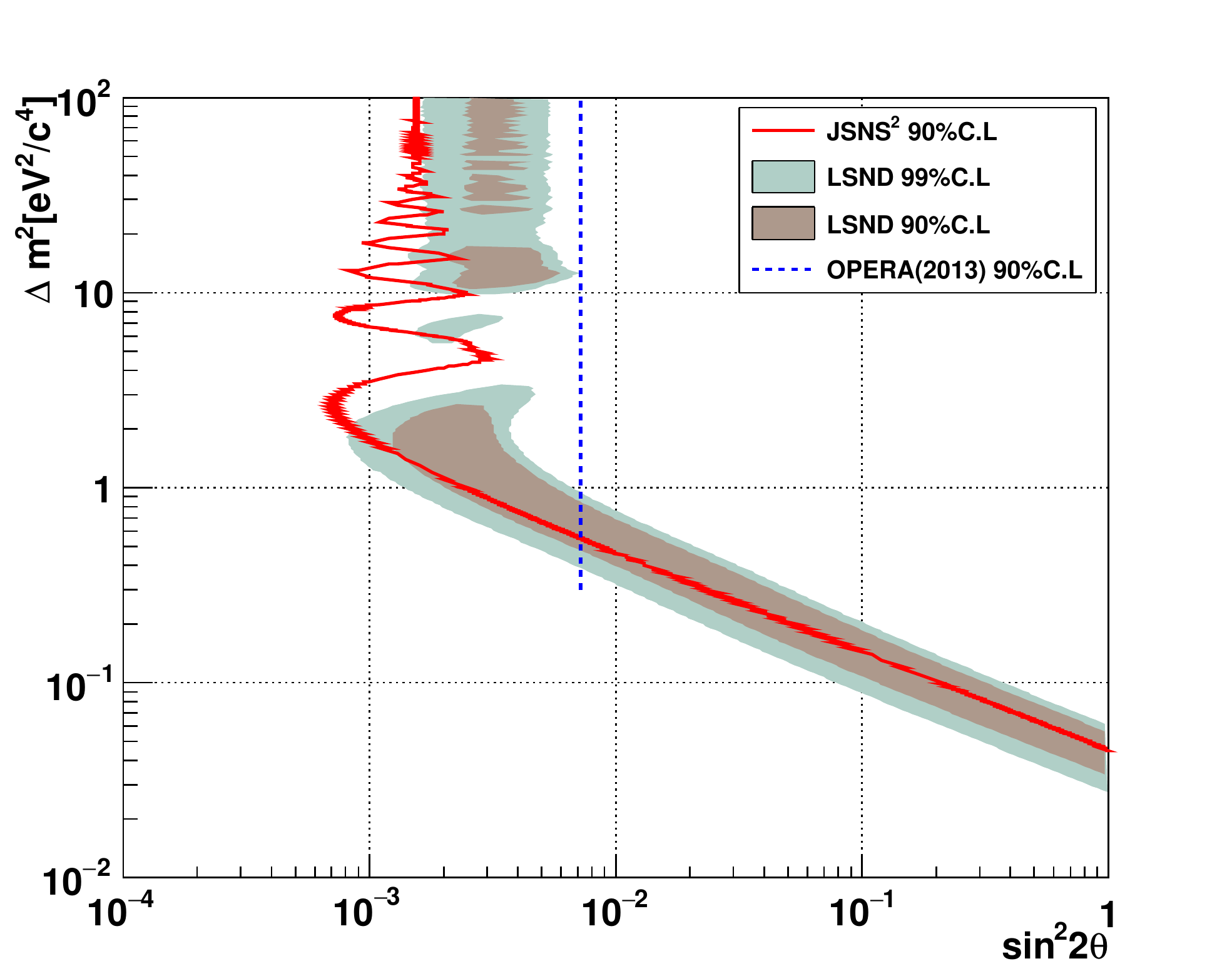}
\caption{\setlength{\baselineskip}{4mm}
Sensitivity of the JSNS$^2$ experiment with the latest configuration 
(1 MW $\times$ 3 years $\times$ 1 detector).
The red line shows the 90$\%$ C.L..  
The exclusion line of the OPERA experiment is also shown~\cite{CITE:OPERA}. 
The region to the right of the line is excluded with 90$\%$ confidence. 
}
 \label{FIG:senswP56}
\end{figure}

Figure~\ref{FIG:senswP56} shows 
the 90$\%$ C.L sensitivities with this condition. \\

We expect to have the preliminary result in 2021 because our
background and energy reconstruction uncertainty is small using the
$DAR$ neutrino flux and IBD signal.  The calibration
scheme is relatively straight-forward (with only 200 PMTs).
Compared to the world experiments, especially, the SBN program,
it is possible to have competitive results from JSNS$^2$.

\section{Summary}
\indent

The JSNS$^2$ experiment can provide timely, competitive results
in the sterile neutrino search via the $\bar{\nu_{\mu}} \to \bar{\nu_{e}}$
mode by utilizing the best existing facility (the J-PARC
MLF) and established detector techniques.
JSNS$^2$ is a direct test of LSND and can have a large impact on our current picture of neutrino physics. JSNS$^2$ can also provide measurements of other important physics processes including cross section measurements using neutrinos from $\mu$DAR and KDAR.

Most of the hardware components of the detector are reliable to use because they have been used in other reactor experiments, and are ready to order. They need about a half of year to be produced. We plan to produce the stainless tank and
the acrylic vessel at first in JFY2017, and plan to install the vessel, PMTs, the reflection sheet by the middle of JFY2018. Meanwhile, the Gd-LS and LS will be produced in Korean facility site, and the electronics will be produced by CAEN.
The timescale to start the experiment is around of end of JFY2018.   
In pursuit of these goals, the JSNS$^2$ collaboration continues to work hard to realize this important experiment.
 
\section{Requests to J-PARC PAC}
\indent

We request the stage-2 approval to the J-PARC PAC.

\section{Acknowledgements}
\indent

We warmly thank the MLF people, especially, MLF Division leader, the neutron source group, muon group and user facility group for the various kinds of supports.
This work is also supported by the JSPS grant-in-aids (Grant Number 16H06344, 16H03967), Japan, and the National Research Foundation of Korea (NRF) Grants 2009-0083526 and 2017K1A3A7A09015973.
Finally, we thank the continuous supports from J-PARC and KEK.

\newpage

\appendix
\section{Strength Calculation around the Stainless Tank}
\indent

In this section, the analysis for the strength calculation for the
stainless tank, the tank support structure, the lifting structure
during the earthquake as well as the static status.
This includes how to fix the tank to the building with the anchors.
Note that these calculations are done by the Mitsui-Zosen company,
and the cross checked between hand calculation and FEM analysis each other.

\subsection{Thickness of the Stainless Tank for the Static Operation}

\subsubsection{Wall}
\indent

Table~\ref{Tab:specSS} shows the inputs for the calculation for the stainless
tank.

\begin{table}[htbp]
\begin{center}
	\begin{tabular}{|c|c|}
	\hline
	Parameter Name     & Parameter   \\ \hline \hline
        Diameter (D)          & 4600 mm     \\
        Height   (H)          & 3500 mm     \\
        Maximum liquid pressure (P$_1$) & 0.35 kg/cm$^2$ \\
        Internal Pressure (P$_2$) & 1.0 kg/cm$^2$ \\
        Design Pressure (P) & 1.37 kg/cm$^2$ \\  
        Material           & SUS301      \\
        Maximum allowable stress of SUS301 ($\sigma_{a}$) & 13.1 kg/mm$^2$ \\
        Liquid inside tank & Water (for conservative calc.) \\
        Density of water            & 1g/cm$^3$ \\
        Efficiency of Welding ($\eta$) & 0.70 \\
        Corrosion factor (C)  & 0.0 \\
	\hline
	\end{tabular}
	\caption{Parameters of the inputs of the strength calculations.}
        \label{Tab:specSS}
\end{center}
\end{table}

The minimum thickness (t) of the stainless tank wall is given by
the following formulae;
\begin{equation}
  t = \frac{P \cdot D}{200 \sigma_a \cdot \eta - 1.2 P} + C 
\label{Eq:wall}
\end{equation}
and it is t = 3.4 mm. The minimum commercial (available) thickness of the
stainless sheet is
5.0 mm and this calculation contains the safety factors (4.3) due to the
liquid density (LAB 0.86 g/cm$^3$) and internal pressure of 1.0 kg/cm$^2$
(we will use N$_2$ gas with 0.0 level internal pressure but not in the
such high pressure). The thickness of 5.0 mm is safe with the safety factor of
6.3.

\subsubsection{Top}
\indent

In this calculation, we need another parameter, "curvature radius of the
top roof (flange) of the tank (R) ", which we assume 12,000 mm.
If we use similar equation as the Eq.~\ref{Eq:wall} as follows,
\begin{equation}
  t = \frac{P \cdot R}{200 \sigma_a \cdot \eta - 0.2 P_2 } + C 
\end{equation}

we obtain the minimum thickness of the top flange. It is 6.5 mm.
We use 10mm thickness of stainless for this part, and also the internal
pressure is almost 0.0, therefore the safety factor is very large.

\subsection{Strength of Stainless Tank during the Earthquake}

\subsubsection{Wall Part}
\indent

When the earthquake is occurred, the detector must not have buckling effects.
We assume 0.25G is loaded to the detector wall during the earthquake.
Note that 0.25G is the same assumption of the weight load as the other MLF
building materials during the earthquake.

The total weight of the stainless tank and the first "Anti oil-leak tank" is
about 58.5 tons, therefore the maximum horizontal weight
load during the
earthquake is 0.25$\times$58.5 = 14.6 tons.

The center of gravity for the height is 1.75 meter, then the
corresponding bending moment ($M_h$) is 25.6 ton $\cdot$ m.

The cross sectional second moment for the detector wall is \\
\begin{equation}
  I = \pi r_m^{3} t = 1.90 \times 10^{11} mm^{4}
\end{equation}
where, $t$ is the thickness of the detector wall (5mm) and $r_m^{3}$ is the
average radius of the detector (2297.5 mm).
The cross section coefficient ($Z = I/r_m$) is 8.29$\times$10$^{7} mm^3$.
\textbf{The expected stress from the bend compression ($\sigma_m = M_h/Z$)
  during the earthquake is 0.31 kg/mm$^2$}.   

On the other hand, the maximum compression stress from the theory is expressed
as:
\begin{equation}
  \sigma_{cr} = \frac{E}{\sqrt{3(1-\nu^{2})}} \cdot \frac{t}{r} \sim 0.6 E \frac{t}{r}
\end{equation}
where, $\nu$ is Poisson ratio (0.3 at this case) and $E$ is the elastic
coefficient which is 21000 kg/mm$^2$.
Then the maximum stress is 28.0 kg/mm$^2$, therefore this value is
much larger than the expected stress. \textbf{The detector is not broken
  during the earthquake.}

\subsubsection{Welding Condition to Prohibit Sliding of the Detector}
\indent

In this subsection, we consider the welding condition not to have sliding
between the detector tank and the support structure.

The maximum horizontal weight load during the earthquake as shown in the
previous subsection is 0.25$\times$68.5 = 17.1 tons.
Although the friction coefficient of the detector tank is 0.1$\sim$0.3,
here we assume all 17.1 tons acts as the sliding force because of the
safety reason.   

The welding part can stand up to 7.24 kg/mm$^2$ for shearing force at
most\footnote{This value is based on the Japanese law},
thus the needed welding area to prohibit the sliding is
17.1 tons / 7.24 kg/mm$^2$ = 2363.7 mm$^2$.

\begin{figure}[htbp]
 \centering
 \includegraphics[width=0.5 \textwidth]{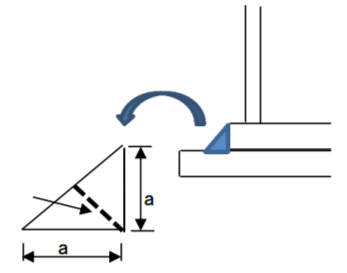}
 \caption{\setlength{\baselineskip}{4mm} 
   The length of "a" is assumed to be 5 mm, which is same as the thickness
   of the stainless sheet. In this case, corresponding resistance area
   against the shearing force is a$\times \cos \pi/4$ = 3.5mm.
}
 \label{fig:weld}
\end{figure}

According to Fig.~\ref{fig:weld}, the resistance area of against
the shearing force is a$\times \cos \pi/4$ = 3.5mm, therefore
the required length of the welding region is 2363.7 / 3.5 = 668.6 mm.
\textbf{Our welding design satisfies this requirements.}

\subsection{FEM calculation}
\indent

\begin{figure}[htbp]
 \centering
 \includegraphics[width=0.8 \textwidth]{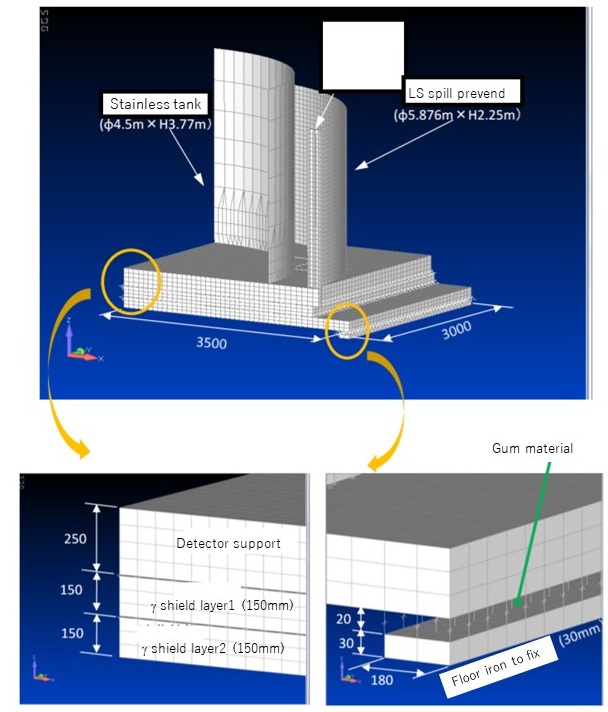}
 \caption{\setlength{\baselineskip}{4mm} 
   The setup of the FEM calculation. The quarter of the detector is
   shown for the simplicity hereafter unless it is noticed.
}
 \label{fig:FEMmodel}
\end{figure}

In this subsection, the results using the FEM calculation is shown.
Figure~\ref{fig:FEMmodel} shows the setup of the FEM calculation. 

Tables~\ref{Tab:material} and \ref{Tab:size} show
the parameter for the materials
and size of the relevant detector components.
\begin{table}[htbp]
\begin{center}
	\begin{tabular}{|c|c|c|c|c|}
	\hline
	Material &  Elastic coefficient & Poisson Ratio & density & comments \\
                 &  (N/m$^2$)           &               & (kg/m$^3$)& \\ \hline \hline
        Steel  & 2.06$\times 10^{11}$ & 0.3 & 7850 & \\
        Stainless  & 1.96$\times 10^{11}$ & 0.3 & 7930 & \\
        Gum  & 3.31$\times 10^{5}$ & -- & -- & modeled as the springs\\ 
	\hline
	\end{tabular}
	\caption{Parameters of materials in the FEM .}
        \label{Tab:material}
\end{center}
\end{table}
\begin{table}[htbp]
\begin{center}
	\begin{tabular}{|c|c|c|c|c|}
	\hline
	Component &  thickness & length & width & comments  \\ 
                  & (mm)       & (mm)   & (mm)  &           \\ \hline \hline
        $\gamma$ shield layer 2 & 150 & 7000 & 6000 & \\
        $\gamma$ shield layer 1 & 150 & 5876 & 6000 & \\
        iron sheet to fix the detector  & 30 & 6000 & 200 & \\
        Gum (on the iron sheet)  & 20 & 6000 & 200 & \\ \hline
        Support (top sheet) & 9.0 & 5876 & 5876 & \\
        Support (bottom sheet) & 9.0 & 5876 & 5876 & \\
        Support (H type beam) & 6.0 & 5876 & 250 & every 750 mm \\ \hline
        Stainless tank & 5.0 & -- & -- & 4.5m dia. $\times$ 3.77 height \\
        LS Spill Prevention & 4.5 & -- & -- & 5.876m dia. $\times$ 2.25 height \\
	\hline
	\end{tabular}
	\caption{Size of the detector components in the FEM .}
        \label{Tab:size}
\end{center}
\end{table}

\subsubsection{Static Operation}
\indent

The static operation case is considered at first. The maximum stress, the
maximum displacement and the maximum tilting are calculated by the
FEM analysis.
The expected load parameters are shown in Table~\ref{Tab:load}
during the static operation case, 
\begin{table}[htbp]
\begin{center}
	\begin{tabular}{|c|c|c|}
	  \hline
        Parameters & Weight load & comments \\ \hline \hline    
	Vertical Pressure &  34293 N/m$^2$ & liquid+top/bottom tank structure\\
        Material self-weight &  9.8 m/s$^2$ & inertia force due to mass \\ 
        Vertical pressure due to liquid & 15148 N/m$^2$ & average of height direction\\
	\hline
	\end{tabular}
	\caption{Weight Load Parameters}
        \label{Tab:load}
\end{center}
\end{table}

\begin{figure}[htbp]
 \centering
 \includegraphics[width=0.6 \textwidth]{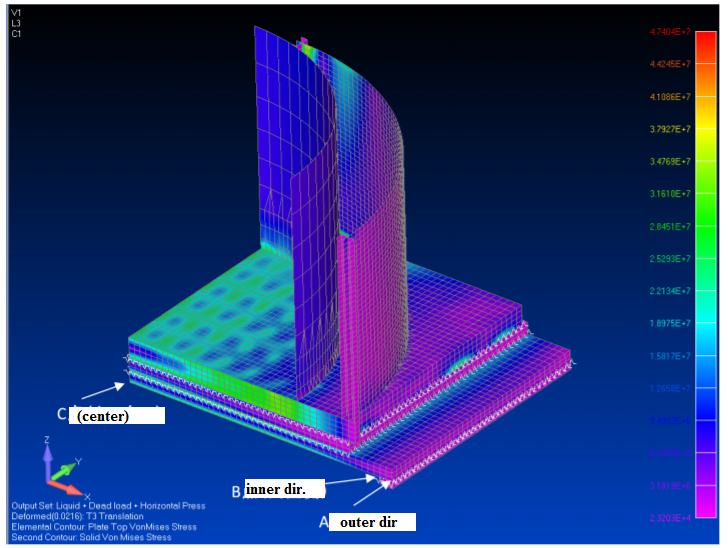}
 \caption{\setlength{\baselineskip}{4mm} 
   The results for bending and stress calculations.   
}
 \label{fig:FEMstat}
\end{figure}

Figure~\ref{fig:FEMstat} shows the results visually.

As expected, the central part of the detector is bent largely. For the
quantitative statements, Table~\ref{Tab:result1} shows the comparison
between the FEM results and the allowed parameters.
Note that allowances are based on the standard values from the Design.
("Kou-kouzou design standard values")
All numbers are safe within the standard design, therefore the
hand calculation is proved by the FEM analysis as well.

\begin{table}[htbp]
\begin{center}
	\begin{tabular}{|c|c|c|}
	  \hline
        Parameters & Output & Allowance Values  \\ \hline \hline    
	Maximum Stress  &  73.2 MPa & $<$158.6 MPa \\
        Maximum Displacement &  7.8 mm & -- \\ 
        Maximum Tilting & 1/410 & $<$ 1/300 \\
	\hline
	\end{tabular}
	\caption{\setlength{\baselineskip}{4mm}
        Comparison between FEM calculation and allowed parameters.
        Allowances are based on the standard values from the Design.
        ("Kou-kouzou design standard values')
        }
        \label{Tab:result1}
\end{center}
\end{table}

\begin{figure}[htbp]
 \centering
 \includegraphics[width=0.8 \textwidth]{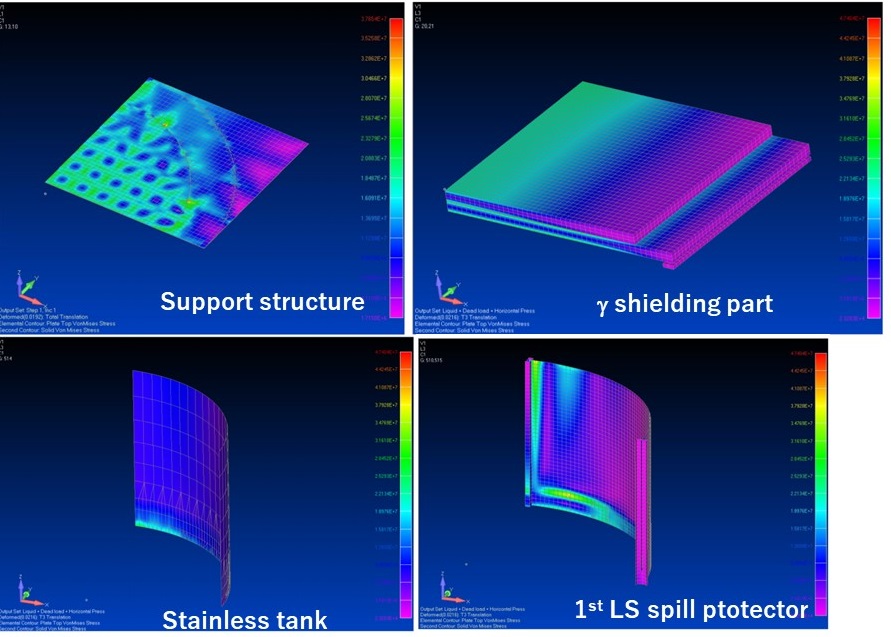}
 \caption{\setlength{\baselineskip}{4mm} 
 the results of bending and stress
for each detector components using the FEM analysis.   
}
 \label{fig:FEMcomp}
\end{figure}

Figure~\ref{fig:FEMcomp} shows the results of bending and stress
for each detector components using the FEM analysis.

\subsubsection*{Bearing Pressure for Concrete Floor}
\indent

The concrete floor has the allowed maximum bearing pressure.
It is 6000000 N/m$^3$, which is defined by the standard value of the concrete
design in Japan.
Thus, we have to check if this detector structure does not break the concrete
with the local weights.

\begin{figure}[htbp]
 \centering
 \includegraphics[width=0.8 \textwidth]{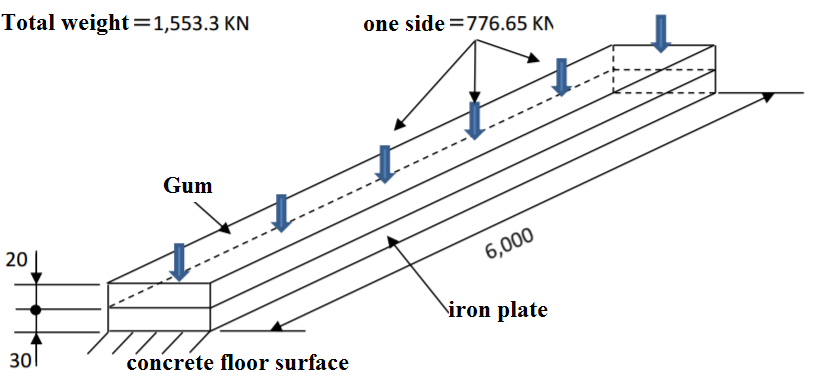}
 \caption{\setlength{\baselineskip}{4mm} 
   The structure to receive the detector weight made by iron plates and gums to
   reduce the local weight from the detector.
  }
 \label{fig:Iron}
\end{figure}

To reduce the local weight,
we have structure made by iron plates and gums
as shown in Fig.~\ref{fig:Iron}.
The size of the structure is already shown in Table~\ref{Tab:size}.

Using the structure, the FEM calculation shows that the average force
to the floor concrete is small enough compared to the allowance by factor
of more than 200.
\textbf{The concrete is not broken using the structure.}

\subsubsection{Earthquake case}
\indent

In this subsection, we calculate the detector's own endemic oscillation mode
including whole supports, the $\gamma$ shields, the "anti oil-leak tank"
using FEM analysis.
The details are not shown here, but the conclusions from the FEM
analysis are as follows:
\begin{itemize}
\item The vertical vibration affects to the detector dominantly. Effects
  from horizontal directional waves are negligibly small.
\item The short cycle period of the earthquake wave ($>$7.85 Hz =
  cycle period $<$ 0.127 sec) only affects to
  the detector. Effects from other frequency regions are also
  negligibly small.
\end{itemize}
This is because the detector support, $\gamma$ shields and iron plates
structure has low height weight center, and the wide strcuture for
the horizontal directions.

\begin{figure}[htbp]
 \centering
 \includegraphics[width=0.9 \textwidth]{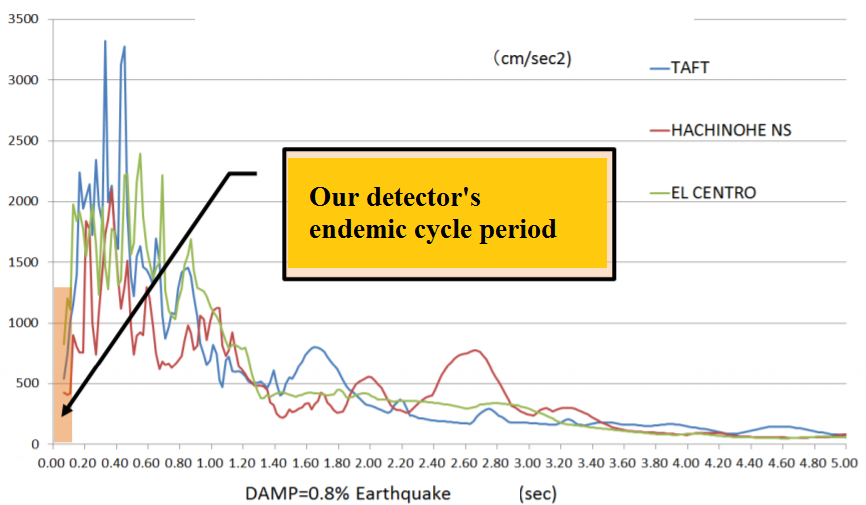}
 \caption{\setlength{\baselineskip}{4mm} 
The endemic oscillation cycle period of our detector (orange region),
and the cycle period spectrum of the typical earthquakes
(Taft, El Centro, Hachinohe). Compared to the earthquake spectrum peaks, our
detector's endemic oscillation cycle period is very low. 
  }
 \label{fig:EQ}
\end{figure}

Figure~\ref{fig:EQ} shows the endemic oscillation cycle period of our detector,
and the cycle period spectrum of the typical earthquakes
(Taft, El Centro, Hachinohe). Compared to the earthquake spectrum peaks, our
detector's endemic oscillation cycle period is very low. 
\textbf{We conclude that the usual 0.25G static force to the vertical direction
only is good enough to consider the earthquake case from the result.}

\subsubsection*{Comparison the Results and Allowed Values}
\indent

Based on the discussion in the previous subsection, we only need to compare
static force and the maximum allowance values.
For the safety reason, we also compare the horizontal directional case.

Table~\ref{Tab:EQ2} shows the judgements for the earthquake cases.
As expected, \textbf{the earthquake does not provide any breaking of the
JSNS$^2$ detector.}
\begin{table}[htbp]
\begin{center}
	\begin{tabular}{|c|c|c|}
	  \hline
        Parameters & Output & Allowance Values  \\ \hline \hline    
	Maximum Stress  &  91.5 MPa & $<$237.9 MPa \\
        Maximum Displacement &  9.75 mm & -- \\ 
        Maximum Tilting & 1/320 & $<$ 1/200 \\
	\hline
	\end{tabular}
	\caption{\setlength{\baselineskip}{4mm}
        Comparison between the maximum allowance parameters and the
        extreme stress during the earthquake, 
        Allowances are based on the standard values from the Design.
        ("Kou-kouzou design standard Values")
        }
        \label{Tab:EQ2}
\end{center}
\end{table}

\subsubsection{How to fix the Detector to the Floor}
\indent

\begin{figure}[htbp]
 \centering
 \includegraphics[width=0.8 \textwidth]{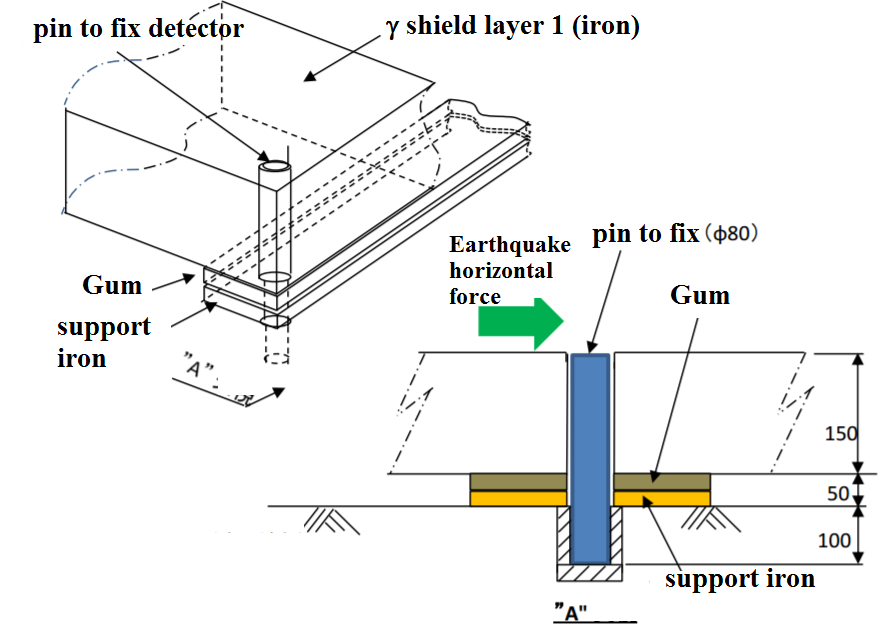}
 \caption{\setlength{\baselineskip}{4mm} 
The structure to fix the detector, which consists of iron, gum and pins.
Basically, pins stop the sliding the detector during the earthquake.
  }
 \label{fig:EQ_support}
\end{figure}

In this subsection, how to fix the detector is described.
Figure~\ref{fig:EQ_support} shows the structure to fix the detector,
which consists of iron, gum and pins.
Basically, pins stop the sliding the detector during the earthquake.

To calculate the strength of the pins, following assumptions are made:\\
(1) 0.25G is assumed for the earthquake. \\
(2) The friction coefficient between iron plates is assumed to be 0. (this
   include large safety factors because the real coefficient is 0.2$\sim$0.3).\\
(3) The friction coefficient between iron and gum is more than 0.7, therefore
   we assume the iron is not moved in this attached areas. \\

   Table~\ref{Tab:EQ3}
   shows the strength of the pins to fix the detector (including
   the detector support, the $\gamma$ shield layers, gums, the support
   structure to fix made by iron.)
   All pins with this design satisfy the requirements.
\begin{table}[htbp]
\begin{center}
	\begin{tabular}{|c|c|c|c|c|c|c|}
	  \hline
          Parts & Weights & Horizontal & $\#$ of & Diameter & Shearing & Tolerance  \\ 
          &  & Force & pins & of pin & Force  &  \\ 
          & (kg)    & (kg)  &    & (mm) & (kg/cm$^2$) & (kg/cm$^2$) \\ \hline \hline 
	Detector/$\gamma$ shield 1 &  68,243 & 17,061 & 4 & 60 & 150.8 & 1086 \\
        $\gamma$ shield 1 / 2 &  41,514 & 27,439 & 4 & 70 & 178.2 & 1086 \\
        $\gamma$ shield 2 / Gum &  49,455 & 39803 & 8 & 80 & 99.0  & 1086 \\  
	\hline
	\end{tabular}
	\caption{\setlength{\baselineskip}{4mm}
   The strength of the pins to fix the detector (including
   the detector support, the $\gamma$ shield layers, gums, the support
   structure to fix made by iron.)
        }
        \label{Tab:EQ3}
\end{center}
\end{table}

\begin{figure}[htbp]
 \centering
 \includegraphics[width=0.8 \textwidth]{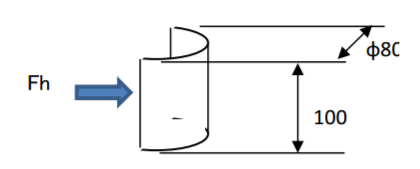}
 \caption{\setlength{\baselineskip}{4mm} 
  The force from the pins to the hole in the concrete floor. 
 }
 \label{fig:pin_concrete}
\end{figure}

The bearing pressure of the concrete to fix the pins is also calculated.
As shown in Fig.~\ref{fig:pin_concrete}, the force is transported from the
pins to the concrete hole.  
The bearing pressure distribution of the concrete which
receives the horizontal act force (F$_h$) is distributed as cosine shape.
The maximum bearing pressure ($\sigma$) is calculated as;
\begin{equation}
  \sigma = F_h/(D/h) = 62.2 kg/cm^2
\end{equation}
where, $F_h$= 4975 kg, D = 80 mm, h = 100 mm.

The tolerance of the concrete for the bearing pressure is
91.8 kg/cm$^2$, therefore this design is permitted.

\newpage

\end{document}